\newcommand{\slh}{\!\!\!\slash}

\documentclass[twocolumn,showpacs,preprintnumbers,amsmath,amssymb,letterpaper]{revtex4}
\usepackage{epsf}

\begin{document}

\title{
Continuous quantum phase transitions beyond Landau's paradigm\\
in a large-$N$ spin model
}
\author{Ying Ran}
\author{Xiao-Gang Wen}
\homepage{http://dao.mit.edu/~wen}
\affiliation{Department of Physics, Massachusetts Institute of Technology,
Cambridge, Massachusetts 02139}
\date{August 2006}

\begin{abstract}
We study a large-$N$ generalization of $J_1$-$J_2$ Heisenberg model on square
lattice -- an $Sp(2N)$ spin model.  The possible quantum spin liquid phases of
the $Sp(2N)$ model are studied using the $SU(2)$ projective construction.
We find several spin liquid states at least in large $N$ limit, which include
$SU(2)$ $\pi$-flux state, $SU(2)$ chiral spin state and $Z_2$ spin liquid
states.  All those spin liquid states have non-trivial quantum orders. 
We show how projective symmetry group, which characterizes quantum order,
protects the stability of even gapless spin liquids.
We also study the continuous quantum 
phase transition from the $SU(2)$ $\pi$-flux state to the $SU(2)$
chiral spin liquid state, and from the $SU(2)$ $\pi$-flux state to the 
$Z_2$ spin
liquid state.  We show that those phase transitions are beyond the
paradigm of Landau symmetry-breaking theory.  The first phase transition,
although a $Z_2$ symmetry breaking transition, contains critical exponents that
are different from those obtained from the Ginzburg-Landau theory of Ising
universality class.  The second transition does not even involve a change of
symmetry and has no symmetry breaking order parameter.
\end{abstract}

\maketitle

\section{Introduction}

The $SU(2)$ projective construction (or, more generally, slave-boson theory
\cite{BZA8773,BA8880}) for 2D spin
liquids in spin-1/2 systems has been studied over ten years
\cite{AZH8845,DFM8826}.  Upon doping, the $SU(2)$ projective construction
provides a quite complete theory for underdoped high $T_c$ superconductors
\cite{WLsu2,LNNWsu2}.  The theory explains the strange Fermi surfaces
\cite{WLsu2,LNNWsu2}, the strong $(\pi,\pi)$ spin fluctuations
\cite{KL9930,RWspin}, and the temperature dependence of superfluid density
\cite{LW9711,WL9893} for underdoped samples.

At half filling, the $SU(2)$ projective construction can also be used to
describe various spin liquids.  In fact, the $SU(2)$ projective construction
(or more generally, the slave boson theory) have predict many different spin
liquids, such as the algebraic spin liquid \cite{MA8938,Wqoslpub,RWspin},
chiral spin liquid \cite{KL8795,WWZcsp}, $Z_2$ spin liquids
\cite{RS9173,Wsrvb,MF9400,SF0050}.  It was also shown that the $SU(2)$
projective construction is capable of describing hundreds of different spin
liquids that have the same symmetry but different quantum orders
\cite{Wqoslpub}.

The above predictions of spin liquids and the classification of spin liquids
were based on mean-field calculations.  At the mean-field level, it is not
very hard to design a spin Hamiltonian that realizes a spin liquid that has
one of a few hundreds quantum orders. It is also not hard to find the
mean-field ground state for a given spin Hamiltonian.  The real issue is
whether we should trust the mean-field results.  It was argued
\cite{Wsrvb,Wqoslpub} that, if the obtained mean-field ground state is
unstable (i.e. if the mean-field fluctuations cause diverging interactions at
low energies), then the mean-field result cannot be trusted and the mean-field
state does not correspond to any real physical spin state.  It was also argued
that, if the mean-field ground state is stable (if the mean-field fluctuations
cause vanishing interactions at low energies), then the mean-field result can
be trusted and the mean-field state does correspond to a real physical spin
liquid state.

However, the above statement about stable mean-field states is too optimistic.
A `stable mean-field state' does not have diverging fluctuations at low
energies. So it does not have to be unstable. On the other hand, it does not
have to be stable either.  This is because short-distance fluctuations, if
strong enough, can also cause phase transitions and instabilities. Therefore,
in order for a mean-field result to be reliable, the mean-field state must be
stable ({\it i.e.} no infrared divergence) \emph{and} the short-distance
fluctuations must be weak.  As we do not have any small parameters in the
$SU(2)$ slave-boson theory for spin-1/2 systems, the short-distance
fluctuations are not weak, even for stable mean-field states. Because of this,
it is not clear if the mean-field results, even for the stable states, can be
applied to the spin-1/2 model or not.

In this paper, we will generalize the spin-1/2 model to a large-$N$
model.  The large-$N$ model can be solved approximately using the  $SU(2)$
projective construction.  We will show that the $SU(2)$ projective
construction for the large-$N$ model have weak short-distance fluctuations.
Thus, the stable $SU(2)$ mean-field states for the large-$N$ model do
correspond to real physical spin liquid states.  The $SU(2)$ mean-field
results for the stable states, such as fractionalization, emergent gauge
structures and emergent Fermi statistics, can be applied to the large-$N$
model.

We concentrate on a model with nearest-neighbor $J_1$ coupling and
next-nearest-neighbor $J_2$ coupling.  We find several mean-field phases that
include $SU(2)$ $\pi$-flux state, $SU(2)$ chiral spin state and $Z_2$ spin
liquid state.

The $SU(2)$ $\pi$-flux state is described by a low energy effective theory that
includes gapless Dirac fermions coupled to $SU(2)$ gauge fields.  Due to the
non-vanishing interaction between the fermions and the gauge bosons down to zero
energy, there is no free fermioninc or bosonic low energy quasiparticles in the
$SU(2)$ $\pi$-flux state.  Despite this, we show that the $SU(2)$ $\pi$-flux
state is a stable spin liquid state. It is a realization of algebraic spin
liquids. \cite{Wqoslpub,RWspin} 

The $SU(2)$ chiral spin state and
$Z_2$ spin liquid state are both gapped and, thus, naturally stable.
Both state carry non-trivial topological orders.

Within our $J_1$-$J_2$ model, the quantum transition (i.e. the zero-temperature
transition) between the $SU(2)$ $\pi$-flux state and the $SU(2)$ chiral spin
state turns out to be a continuous transition. The transition breaks the
time-reversal symmetry and has a well defined $Z_2$ order parameter.  However,
we show that the critical properties of the transition are not described by the
3D Ising universality class of the Ginzburg-Landau theory.  This is because the
transition not only break the $Z_2$ symmetry, it also changes the
quantum/topological order.  The continuous quantum transition between the
$SU(2)$ $\pi$-flux state and the $SU(2)$ chiral spin state is a new class of
quantum transition.

It has been shown that continuous quantum transitions are possible between two
states with the same symmetry (but different topological orders).
\cite{WWtran,CFW9349,SMF9945,RG0067,Wctpt} The  continuous quantum transitions
are also possible between two states with the \emph{incompatible}\footnote{If
the symmetry group of one phase is not a subgroup of the other phase and vice
versa, then the two phases are said to have incompatible symmetries.  According
to Landau's symmetry breaking theory, two phases with incompatible symmetries
cannot have continuous phase transition between them.}
symmetries.\cite{SVB0490}.  In this paper, we show that even symmetry-breaking
transitions with well defined order parameters, sometimes are not described
by Landau's symmetry breaking theory.  So it appear that most quantum
continuous transitions are not described by Ginzburg-Landau theory, regardless
if they have symmetry breaking and order parameter or not.

We also study the transitions from the $SU(2)$ $\pi$-flux state to a
$Z_2$-linear spin liquid state, and from the $SU(2)$ $\pi$-flux state to a
$U(1)$-linear spin liquid state.  The transtion between the $SU(2)$ $\pi$-flux
state to the $Z_2$-linear state is found to be a continuous quantum transition
that does not break any symmetry.  The transtion between the $SU(2)$
$\pi$-flux state to the $U(1)$-linear state, on the other hand, is  a
continuous quantum transition that breaks lattice rotaion and translation
symmetry. What is surprising is that the two seemingly very different
transitions are described by the same critical point with the same set of
critical exponents.


\section{Formulation of $SU(2)$ projective construction}

First we would like to briefly review the $SU(2)$ projective construction.  We
will mainly follow the notation of Ref. \cite{Wqoslpub,Wen04}. Let us start
with spin-1/2 Heisenberg Model, and introduce the fermion representation of
spins:
\begin{align}
\label{heisenberg}
H&=\sum_{\<\v i\v j\>}J_{\v i\v j}\,\mathbf{S}_{\v i} \cdot \mathbf{S}_{\v j}\\
\mathbf{S}_{\v i}&=\frac{1}{2}f_{\v i\alpha}^{\dag}\boldsymbol{\sigma}_{\alpha\beta}
f_{\v i\beta}
\end{align}
If we regard $H$ as an operator acting on the fermion Hilbert space, then we
have enlarged the Hilbert space. Therefore we have to add extra constraints to
reduce the enlarged Hilbert space to the original one for the spin system. The
constraints are the one-particle per site constraints:
\begin{align}
f_{\v i\alpha}^{\dag}f_{\v i\alpha}&=1\label{E:con}\\
\Rightarrow f_{\v i\alpha}f_{\v i\beta}\epsilon_{\alpha\beta}&=0,\quad
f_{\v i\alpha}^{\dag}f_{\v i\beta}^{\dag}\epsilon_{\alpha\beta}=0\label{E:extracon}
\end{align}
The two extra constraints in Eq.~(\ref{E:extracon}) are results of the first
one in Eq.~(\ref{E:con}). However in path integral formalism, if we enforce all
constraints simultaneously by introducing some Lagrangian multipliers, a
lattice gauge theory with $SU(2)$ gauge group can be derived. Let us
introduce some notations:
\begin{align}
\hat\eta_{\v i\v j}^{\dag}&=f_{\v i\alpha}^{\dag}\epsilon_{\alpha\beta}f_{\v j\beta}^{\dag}\\
\hat\chi_{\v i\v j}&=f_{\v i\alpha}^{\dag}\delta_{\alpha\beta}f_{\v j\beta}.\label{Eq:chieta}
\end{align}
These are the only singlet bilinear forms of the pairing between
site $i$ and $j$. 
After some rearrangements, one has:
\begin{align*}
\mathbf{S}_{\v i}
\cdot\mathbf{S}_{\v j}=
-\frac{1}{4}\hat\eta_{\v i\v j}^{\dag}\hat\eta_{\v i\v j}
-\frac{1}{4}\hat\chi_{\v i\v j}^{\dag}\hat\chi_{\v i\v j}+\frac{1}{4}.
\end{align*}
Ignoring the irrelevant constant, the path integral Lagrangian of
the Heisenberg model Eq.~(\ref{heisenberg}) turn out to be:
\begin{align*}
L'&=\sum_{\v i}f_{\v i\alpha}^\dagger i\partial_t f_{\v i\alpha}\notag\\
&-\sum_{\v i}\left(\frac{1}{2}a_{0,\v i}^{-}\hat \eta_{\v i\v i}^{\dagger }
+\frac{1}{2}a_{0,\v i}^{+}\hat \eta_{\v i\v i}
+\frac{1}{2}a_{0,\v i}^{3}(\hat \chi_{\v i\v i}^\dagger -1)\right)\notag\\
&+\frac{1}{4}\sum_{\<\v i\v j\>}J_{\v i\v j}
\left(\hat \eta_{\v i\v j}^{\dagger }\hat \eta_{\v i\v j}
+\hat \chi_{\v i\v j}^\dagger \hat\chi_{\v i\v j}\right)
\end{align*}
The constraint of one particle per site
has been encoded by the Lagrangian multipliers in the second line:
\begin{align*}
\left( \begin{array}{c}
a_{0,\v i}^{-}\\
a_{0,\v i}^{+}\\
a_{0,\v i}^{3}
\end{array}\right)=
\left( \begin{array}{c}
\frac{1}{2}(a_{0,\v i}^1-ia_{0,\v i}^2)\\
\frac{1}{2}(a_{0,\v i}^1+ia_{0,\v i}^2)\\
a_{0,\v i}^3
\end{array}\right)
\end{align*}
If we do a particle-hole transformation of the spin-down fermions
$f_{\v i\downarrow}$
together
with a Hubbard-Stratonovich transformation, the Lagrangian can be written
in a form with the explicit $SU(2)$ gauge invariance:
\begin{align}
L&=\sum_{\v i}\left[\psi_{\v i}^\dagger(i\partial_t-a^l_{0,\v i}\tau^l)\psi_{\v i}\right]
\notag\\\
&-\sum_{\<\v i\v j\>}\frac{1}{4}J_{\v i\v j}\left[\frac12 
\text{Tr} U_{\v i\v j}U_{\v i\v j}^\dag +\left(\psi_{\v i}^\dagger U_{\v i\v j}\psi_{\v j}
+h.c.\right)\right]\label{E:single}
\end{align}
where
\begin{align}
\psi_{\v i}&=\left(\begin{array}{c}
f_{\v i\uparrow }\\
f_{\v i\downarrow }^{\dag}
\end{array}\right)\label{Eq:p-h}\\
U_{\v i\v j}&=\left(\begin{array}{cc}
\chi_{\v i\v j}&\eta_{\v i\v j}\\
\eta_{\v i\v j}^*&-\chi_{\v i\v j}^* .
\end{array}\right)
\end{align}
The path integral that describes the spin-1/2 system is given by
\begin{equation*}
 Z=\int D(\psi)D(a^l_{0,\v i})D(U_{\v i\v j}) e^{i\int dt\; L}
\end{equation*}

The $SU(2)$ gauge transformation is given by
\begin{align*}
\psi_{\v i}&\rightarrow \psi '_{\v i}=W_{\v i}\psi_{\v i},\\
a^l_{0,\v i}\tau^l&\rightarrow
a^{'l}_{0,\v i}=W_{\v i}a^l_{0,\v i}\tau^l W_{\v i}^{\dag}
+(i\partial_{t}W_{\v i})W_{\v i}^{\dag}\\
U_{\v i\v j}&\rightarrow W'_{\v i\v j}=W_{\v i}U_{\v i\v j}W_{\v j}^{\dag}
\end{align*}
where all $W_{\v i}\in SU(2)$.

There are some other useful relations. For each site, the conjugate
of fundamental representation of $SU(2)$ is equivalent to the fundamental
representation itself. One therefore can introduce another $SU(2)$
doublet (neglecting the site label):
\begin{align}
\widehat{\psi}=i\sigma_2
\psi^*=\left(\begin{array}{c}
f_{\downarrow}\\
-f_{\uparrow}^{\dag}
\end{array}\right)
\label{Eq:hatpsi}
\end{align}
Thus there are only three $SU(2)$ gauge invariant bilinear forms
for $\psi$ fields on the same site
\begin{align}
S^+&=\frac{1}{2}\psi_{\alpha}^{\dag}\widehat{\psi}_{\alpha}=f_{\uparrow}^{\dag}f_{\downarrow}\label{E:s1}\\
S^-&=\frac{1}{2}\widehat{\psi}_{\alpha}^{\dag}\psi_{\alpha}=f_{\downarrow}^{\dag}f_{\uparrow}\label{E:s2}\\
S^3&=\frac{1}{2}\left(\psi_{\alpha}^{\dag}\psi_{\alpha}-1\right)=\frac{1}{2}\left(1-\widehat{\psi}_{\alpha}^{\dag}\widehat{\psi}_{\alpha}\right)\notag\\
&=\frac{1}{2}\left(f_{\uparrow}^{\dag}f_{\uparrow}-f_{\downarrow}^{\dag}f_{\downarrow}\right)\label{E:s3}
\end{align}
They turned out to be the generators of the spin rotation
symmetry.

In the zeroth order approximation, or at the mean-field level, one assumes
that the boson fields $a^l_{0,\v i}$ and $U_{\v i\v j}$ get condensed, or more
specifically, can be replaced by some time-independent c-numbers. In this
approximation, the system is described by the following mean-field Hamiltonian
\begin{align}
H_{mean}&=\sum_{\v i}\psi_{\v i}^{\dag}a^l_{0,\v i}\tau^l\psi_{\v i} \notag\\
&+\sum_{\<\v i\v j\>}\frac{1}{4}J_{\v i\v j}\left[\frac12 
\text{Tr} U_{\v i\v j}U_{\v i\v j}^\dag +\left(\psi_{\v i}^{\dag}U_{\v i\v j}\psi_{\v j}+h.c.\right)\right]
\label{Eq:Hmean}
\end{align}

\section{A large-$N$ limit of $SU(2)$ projective construction}

The above mean-field approximation is a good one only when there are some
reasons to suppress the fluctuations of the boson fields $a^l_{0,\v i}$ and
$U_{\v i\v j}$. One way to suppress these fluctuations is to go to a large-$N$
limit. Actually the mean-field result is exact when $N=\infty$. In this
section, we try to answer the following questions:
\begin{itemize}
\item What is the lattice spin model that the large-$N$ limit corresponds to?
\item What is ground state of this lattice spin model?
\end{itemize}

Here we present our answers of these problems first. Later we will see the detailed derivation of these answers:

\begin{itemize}
\item 
The lattice spin model is a $Sp(2N)$ spin model described by
Eq. \ref{SpNH}. The $Sp(2N)$ spin model  is a generalization of the
usual $SU(2)$ spin model (which corresponds to $N=1$ case).

\item 
The ground states of the model are $Sp(2N)$ singlets.  For small $J_2$
(the next-nearest-neighbor coupling), The system is in an algebraic spin liquid
state.\cite{Wqoslpub,RWspin}  The excitations are the fractionalized particles
(spinons) coupled to $SU(2)$ gauge fields. The gauge field is deconfined. 
For larger $J2/J1$, there is a continuous quantum phase transition and the
system goes into the chiral spin state which breaks the time reversal
symmetry.\cite{WWZcsp} 
\end{itemize}

It was debated for long time the existence of featureless Mott insulator if the
unit cell does not have even number of electrons.  In that case, it seems
that in order to have a
Mott insulator state, the ground state must break translation symmetry to enlarge the unit cell
such that the number of electrons per unit cell becomes even.  Here we present a counter-example. The ground state of
$Sp(2N)$ when $N$ is large can be featureless Mott insulator with odd numbers of
electrons per unit cell.

We want to introduce a large-$N$ limit and maintain the $SU(2)$ gauge
structure.  The simplest way to do this is to introduce $N$ flavors of
fermions (later in this paper we also denote $N$ as $N_f$ to emphasize it represents the number of flavors of fermions):
\begin{align*}
\psi_{\v i}^a=\left(\begin{array}{c}
f_{\v i\uparrow}^a\\
f_{\v i\downarrow}^{a\dag}
\end{array}\right),\quad a=1,2,\cdots N.
\end{align*}
Then the Lagrangian of the $N$-flavor model is:
\begin{align*}
L&=\sum_{\v i} \psi_{\v i}^{\dagger a}(i\partial_t -a^l_{0,\v i}\tau^l)\psi_{\v i}^a\notag\\
&+\sum_{\<\v i\v j\>}\frac{1}{4}J_{\v i\v j}\left[\frac N2 \text{Tr} U_{\v i\v j}U_{\v i\v j}^\dag 
+(\psi_{\v i}^{\dagger a}U_{\v i\v j}\psi_{\v j}^a+h.c.)\right]
\end{align*}
where the repeated index $a$ is summed.  
The large-$N$ model is clearly invariant under the $SU(2)$ gauge transformation.
Comparing with the original model
Eq.~(\ref{E:single}), after integrating out the fermions, the effect of the
$N$-flavor is to put a factor of $N$ in front of the boson fields Lagrangian,
which controls the fluctuation of them when $N$ is large.

Now we want to find out the corresponding spin model for this large-$N$ theory.
It should be some generalization of the spin-1/2 Heisenberg model. To do so, we
need to integrate out the boson fields. If we go back to the $f$ picture, we
have:
\begin{align*}
L&=\sum_{\v i}f_{\v i\alpha}^{\dagger a}i\partial_t f_{\v i\alpha}^a\notag\\
&-\sum_{\v i}\left[\frac{1}{2}a_{0,\v i}^- \hat\eta_{\v i\v i}^{aa\dag}
             +\frac{1}{2}a_{0,\v i}^+ \hat\eta_{\v i\v i}^{aa}
             +\frac{1}{2}a_{0,\v i}^3(\hat\chi_{\v i\v i}^{aa\dag}-N)\right]\notag\\
&+\sum_{\<\v i\v j\>}\frac{J_{\v i\v j}}{4N}\left(\hat\eta_{\v i\v j}^{aa\dag}\hat\eta_{\v i\v j}^{bb}
+\hat\chi_{\v i\v j}^{aa\dag}\hat\chi_{\v i\v j}^{bb}\right),
\end{align*}
where
\begin{align*}
\hat \eta_{\v i\v j}^{ab\dag}&=
f_{\v i\alpha}^{\dagger a}\epsilon_{\alpha\beta}f_{\v j\beta}^{b\dag}
=\psi_{\v i1}^{\dagger a}\psi_{\v j2}^b-\psi_{\v i2}^a\psi_{\v j1}^{b\dag}\\
\hat \chi_{\v i\v j}^{ab\dag}&=
f_{\v i\alpha}^{\dagger a}\delta_{\alpha\beta}f_{\v j\beta}^{b\dag}
=\psi_{\v i1}^{\dagger a}\psi_{\v j1}^b+\psi_{\v i2}^a\psi_{\v j2}^{b\dag}
\end{align*}
Here one immediately see the constraints become
$\chi_{\v i\v i}^{aa}=N$ and $\eta_{\v i\v i}^{aa}=0$, or
\begin{align}
f_{\v i\alpha}^{a\dag}f_{\v i\alpha}^a&=N
&f_{\v i\alpha}^{a\dag}\epsilon_{\alpha\beta}f_{\v i\beta}^{a\dag}&=h.c.=0
\label{constraints}
\end{align}
for each site. 
The Hamiltonian under these constraints are: \begin{align}
H&=-\sum_{\<\v i\v j\>}\frac{J_{\v i\v j}}{4N}\left(\hat\eta_{\v i\v j}^{aa\dag}
\hat\eta_{\v i\v j}^{bb}+\hat\chi_{\v i\v j}^{aa\dag}\hat\chi_{\v i\v j}^{bb}\right)\\
&=\sum_{\<\v i\v j\>}-\frac{J_{\v i\v j}}{4N}
\left(-f_{\v i\alpha}^{a\dag}f_{\v i\beta}^b 
f_{\v j\beta}^{a\dag}f_{\v j\alpha}^b\right.\notag\\
&\left. -f_{\v i\alpha}^{a\dag}f_{\v i\beta}^b
f_{\v j\beta}^{b\dag}f_{\v j\alpha}^a+f_{\v i\alpha}^{a\dag}
f_{\v i\alpha}^bf_{\v j\beta}^{a\dag}f_{\v j\beta}^b+N\right)
\label{Hamilton}
\end{align}

We want to understand the symmetry of this model and try to rewrite it in
a form of spin coupling. Here, the symmetry that we are studying is not the
local gauge invariance, but some global physical symmetry in analogy with the
spin rotation symmetry for the $N=1$ model. Motivated by
Eq.~(\ref{E:s1}-\ref{E:s3}), we construct all the gauge invariant bilinear
forms of $\psi$ for each site (neglecting the site label). Let
\begin{equation*}
\widehat{\psi}^a \equiv
i\sigma_2\psi^{\dagger a}=\left(\begin{array}{c} f_{\downarrow}^a\\
-f_{\uparrow}^{a\dag}
\end{array}\right) 
\end{equation*}
The $SU(2)$ gauge invariant bilinears are
\begin{align*}
S^{ab+}&\equiv
\frac{1}{2}\psi_{\alpha}^{a\dag}\widehat{\psi}_{\alpha}^{b}
=\frac{1}{2}\left(f_{\uparrow}^{a\dag}f_{\downarrow}^b
+f_{\uparrow}^{b\dag}f_{\downarrow}^a\right)\\
S^{ab-}&\equiv
\frac{1}{2}\widehat{\psi}_{\alpha}^{a\dag}\psi_{\alpha}^{b}
=\frac{1}{2}\left(f_{\downarrow}^{a\dag}f_{\uparrow}^b
+f_{\downarrow}^{b\dag}f_{\uparrow}^a\right)\\
S^{ab3}&\equiv\frac{1}{2}\left(\psi_{\alpha}^{a\dag}
\psi_{\alpha}^b-\delta^{ab}\right)
=\frac{1}{2}\left(\delta^{ab}
-\widehat{\psi}_{\alpha}^{b\dag}\widehat{\psi}_{\alpha}^a\right)\notag\\
&=\frac{1}{2}\left(f_{\uparrow}^{a\dag}
f_{\uparrow}^b-f_{\downarrow}^{b\dag}f_{\downarrow}^a\right)
\end{align*}
What is the group generated by these $S$ operators? First let us
count how many of them there are. For $S^{ab+}$ or $S^{ab-}$, the
label is symmetric for $a$ and $b$, so there are
$\frac{N(N+1)}{2}$ operators of each type. For $S^{ab3}$, the
labels are not symmetric, so there are simply $N^2$ of them.
Totally we have $N(N+1)+N^2=2N^2+N$ of them. One can further
examine their commutation relations:
\begin{align*}
\left[S^{ab-},\;S^{cd-}\right]&=0, \quad
\left[S^{ab+},\;S^{cd+}\right]=0\\
\left[S^{ab3},\;S^{cd3}\right]&=\frac{1}{2}\left(\delta^{bc}S^{ad3}-\delta^{ad}S^{cb3}\right)\\
\left[S^{ab3},\;S^{cd+}\right]&=\frac{1}{2}\left(\delta^{bc}S^{ad+}+\delta^{bd}S^{ac+}\right)\\
\left[S^{ab3},\;S^{cd-}\right]&=-\frac{1}{2}\left(\delta^{ad}S^{bc-}+\delta^{ac}S^{bd-}\right)\\
\left[S^{ab+},\;S^{cd-}\right]&=\frac{1}{2}\left(\delta^{ac}S^{bd3}+\delta^{ad}S^{bc3}\right.\notag\\
&\quad\quad\left.+\delta^{bc}S^{ad3}+\delta^{bd}S^{ac3}\right)
\end{align*}
These are the relations for $SP(2N)$ algebra, so all the $S$
operators are the $2N^2+N$ generators which generate an $SP(2N)$
group. When $N=1$, $SP(2)$ is isomorphic to $SO(3)$. After some
rearrangements, the Hamiltonian Eq.~(\ref{Hamilton}) can be
rewritten as:
\begin{align}
\label{SpNH}
H=\sum_{\<\v i\v j\>}\frac{J_{\v i\v j}}{N}\left[\frac{1}{2}S_{\v i}^{ab+}S_{\v j}^{ab-}+\frac{1}{2}S_{\v i}^{ab-}S_{\v j}^{ab+}+S_{\v i}^{ab3}S_{\v j}^{ba3}\right]
\end{align}
If we define:
\begin{align*}
S_{\v i}^{ab1}&=S_{\v i}^{ba1}=\frac{1}{2}\left(S_{\v i}^{ab+}+S_{\v i}^{ab-}\right)\\
S_{\v i}^{ab2}&=S_{\v i}^{ba2}=\frac{1}{2i}\left(S_{\v i}^{ab+}-S_{\v i}^{ab-}\right),
\end{align*}
then the vector
\begin{align*}
\boldsymbol{S}^{ab}_{\v i}=\left(S_{\v i}^{ab1},S_{\v i}^{ab2},S_{\v i}^{ab3}\right)
\end{align*}
can simplify the Hamiltonian to:
\begin{align}
\label{SPNHam}
H_{Sp(2N)}=\sum_{\<\v i\v j\>}\frac{J_{\v i\v j}}{N}\;\mathbf{S}_{\v i}^{ab}\cdot\mathbf{S}_{\v j}^{ba}
\end{align}
Here we have to mention that the three components of
$\boldsymbol{S}^{ab}_{\v i}$ are actually not on the same footing,
since the first two are symmetric with respect to the flavor
labels but the third one is not. It is a simple task to check that
all the generators commute with the Hamiltonian. And one can even
check that the $SP(2N)$ is indeed the full symmetry group of it.

Now we want to know what the physical Hilbert space is. It should be the
subspace of the enlarged Hilbert space in which the constraints
Eq.~(\ref{constraints}) are satisfied. When $N=1$, the physical Hilbert space
is the spin-1/2 Heisenberg's, where we have only two states on each site:
\begin{align*}
|\uparrow\rangle,\quad |\downarrow\rangle
\end{align*}
When $N=2$, we have 5 states on each site:
\begin{align*}
&|1_{\uparrow}2_{\uparrow}\rangle,
\quad|1_{\uparrow}2_{\downarrow}\rangle,
\quad|1_{\downarrow}2_{\uparrow}\rangle,
\quad|1_{\downarrow}2_{\downarrow}\rangle,\\
&\frac{1}{\sqrt{2}}
\left(|1_{\uparrow\downarrow}2_{0}\rangle
-|1_{0}2_{\uparrow\downarrow}\rangle\right)
\end{align*}
Here, for example, $|1_{\uparrow}2_{\downarrow}\rangle$ represents a fermion of flavor $1$ and spin up and another fermion of flavor $2$ and spin down; and
$|1_{\uparrow\downarrow}2_{0}\rangle$  represents a fermion of flavor $1$ and spin up,
another fermion of flavor $1$ and spin down, and no fermion of flavor $2$.
One can check the dimension of the Hilbert space on each site
for larger N's:
\begin{align*}
&\mbox{N=3 }\quad \mbox{dimension}=14,\\
&\mbox{N=4 }\quad \mbox{dimension}=43,\\
&\mbox{N=5 }\quad \mbox{dimension}=142,\\
&\mbox{N=6 }\quad \mbox{dimension}=429,\cdots
\end{align*}

This Hilbert space turns out to be an irreducible representation
of the $SP(2N)$ symmetry group. If we label an irreducible
representation by its highest weight state for a particular Cartan
basis. The Cartan basis for $SP(2N)$ can be chosen to be the
z-component spins of each flavor:
\begin{align*}
S^{aa3},\mbox{ where }a=1,2,\cdots,N,
\end{align*}
with $N$ generators. Then the highest weight state in our Hilbert
space is simply:
\begin{align*}
|1_{\uparrow}2_{\uparrow}3_{\uparrow}\dots N_{\uparrow}\rangle
\end{align*}

\section{Phase diagram of the $Sp(2N)$ model on 2D square lattice}

In this section we would like to calculate the zero temperature phase diagram
for the $Sp(2N)$ model Eq.~(\ref{SPNHam}) on 2D square lattice.
We assume that only the nearest-neighbor couplings
and the next-nearest-neighbor couplings are non-zero
\begin{align*}
 J_{\v i,\v i+\v x}&= J_{\v i,\v i+\v y}=J_1, \nonumber\\
 J_{\v i,\v i+\v x+\v y}&= J_{\v i,\v i+\v x- \v y}=J_2.
\end{align*}
We also assume $J_1+J_2=1$.

There are two mean-field approaches to the $Sp(2N)$ model.
In the first mean-field approach, we 
use the ground state $|\Phi^{(\v m_{\v i}^{ab})}\>$ of a trial Hamiltonian
\begin{align}
\label{SPNmean}
H_{trial}=\sum_{\v i\v j}\frac{J_{\v i\v j}}{N}\;\mathbf{m}_{\v i}^{ab}\cdot\mathbf{S}_{\v j}^{ba}
\end{align}
as the trivial wave function and obtain the mean-field ground state
of the $Sp(2N)$ model Eq.~(\ref{SPNHam}) by minimizing
\begin{equation*}
\<\Phi^{(\v m_{\v i}^{ab})}|H_{Sp(2N)}|\Phi^{(\v m_{\v i}^{ab})}\>
\end{equation*}
as we vary the variational parameters $\v m_{\v i}^{ab}$.
The obtained ground state corresponds to a $Sp(2N)$ spin polarized state.
One can show that the ground state energy obtained in this mean-field
approach is always of order $1$ per site in the large $N$ limit.

In the second mean-field approach (the projective
construction),\cite{BZA8773,Wen04} we start with a fermion trial Hamiltonian
\begin{align}
\label{SPNproj}
H_{trial}=\sum_{\<\v i\v j\>}  \psi^{a\dag}_{\v i} u_{\v i\v j} \psi^a_{\v j}
\end{align}
where $u_{\v i\v j}=u^\mu_{\v i\v j} \tau^\mu$ are two by two complex matrices that
satisfy
\begin{equation*}
u_{\v i\v j}^\dag=u_{\v j\v i},\ \ \ \ \ \ \
u^0_{\v i\v j}=\text{imaginary},\ \ \ \ \ \ \
u^l_{\v i\v j}=|_{l=1,2,3} \text{real}
\end{equation*}
Let $|\Phi_{mean}^{(u_{\v i\v j})}\>$ be the ground state of the above 
trial Hamiltonian.
$u_{\v i\v i}^l\equiv a_{0,\v i}^l$ are chosen such that
the constraints $\psi^{a\dag}_{\v i}\tau^l\psi^b_{\v i}=0$ are satisfied on average:
\begin{equation*}
\<\Phi_{mean}^{(u_{\v i\v j})}|\psi^{a\dag}_{\v i}\tau^l\psi^b_{\v i}|\Phi_{mean}^{(u_{\v i\v j})}\>=0
\end{equation*}
We then project $|\Phi_{mean}^{(u_{\v i\v j})}\>$ to the physical Hilbert
space and obtain $P|\Phi_{mean}^{(u_{\v i\v j})}\>$.  The physical Hilbert
space is formed by states $|phys\>$ that satisfy the constraints
Eq.~(\ref{constraints}) or
\begin{equation*}
 \psi^{a\dag}_{\v i}\tau^l\psi^b_{\v i}|phys\>=0
\end{equation*}
The projected wave function $P|\Phi_{mean}^{(u_{\v i\v j})}\>$ is our trial
wave function with $u_{\v i\v j}$ as variational parameters.  Minimzing
\begin{equation*}
\<\Phi_{mean}^{(u_{\v i\v j})}|P H_{Sp(2N)}P|\Phi_{mean}^{(u_{\v i\v j})}\>
\end{equation*}
by varying $u_{\v i\v j}$, we obtain the approximated ground state.

Since in the large-$N$ limit, the fluctuations of the Lagrangian multiplier
$a_{0,\v i}^l(t)$ are weak, we expect that removing the projection $P$ only
causes an error of order $1/N$. Also, other mean-field fluctuations are weak
in the large-$N$ limit, so we expect that the minimized mean-field energy
\begin{align*}
E&=\<\Phi_{mean}^{(u_{\v i\v j})}| H_{Sp(2N)}|\Phi_{mean}^{(u_{\v i\v j})}\> \nonumber\\
&= -\sum_{\<\v i\v j\>}\frac{J_{\v i\v j}}{4N}\left(\eta_{\v i\v j}^{aa\dag}\eta_{\v i\v j}^{bb}
+\chi_{\v i\v j}^{aa\dag}\chi_{\v i\v j}^{bb}\right) + O(1)
\end{align*}
to be the true ground state energy at the leading order in the large-$N$
expansion.  Here
\begin{equation*}
 \chi_{\v i\v j}^{ab}= 
\<\Phi_{mean}^{(u_{\v i\v j})}|\hat\chi_{\v i\v j}^{ab} |\Phi_{mean}^{(u_{\v i\v j})}\>
,\ \ \ \ \ \ \ \ \
 \eta_{\v i\v j}^{ab}= 
\<\Phi_{mean}^{(u_{\v i\v j})}|\hat\eta_{\v i\v j}^{ab} |\Phi_{mean}^{(u_{\v i\v j})}\>
\end{equation*}
note that the above minimized energy is of order $-N$ per site.  Thus the
states obtained in the first mean-field approximation Eq. \ref{SPNmean} cannot
be the ground state.

\begin{figure}[tb]
\centerline{
\includegraphics[width=2.2in]{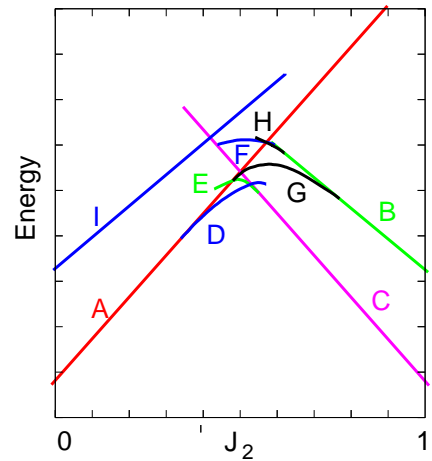}
}
\caption{
The mean-field energies for various phases in a $J_1$-$J_2$ spin system.
(A) the $\pi$-flux state  (the $SU(2)$-linear state SU2B$n0$). 
(B) the $SU(2)\times SU(2)$-gapless state. 
(C) the $SU(2)\times SU(2)$-linear state. 
(D) the chiral spin state (an $SU(2)$-gapped state). 
(E) the $U(1)$-linear state \Eq{u1l90b} which breaks $90^\circ$ 
rotation symmetry. 
(F) the $U(1)$-gapped state U1C$n00x$. 
(G) the $Z_2$-linear state Z2A$zz13$.
(H) the $Z_2$-linear state Z2A$0013$.
(I) the uniform RVB state  (the $SU(2)$-gapless state SU2A$n0$). 
}
\label{phaseJ12}
\end{figure}

Within the $SU(2)$ projective construction and using the translation invariant
ansatz $u_{\v i\v j}$ with only nearest- and next-nearest-neighbor couplings,
we find many local minima of $-\sum_{\<\v i\v j\>}
(
\eta_{\v i\v j}^{aa\dag}\eta_{\v i\v j}^{bb}
+\chi_{\v i\v j}^{aa\dag}\chi_{\v i\v j}^{bb}
)$ as we vary $u_{\v i\v j}$.  We plot those
minima in Fig. \ref{phaseJ12} as functions of $J_2$ (note $J_1+J_2=1$).

As we change $J_2$, the energy of the state changes smoothly along each curve.
So there is no quantum phase transition as we move from one point of a curve
to another point on the same curve. The ansatzs on the same curve belong to the
same phase.  However, if two curves cross each other, the crossing point
represents a quantum phase transition. This is because the ground state energy
is not analytic at the crossing point. If the slopes of the curves at the
crossing point are different, the quantum phase transition is first order.  If
the  slopes at the crossing point are the same, the quantum phase transition
is second order.

From Fig. \ref{phaseJ12}, we see second-order (or continuous) phase
transitions (at mean-field level) between the following pairs of phases:
(A,D), (A,G), (B,G), (C,E), and (B,H). We used to believe that all the
second-order phase transitions are caused by symmetry breaking.
So a natural question is what symmetries are broken for the
above five second-order phase transitions?

It turns out that, except phase (D) and phase (E), all other phases have the
same symmetry. In other words, the projected ground state wave functions
$P|\Phi_{mean}^{(u_{\v i\v j})}\>$ for the ansatz $u_{\v i\v j}$ associated
with those phases have identical symmetry. Thus the three continuous
transitions (B,G), (B,H) and (A,G) do not change any symmetries. It was pointed
out that those phases, despite having the same symmetry, contain different
\emph{quantum orders} \cite{Wqoslpub}.  The projective symmetry group (PSG),
defined as the invariant group of the ansatz $u_{\v i\v j}$, is introduced to
describe this new class of orders \cite{Wqoslpub}.

The ansatzs on the same curve have the same PSG and correspond to the same
quantum phase.  On the other hand, the ansatzs on the different curves have
different PSG's.  We see that a quantum phase transition is characterized by a
change in PSG. Those quantum phase transitions represent a new class
of phase transitions beyond the Landau's symmetry breaking theory.
Other examples of phase transitions beyond the Landau's theory
can be found in \Ref{DH8156,WWtran,CFW9349,SMF9945,RG0067,Wctpt}.

In the following we will discuss quantum orders (or PSG's) for mean-field
phases in Fig.  \ref{phaseJ12}. The PSG's for those quantum orders are labeled
by labels which look like Z2A$0013$ \cite{Wqoslpub}.

The phase (A) \cite{AM8874} is the $\pi$-flux state, or the
SU2B$n0$ state 
\begin{align}
\label{SU2lA}
u_{\v i,\v i+{\v x}} =& i\chi , &
u_{\v i,\v i+{\v y}} =& i(-)^{i_x} \chi & a^l_0&=0.
\end{align}
The low energy excitations are described by massless Dirac fermions with a
linear dispersion and gapless $SU(2)$ gauge fluctuations. Therefore we also call
such a state $SU(2)$-linear state.

The phase (B) \cite{Wqoslpub} is a state with two independent uniform RVB
states \cite{BZA8773} on the diagonal links.  The gapless fermions have finite
Fermi surfaces. The fermions interact with
$SU(2)\times SU(2)$ gauge fluctuations.
Such a state is called $SU(2)\times SU(2)$-gapless state ($SU(2)\times SU(2)$
indicates the low energy gauge group and ``gapless'' indicates finite Fermi
surface).  Its ansatz is given by
\begin{align}
\label{su2su2gl}
u_{\v i,\v i+{\v x}+{\v y}} =&   \chi \tau^3 ,  &
u_{\v i,\v i+{\v x}-{\v y}} =&   \chi \tau^3 ,  &
a^l_0 =& 0  .
\end{align}

The phase (C) \cite{Wqoslpub} is a state with two independent $\pi$-flux
states on the diagonal links. It has $SU(2)\times SU(2)$ gauge fluctuations at
low energies and will be called an $SU(2)\times SU(2)$-linear state. Its
ansatz is given by
\begin{align}
\label{su2su2}
u_{\v i,\v i+{\v x}+{\v y}} =&   \chi (\tau^3  + \tau^1) &
u_{\v i,\v i+{\v x}-{\v y}} =&   \chi (\tau^3  - \tau^1) \nonumber\\
a^l_0 =& 0
\end{align}
The low energy excitations are $SU(2)\times SU(2)$ gauge fluctuations
and massless Dirac fermions.

The phase (D) is the chiral spin state\cite{WWZcsp}
\begin{align}
\chi_{\v i,\v i+ x}=& i\chi_1 , &
\chi_{\v i,\v i+ y}=& i\chi_1 (-)^{i_x}, &a^l_0=& 0, \nonumber\\
\chi_{\v i,\v i+\v x+\v y}=& -i\chi_2 (-)^{i_x}, & 
\chi_{\v i,\v i+\v x-\v y}=&  i\chi_2 (-)^{i_x} .
\label{SU2csp}
\end{align}
Both fermionic excitations and $SU(2)$ gauge excitations are gapped.  The gap
of the $SU(2)$ gauge excitations is due to an $SU(2)$ Chern-Simons term.

The phase (E) \cite{Wqoslpub} is
described by an ansatz
\begin{align}
\label{u1l90b}
u_{\v i,\v i+{\v x}+{\v y}} =& \chi_1 \tau^1 + \chi_2 \tau^2 &
u_{\v i,\v i+{\v x}-{\v y}} =& \chi_1 \tau^1 - \chi_2 \tau^2 \nonumber\\
u_{\v i,\v i+{\v y}} =&   \eta \tau^3  &
a^l_0 =& 0
\end{align}
which breaks the $90^\circ$ rotation symmetry. It is a $U(1)$-linear state,
i.e., the low lying excitations are massless $U(1)$ gauge fluctuations
interacting with massless Dirac fermions.

The phase (F) \cite{Wqoslpub} is described by the following U1C$n00x$
ansatz 
\begin{align} 
\label{U1lB}
u_{\v i,\v i+{\v x}} =&   \eta \tau^1  & 
u_{\v i,\v i+{\v y}} =&   \eta \tau^1 \nonumber\\ 
u_{\v i,\v i+{\v x}+{\v y}} =&    \chi \tau^3  & 
u_{\v i,\v i+{\v x}-{\v y}} =&    \chi \tau^3 \nonumber\\ 
a^3_0 =& \la,\ \ \ \ \ \ a^{1,2} = 0
\end{align}
The  U1C$n00x$ state can be a $U(1)$-linear state
where fermions are gapless with a
linear dispersion relation (if $a^3_0$ is small) or a $U(1)$-gapped state
where the fermions are gapped (if $a^3_0$ is large).  The state for phase (F)
turns out to be a $U(1)$-gapped state. The only low energy excitations are
massless $U(1)$ gauge bosons. 

The phase (G) \cite{Wqoslpub} is described by the Z2A$zz13$
ansatz
\begin{align}
\label{Z2lA}
u_{\v i,\v i+{\v x}} =& \chi \tau^1 - \eta\tau^2 ,&
u_{\v i,\v i+{\v y}} =& \chi \tau^1 + \eta\tau^2 ,\nonumber\\
 u_{\v i,\v i+{\v x}+{\v y}} =& - \ga \tau^1, &
 u_{\v i,\v i-{\v x}+{\v y}} =& + \ga \tau^1 ,\nonumber\\
 u_{\v i,\v i+2{\v x}} =& 
 u_{\v i,\v i+2{\v y}} = 0,
 &
 a^{1,2,3}_0  = & 0.
\end{align}
The $SU(2)$ gauge structure is broken down to a $Z_2$ gauge structure.  Hence
there is no gapless gauge fluctuations. The only low energy excitations are
massless Dirac fermions. Such a state is called $Z_2$-linear state.  

The phase (H) \cite{SF0050} is described
by the Z2A$0013$ ansatz
\begin{align}
\label{Z2lC}
a_0^3& \neq 0,\ \ \ \ \ \ a^{1,2}_0=0,  \nonumber\\
u_{\v i,\v i+{\v x}} &= \chi \tau^3 +\eta \tau^1, & 
u_{\v i,\v i+{\v y}} &= \chi \tau^3 -\eta \tau^1,  \nonumber\\
u_{\v i,\v i+{\v x}+{\v y}} &= + \ga \tau^3, &
u_{\v i,\v i-{\v x}+{\v y}} &= + \ga \tau^3.
\end{align}
It is also a $Z_2$-linear state.  

The
phase (I) is the uniform RVB state \cite{BZA8773}, 
i.e., the $SU(2)$-gapless state SU2A$n0$
\begin{align}
\label{SU2gl}
u_{\v i,\v i+\v x} =& i\chi , &
u_{\v i,\v i+\v y} =& i\chi , &
a^l_0 & =0.
\end{align}
It has gapless $SU(2)$ gauge fluctuations and gapless fermionic excitations
that form a finite Fermi surface.

From Fig. \ref{phaseJ12}, we see continuous phase transitions (at mean-field
level) between the following pairs of phases: (A,D), (A,G), (B,G), (C,E), and
(B,H).  For the three continuous transitions (B,G), (B,H) and (A,G) that
do not change any symmetries, we observe that the $SU(2)$ gauge structure in the
phase (A) breaks down to $Z_2$ in the continuous transition from the phase (A)
to the phase (G).  The $SU(2)\times SU(2)$ gauge structure in the phase (B)
breaks down to $Z_2$ in the two transitions (B,G) and (B,H).

\section{A stable algebraic spin liquid -- $SU(2)$-linear spin liquid} \label{su2stable}

In this section, we will study the $SU(2)$-linear state given by
Eq.(\ref{SU2lA}) which describes the phase A in phase diagram Fig.
\ref{phaseJ12}.  In the mean-field theory, the interactions between the
excitations are ignored. In this section, we include those interactions and
study how those interactions affect the low energy properties of the mean-field
$SU(2)$-linear state.  We will show that, after including those interactions,
the gapless excitations in the mean-field $SU(2)$-linear state remain
gapless, which leads to a stable algebraic spin liquid.

%
\begin{figure}
\includegraphics[width=0.2\textwidth]{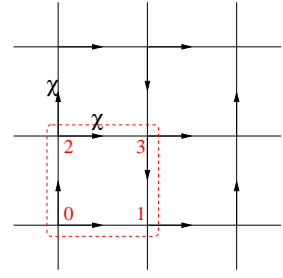}
\caption{We choose the unit cell of $SU(2)$-linear state
to contain \emph{four} lattice sites when we go to the $k$-space.}
\label{F:su2linearcell}
\end{figure}

\subsection{The low energy effective theory of the $SU(2)$-linear state}

To obtain the low energy effective theory of the $SU(2)$-linear state
in the continuum limit,
we choose the unit cell as in Fig. \ref{F:su2linearcell}. It contains 4
sites; each site has spin up and down, so totally 8 fermions. Let us write
down the mean field Hamiltonian:
\begin{align}
&H_{mean}\notag\\
=&\sum_{i}i\chi\left[\psi^{\dag}_{i0}\psi_{i1}+\psi^{\dag}_{i1}\psi_{i+x,0}+\psi^{\dag}_{i2}\psi_{i3}+\psi^{\dag}_{i3}\psi_{i+x,2}\right.\notag\\
&\left.+\psi^{\dag}_{i0}\psi_{i2}-\psi^{\dag}_{i1}\psi_{i3}+\psi^{\dag}_{i2}\psi_{i+y,0}-\psi^{\dag}_{i3}\psi_{i+y,1}\right]+h.c.\notag\\
=&\sum_{k}i\chi\left[\psi_{k0}^{\dag}\psi_{k1}+\psi_{k2}^{\dag}\psi_{k3}+\psi_{k1}^{\dag}\psi_{k0}e^{ik_x}+\psi_{k3}^{\dag}\psi_{k2}e^{ik_x}\right]\notag\\
+&\sum_{k}i\chi\left[\psi_{k0}^{\dag}\psi_{k2}-\psi_{k1}^{\dag}\psi_{k3}+\psi_{k2}^{\dag}\psi_{k0}e^{ik_y}-\psi_{k3}^{\dag}\psi_{k1}e^{ik_y}\right]\notag\\
&+h.c.\notag\\
=&\sum_{k}\left(\psi^{\dag}_{k0},\psi^{\dag}_{k1},\psi^{\dag}_{k2},\psi^{\dag}_{k3}\right)\mathbf{M}\left(\begin{array}{c}
\psi_{k0}\\
\psi_{k1}\\
\psi_{k2}\\
\psi_{k3}
\end{array}\right)
\end{align}
where
\begin{align}
\mathbf{M}=\chi\left(
\begin{array}{l}
\;\;\;\;\;\;0\;\;\;\;\;\;i-ie^{-ik_x}\;\;i-ie^{-ik_y}\;\;\;\;\;\;0\\
-i+ie^{ik_x}\;\;\;0\;\;\;\;\;\;\;\;\;\;\;\;\;\;\;0\;\;\;\;\;\;-i+ie^{-ik_y}\\
-i+ie^{ik_y}\;\;\;0\;\;\;\;\;\;\;\;\;\;\;\;\;\;\;0\;\;\;\;\;\;\;\;\;\;i-ie^{-ik_x}\\
\;\;\;\;\;\;0\;\;\;\;\;\;i-ie^{ik_y}\;\;-i+ie^{ik_x}\;\;\;\;\;\;0
\end{array}\right)
\end{align}
Here we have assumed the lattice constant to be $1/2$, so we have $-\pi<k_x,k_y<\pi$.
Note that here
$\psi_{k,i}$ is actually $SU(2)$-doublet, corresponding to the spin up and
down components in the $f$-formalism; so still totally 8 fermions. After some
rearrangements:
\begin{align}
&\mathbf{M}=2\chi\cdot\notag\\
&\left(\begin{array}{l}
\;\;\;\;\;\;0\;\;\;\;\;\;-e^{-ik_x}\sin k_x\;\;-e^{-ik_y}\sin k_y\;\;\;\;\;\;0\\
-e^{ik_x}\sin k_x\;\;\;\;\;\;0\;\;\;\;\;\;\;\;\;\;\;\;\;\;\;\;\;\;\;\;\;0\;\;\;\;\;\;\;\;e^{-ik_y}\sin k_y\\
-e^{ik_y}\sin k_y\;\;\;\;\;\;0\;\;\;\;\;\;\;\;\;\;\;\;\;\;\;\;\;\;\;\;\;0\;\;\;\;\;\;-e^{-ik_x}\sin k_x\\
\;\;\;\;\;\;0\;\;\;\;\;\;\;\;\;\;e^{ik_y}\sin k_y\;\;\;\;\;\;-e^{ik_x}\sin k_x\;\;\;\;\;\;\;0
\end{array}
\right)
\end{align}
In the continuous limit, the energy spectrum for fermion is characterized by a
single fermi point at $(0,0)$. When $k\approx0$:
\begin{align}
\mathbf{M}=2\chi\left(\begin{array}{cccc}
0&-k_x&-k_y&0\\
-k_x&0&0&k_y\\
-k_y&0&0&-k_x\\
0&k_y&-k_x&0
\end{array}
\right)
\end{align}
We can do an extra rotation to make it the usual form of Dirac fermion:
\begin{align}
\psi\rightarrow \tilde{\psi}&=\mathbf{R}^{\dag}\psi\\
\mathbf{R}&=\left(\begin{array}{cccc}
0&\frac{i}{\sqrt{2}}&\frac{1}{\sqrt{2}}&0\\
\frac{-1}{\sqrt{2}}&0&0&\frac{-i}{\sqrt{2}}\\
\frac{-i}{\sqrt{2}}&0&0&\frac{-1}{\sqrt{2}}\\
0&\frac{-1}{\sqrt{2}}&\frac{-i}{\sqrt{2}}&0
\end{array}
\right)\psi\label{Eq:tildepsi}
\end{align}
Then 
\begin{align}
H_{mean}=\tilde{\psi}^{\dag}\gamma_0\left[ik_x\gamma_1+ik_y\gamma_2\right]\tilde{\psi}
\end{align}
where the $\gamma$ matrices in Euclidean space are:
\begin{align}
\gamma_0=\left(\begin{array}{cc}
\sigma_3&\\
&-\sigma_3
\end{array}\right)\;\gamma_1=\left(\begin{array}{cc}
\sigma_1&\\
&-\sigma_1
\end{array}\right)\;\gamma_2=\left(\begin{array}{cc}
\sigma_2&\\
&-\sigma_2
\end{array}\right)
\end{align}
Here we can also introduce the other two $\gamma$ matrices:
\begin{align}
\gamma_3&=\left(\begin{array}{cc}
&I\\
I&
\end{array}
\right)&
\gamma_5&=\gamma_0\gamma_1\gamma_2\gamma_3=i\left(\begin{array}{cc}
&I\\
-I&
\end{array}
\right)&
\end{align}
where $I$ is the 2 by 2 identity matrix. Notice that both $\gamma_3$ and
$\gamma_5$ anticommute with all space-time components of $\gamma$ matrices:
$\gamma_0$, $\gamma_1$ and $\gamma_2$.  We should also include gauge field
fluctuations above the mean-field theory. The full lagrangian is:
\begin{align}
L=\bar{\tilde{\psi}}\left(\partial_{\mu}-ia_{\mu}^{l}\tau^{l} \right)\gamma_{\mu}\tilde{\psi}+\frac{1}{2g^2}\mbox{Tr}\left[f_{\mu\nu}^l f_{\mu\nu}^l\right]
+\cdots
\label{Eq:SU2lag}
\end{align}
where $\bar{\tilde{\psi}}=\tilde{\psi}^{\dagger}\gamma_0$. So the low energy effective theory of the $SU(2)$-linear state is a QCD$_3$
with a $SU(2)$ gauge field and two massless 4-component Dirac fermions (or
$2N_f$ 4-component Dirac fermions which form $N_f$ $SU(2)$ gauge doublets in
the large $N_f$ limit, notice $N_f=1$ is the physical Heisenberg model's case).

The $\cdots$ in Eq. (\ref{Eq:SU2lag}) represents other terms which may be
generated by the interactions as we integrate out high energy fluctuations.
Understanding those terms is the key to understand the low energy behavior of
the model.  Those terms must be consistent with the underlying lattice
symmetry.  So in the following, we will study the symmetry properties of the
effective theory Eq. (\ref{Eq:SU2lag}). We will show that none of terms allowed by the symmetry are relevant at low energies.  None of those terms
can cause infrared instability. As a result the mean-field $SU(2)$-linear state leads
to a stable algebraic spin liquid.

\subsection{Space translation and rotation symmetry}

\begin{figure}
\begin{center}
$\begin{array}{cc}
\epsfxsize=0.2\textwidth
\epsffile{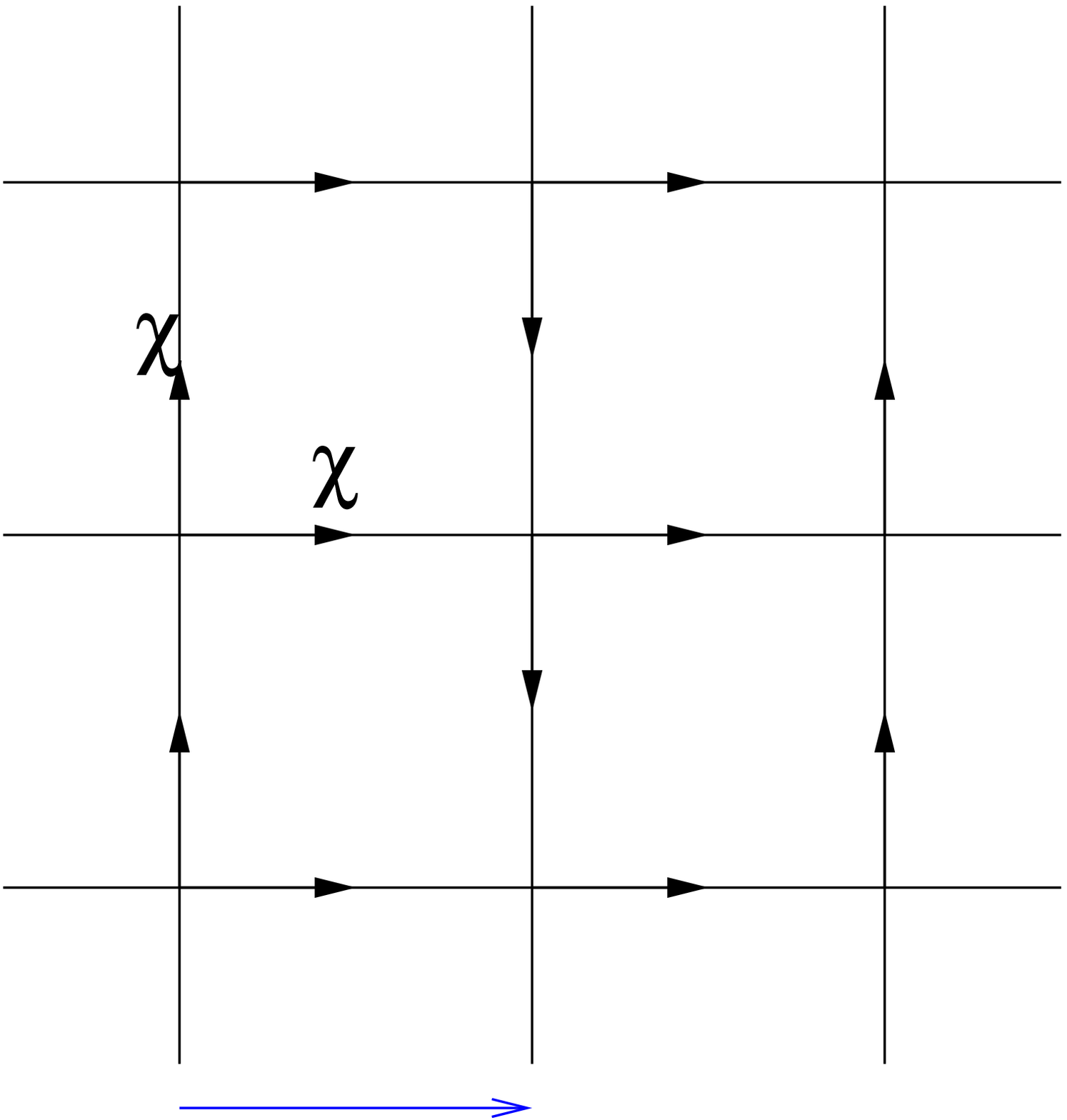}&
\epsfxsize=0.2\textwidth
\epsffile{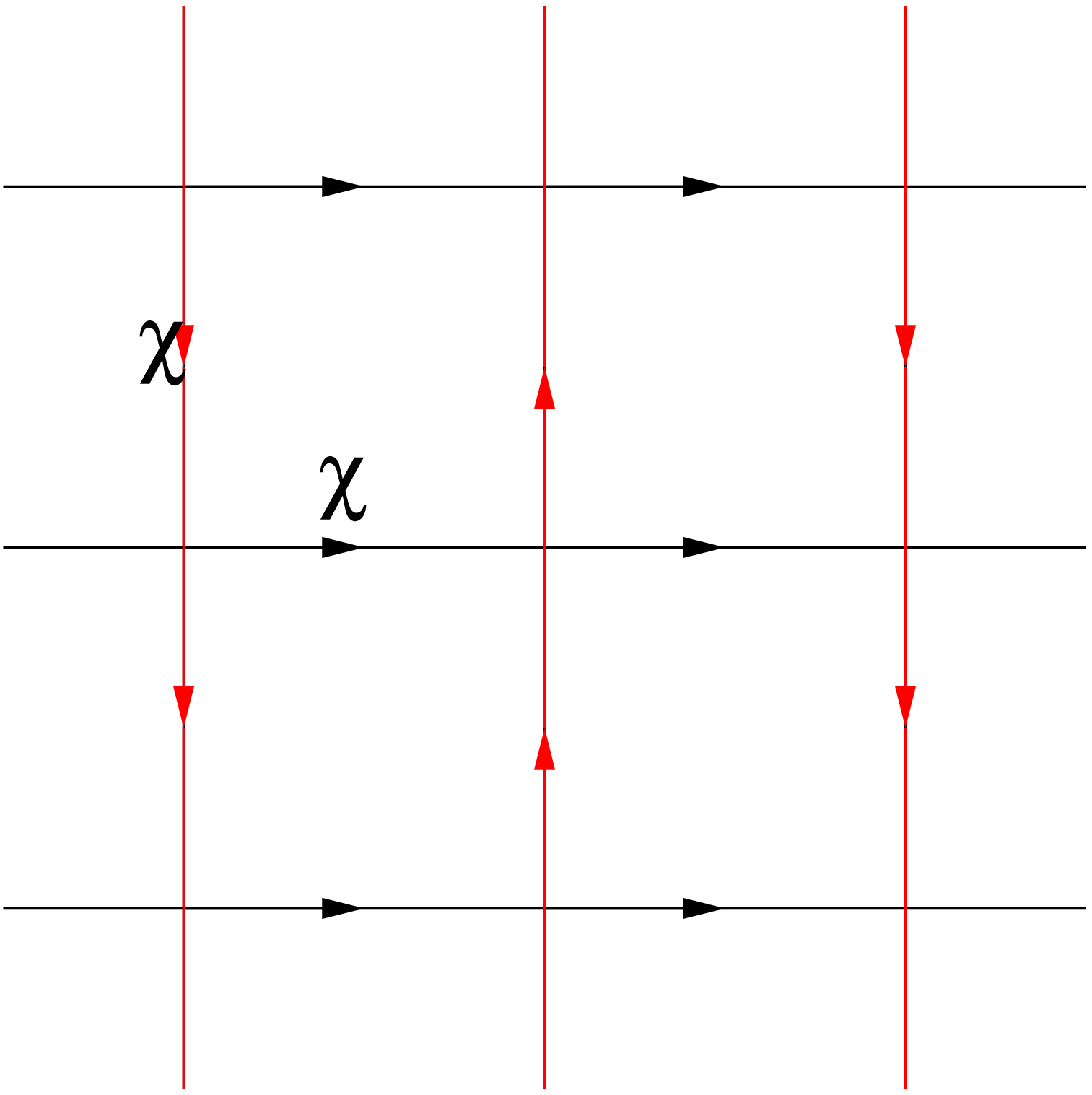}\\
\mbox{(a)}&\mbox{(b)}\\
\epsfxsize=0.2\textwidth
\epsffile{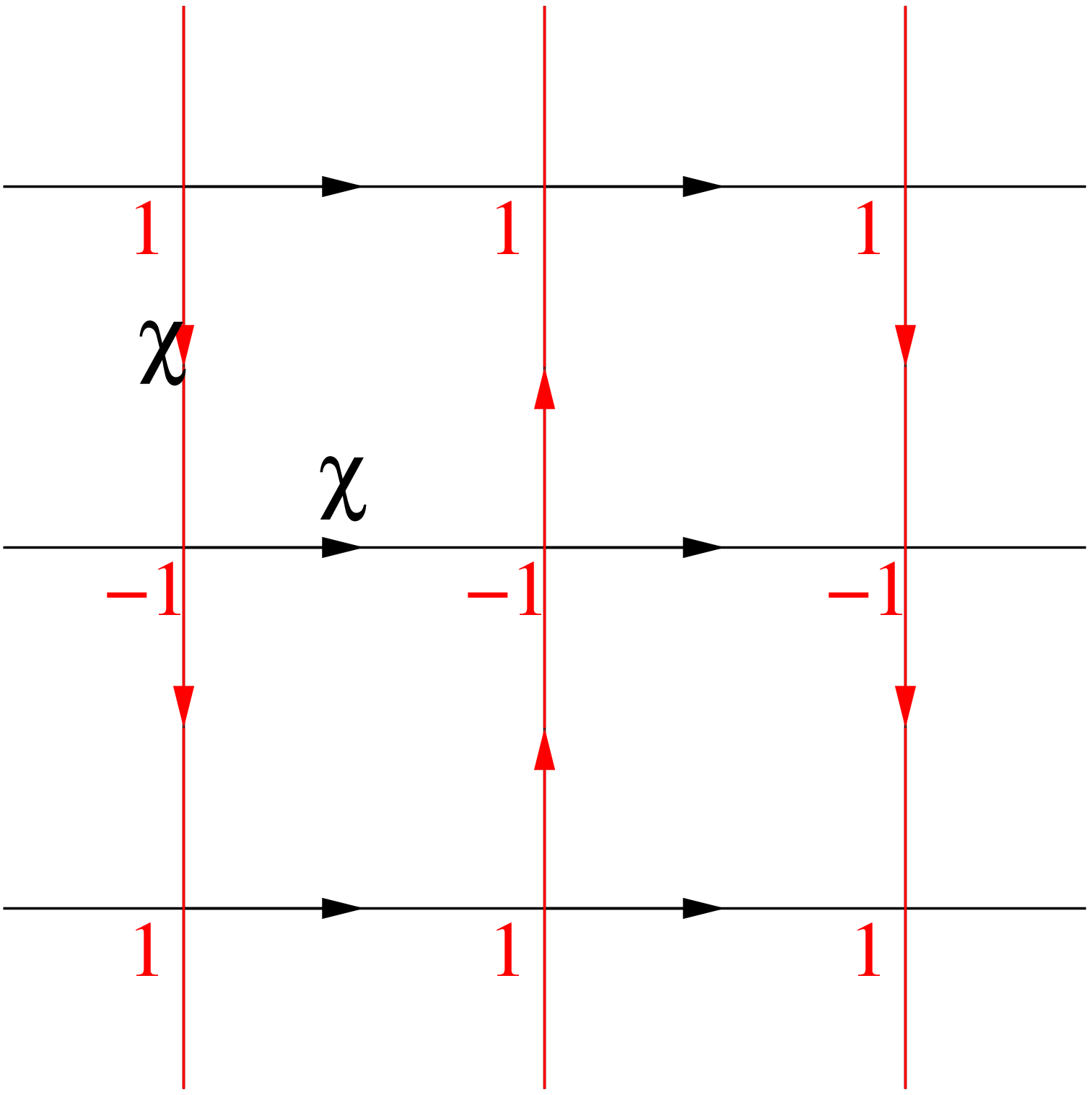}&
\epsfxsize=0.2\textwidth
\epsffile{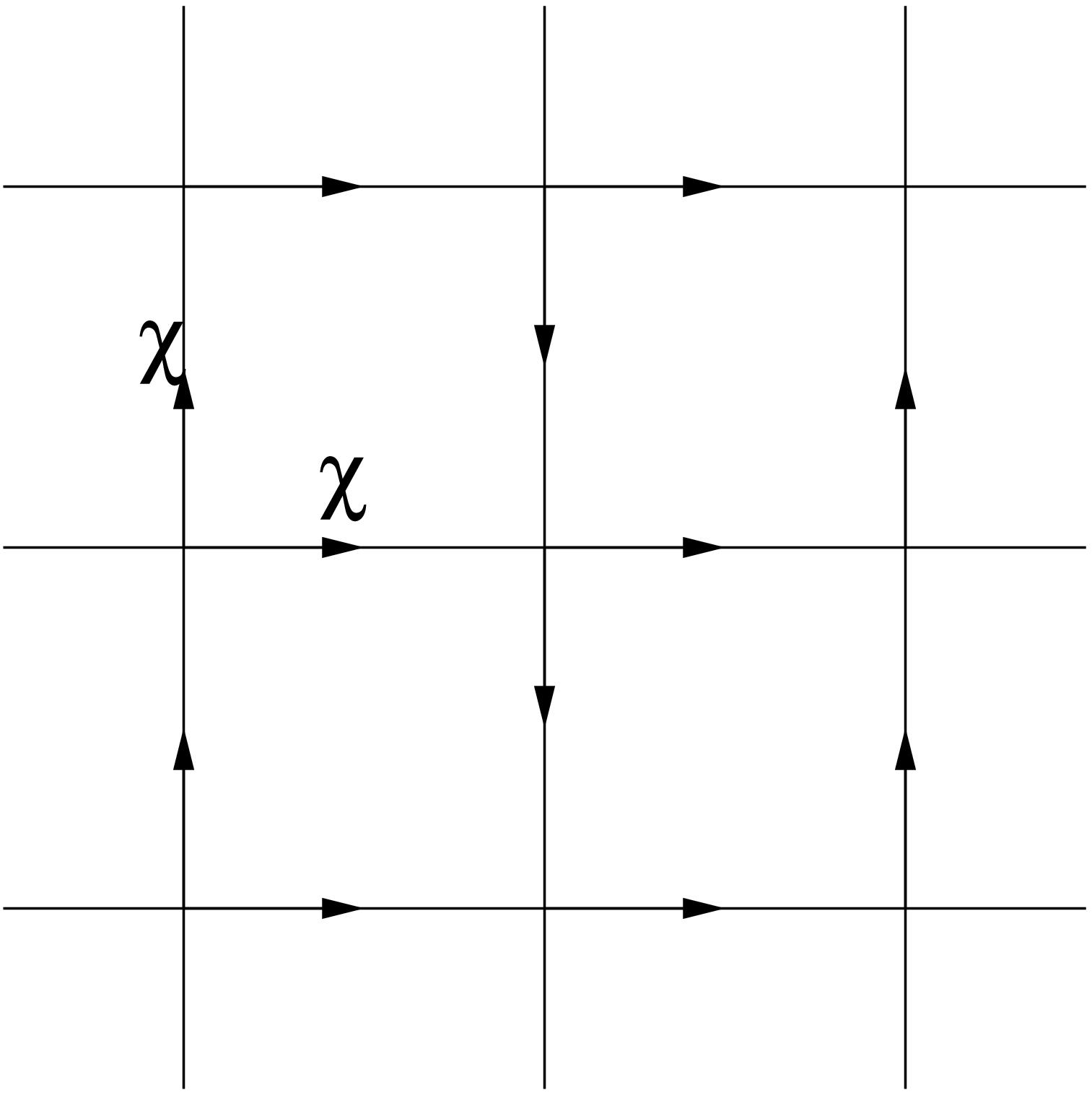}\\
\mbox{(c)}&\mbox{(d)}
\end{array}
$
\end{center}
\caption{Figures (a)--(d) illustrate how the $SU(2)$-linear state respects translation symmetry $T_x$: (a) original ansatz (b) after $T_x$ (c)do $W_{T_x}$ (d) go back to original ansatz.}
\label{F:Tx}
\end{figure}

Now let us think about the corresponding lattice symmetry in continuous limit.
Firstly let us discuss translation by one lattice site along x-direction
$T_x$, in terms of the lattice fields:
\begin{align}
T_x:\lbrace
\begin{array}{rl}
\psi_{\v i}&\rightarrow\psi'_{\v i}=\psi_{\v i-\v x}\\
u_{\v i\v j}&\rightarrow u'_{\v i \v j}=u_{\v i-\v x,\v j-\v x}
\end{array}
\end{align}
It seems that the translation symmetry is broken since $u_{\v i\v j}$ is not
invariant:
\begin{align}
\lbrace
\begin{array}{rl}
u_{\v i,\v i+{\v x}} =& i\chi ,\\
u_{\v i,\v i+{\v y}} =& i(-)^{i_x} \chi
\end{array}
\rightarrow\,\,
\lbrace
\begin{array}{rl}
u'_{\v i,\v i+{\v x}} =& i\chi ,\\
u'_{\v i,\v i+{\v y}} =& -i(-)^{i_x} \chi
\end{array}
\end{align}
But as shown in Fig.\ref{F:Tx} one can do an extra local $SU(2)$ gauge
transformation $W_{T_x}$ to transform $u_{\v i\v j}$ back. Let
\begin{align}
W_{\v i}=(-)^{i_y}
\end{align}
Then
\begin{align}
W_{T_x}:\lbrace
\begin{array}{rl}
\psi'_{\v i}&\rightarrow \psi''_{\v i}=W_{\v i}\psi'_{\v i}=(-)^{i_y}\psi_{\v i-\v x}\\
u'_{\v i\v j}&\rightarrow u''_{\v i \v j}=W_{\v i}u'_{\v i \v j}W_{\v j}^{\dag}
\end{array}
\end{align}
where
\begin{align}
\lbrace
\begin{array}{rl}
u''_{\v i, \v i+\v x}&=u'_{\v i,\v i+{\v x}} = i\chi =u_{\v i,\v i +\v x}\\
u''_{\v i, \v i+\v y}&=-u'_{\v i,\v i+{\v y}}=i(-)^{i_x}\chi=u_{\v i,\v i +\v y} 
\end{array}
\end{align}
Here we point out that the combination of $W_{T_x}$ and $T_x$ is a
transformation leaving $U_{\v i \v j}$ invariant. We call such a 
transformation an element of
PSG. PSG, by definition, is the collection of all transformations leaving the
ansatz $U_{\v i \v j}, a_{0, i}^l$ invariant. Here in $SU(2)$-linear state,
$a_{0, i}^l$ is zero, so it is also invariant. 

After choosing the unit cell as in Fig.\ref{F:su2linearcell}, in terms of the
four component $\psi$ fermion, the combination of $W_{T_x}$ and $T_x$
transforms:
\begin{align}
\left(\begin{array}{c}
\psi_{i0}\\
\psi_{i1}\\
\psi_{i2}\\
\psi_{i3}
\end{array}
\right)\rightarrow
\left(\begin{array}{c}
\psi'_{i0}\\
\psi'_{i1}\\
\psi'_{i2}\\
\psi'_{i3}
\end{array}
\right)=
\left(
\begin{array}{cccc}
0&1&0&0\\
1&0&0&0\\
0&0&0&-1\\
0&0&-1&0
\end{array}
\right)
\left(\begin{array}{c}
\psi_{i0}\\
\psi_{i1}\\
\psi_{i2}\\
\psi_{i3}
\end{array}
\right)
\end{align}
Here we assumed in the continuous limit, $\psi_i$ and $\psi_{i+x}$ are on the
same position.  We can do the extra rotation to transform into
$\tilde{\psi}$, the usual Dirac fermion:
\begin{align}
\tilde{\psi}\rightarrow\tilde{\psi}'&=\mathbf{R}^{\dag}\psi'=\mathbf{R}^{\dag}\left(
\begin{array}{cccc}
0&1&0&0\\
1&0&0&0\\
0&0&0&-1\\
0&0&-1&0
\end{array}
\right)\mathbf{R}\tilde{\psi}\notag\\
&=\left(\begin{array}{cccc}
0&0&-1&0\\
0&0&0&-1\\
-1&0&0&0\\
0&-1&0&0
\end{array}
\right)\tilde{\psi}=-\gamma_3\tilde{\psi}
\end{align}
Eventually we know that the PSG element $W_{T_x}\circ T_x$, which is the
lattice symmetry, if translated into continuous limit, is the internal symmetry
$-\gamma_3$. Note that the minus sign actually corresponds to a global
$SU(2)$ transformation: $W_{i}=-1$. So we have the correspondence:
\begin{align}
W_{T_x}\circ T_x\longleftrightarrow \gamma_3
\end{align}
Similarly one can study the translation by one lattice site along
$y$-direction; we find:
\begin{align}
W_{T_y}\circ T_y\longleftrightarrow \gamma_5
\end{align}
For the reflection $P_x: x\rightarrow -x$:
\begin{align}
W_{P_x}\circ P_x\longleftrightarrow \gamma_1
\end{align}
For the reflection $P_y: y\rightarrow -y$:
\begin{align}
W_{P_y}\circ P_y\longleftrightarrow \gamma_2
\end{align}
For the reflection $P_{xy}: x\rightarrow y, y\rightarrow x$:
\begin{align}
W_{P_{xy}}\circ P_{xy}\longleftrightarrow \frac{1}{2}(\gamma_1-\gamma_2)(\gamma_3+\gamma_5)
\end{align}
Note that $T_x,T_y,P_x,P_y,P_{xy}$ already give the full space-time
symmetry. For example, rotation by 90 degree $R_{90}: x\rightarrow y,
y\rightarrow -x$ is a combination of $P_x$ and $P_{xy}$:
\begin{align}
R_{90}=P_{xy}\circ P_{x}
\end{align}

\subsection{Time-reversal symmetry}

Now let us study the time-reversal symmetry $T$. In terms of spin operator:
\begin{align}
T: \mathbf{S}\rightarrow -\mathbf{S}
\end{align}
One should be cautious that $T$ is not a usual linear operator, instead it is an
anti-linear operator:
\begin{align}
Ti=-iT
\end{align}
What does this operator correspond to in terms of lattice fermion and bond
variable $U_{\v i \v j}$? We know that
\begin{align}
\mathbf{S}=\frac{1}{2}f^{\dag}\boldsymbol{\sigma}f
\end{align}
So the corresponding transformation on $f$ fermion is:
\begin{align}
T: f&\rightarrow i\sigma_2 f,\ \ \ \ \
T: f^\dag\rightarrow  f^\dag (i\sigma_2)^\dag \\
T\mathbf{S}T^{-1}&=Tf^{\dag}T^{-1}T\boldsymbol{\sigma}T^{-1}TfT^{-1}\notag\\
&=f^{\dag}(i\sigma_2)^{\dag}\left(\sigma_1,-\sigma_2,\sigma_3\right)(i\sigma_2)f\notag\\
&=f^{\dag}\left(-\sigma_1,-\sigma_2,-\sigma_3\right)f\notag\\
&=-\mathbf{S}
\end{align}
Here we used the anti-linear property of $T$ operator. Therefore the $T$
transformation on $\psi$ fermion is:
\begin{align}
T: \psi=\left(\begin{array}{c} f_{\uparrow}\\
f_{\downarrow}^{\dag}
\end{array}\right)\rightarrow
\left(\begin{array}{c}
f_{\downarrow}\\
-f_{\uparrow}^{\dag}
\end{array}\right)=i\tau_2\psi^{*}\label{Eq:Tfermion}
\end{align}
Our convention of notation is that in terms of $f$ fermion, we use $\sigma$
to denote Pauli matrices; while in terms of $\psi$ fermion, we use $\tau$ to
denote them.

What is the time-reversal transformation on $U_{\v i\v j}$? Here we notice that
$U_{\v i \v j}$ has two meanings: in Eq.(\ref{E:single}) it means the
operator defined as in Eq.(\ref{Eq:chieta}); while in Eq.(\ref{Eq:Hmean}) it
means the average value of the operator on mean-field ground state. Let us
take a look at how the operator transform under time-reversal:
\begin{align}
T\hat{\chi_{\v i\v j}}T^{-1}&=Tf_{\v i \alpha}^{\dag}f_{\v j \alpha}T^{-1}=Tf_{\v i \alpha}^{\dag}T^{-1}Tf_{\v j \alpha}T^{-1}\notag\\
&=f_{\v i}^{\dag}(i\sigma_2)^{\dag}(i\sigma_2)f_{\v j}=f_{\v i}^{\dag}f_{\v j}=\hat{\chi_{\v i\v j}}\\
T\hat{\eta}_{\v i\v j}^{\dag}T^{-1}&=Tf_{\v i \alpha}^{\dag}\epsilon_{\alpha\beta}f_{\v j \beta}^{\dag}T^{-1}=Tf_{\v i \alpha}^{\dag}T^{-1}\epsilon_{\alpha\beta}Tf_{\v j \beta}^{\dag}T^{-1}\notag\\
&=f_{\v i\gamma}^{\dag}(i\sigma_2)_{\gamma\alpha}^{\dag}\epsilon_{\alpha\beta}(i\sigma_2)^{\dag}_{\delta \beta}f_{\v j\delta}=f_{\v i\gamma}^{\dag}\epsilon_{\gamma\delta}f_{\v j\delta}=\hat{\eta}^{\dag}_{\v i\v j}
\end{align}
Therefore we know that the $\hat{U}_{\v i \v j}$, as an operator, is invariant
under $T$. Here it is helpful to write down $\hat{U}_{\v i \v j}$ in terms of
$\psi$ operators:
\begin{align}
\hat{U}_{\v i \v j}=\left(
\begin{array}{cc}
\hat{\chi}_{\v i\v j}&\hat{\eta}_{\v i\v j}\\
\hat{\eta}_{\v i\v j}^{\dag}&-\hat{\chi}_{\v i\v j}^{\dag}
\end{array}
\right)=-\psi_{\v j}\psi_{\v i}^{\dag}-(i\tau_2\psi_{\v j}^{*})(i\tau_2\psi_{\v i}^{*})^{\dag}\label{Eq:Uinpsi}
\end{align}
Notice Eq.(\ref{Eq:Tfermion}), under $T$ transformation, the first term
transforms into the second term, and the second transforms into the first. So
the whole $\hat{U}_{\v i\v j}$ is invariant. One can also check that together
with transformation Eq.(\ref{Eq:Tfermion}), the Lagrangian Eq.(\ref{E:single})
is invariant under $T$. This is expected since the original Heisenberg
Hamiltonian Eq.(\ref{heisenberg}) is $T$ invariant.

But things are different if $\hat{U}_{\v i\v j}$ condense, more
specifically, if it has some non-zero average value: $U_{\v i\v
j}=\langle\Psi\vert\hat{U}_{\v i\v j}\vert\Psi\rangle$. Because $T$ is an
anti-linear operator, it transforms $U_{\v i\v j}$ into $U_{\v i\v j}^{*}$. We
know that for two arbitrary states $\vert\alpha\rangle$, $\vert\beta\rangle$,
the anti-linear property of $T$ gives
\begin{align}
\langle\alpha\vert\beta\rangle=\langle T\beta\vert T\alpha\rangle
\end{align}
Therefore for an arbitrary linear operator $\hat{O}$,
\begin{align}
\langle\alpha\vert\hat{O}\vert\beta\rangle=\langle\hat{O}^{\dag}\alpha\vert\beta\rangle=\langle T\beta\vert T\hat{O}^{\dag}\alpha\rangle=\langle T\beta\vert T\hat{O}^{\dag}T^{-1}\vert T\alpha\rangle
\end{align}
let $\hat{O}=\hat{\chi}_{\v i\v j}^{\dag}$, and
$\vert\alpha\rangle=\vert\beta\rangle=\vert\Psi\rangle$, one immediately see
that under $T$ transformation, the average value of $\hat{\chi}_{\v i\v j}$
operator transforms as:
\begin{align}
\chi_{\v i\v j}=\langle\Psi\vert\hat{\chi}_{\v i\v j}\vert\Psi\rangle\rightarrow \chi'_{\v i\v j}&=\langle T\Psi\vert\hat{\chi}_{\v i\v j}\vert T\Psi\rangle\notag\\
&=\langle\Psi\vert\hat{\chi}_{\v i\v j}^{\dag}\vert\Psi\rangle=\chi_{\v i\v j}^{*}
\end{align} Here we know that the mean-field variable $U_{\v i\v j}$, as the
average value of $\hat{U}_{\v i \v j}$ operator, transforms into $U_{\v i\v
j}^{*}$ under $T$. This is consistent with our understanding of anti-linear
$T$ operator, since it transforms any $c$-number into its complex conjugate.
\begin{align}
T: U_{\v i\v j}\rightarrow TU_{\v i\v j}T^{-1}=U_{\v i\v j}^*
\end{align} In general, $U_{\v i\v j}$ can be written as:
\begin{align}
U_{\v i\v j}=u_{\v i\v j}^{0}\tau_0+u_{\v i\v j}^{1}\tau_1+u_{\v i\v j}^{2}\tau_2+u_{\v i\v j}^{3}\tau_3\label{Eq:u0123}
\end{align} where $\tau_0$ is identity matrix. For spin rotation invariant
system, one can show that $u_{\v i\v j}^0$ is pure imaginary, while $u_{\v i\v
j}^{i}$ $i=1,2,3$ are pure real. 

How does mean-field Hamiltonian transform under $T$? Here we need combine
transformations on $\psi$ and $U_{\v i\v j}$:
\begin{align}
\psi_{i}^{\dag}U_{\v i\v j}\psi_{\v j}\rightarrow &\psi_{\v i}^{T}(i\tau_2)^{\dag}U_{\v i\v j}^{*}(i\tau_2)\psi_{\v j}^{*}\notag\\
\mbox{(using Eq.(\ref{Eq:u0123}))  }=&\psi_{\v i}^T(-U_{\v i\v j})\psi_{\v j}^*\notag\\
=&\psi_{\v j}^{\dag}(U_{\v i\v j})^{T}\psi_i\notag\\
\mbox{($U_{\v i\v j}=U_{\v j\v i}^{\dag}$)  }=&\psi_{\v j}^{\dag}U_{\v j\v i}^{*}\psi_{\v i}
\end{align} 
Comparing with the term in original mean-field Hamiltonian
$\psi_{\v j}^{\dag}U_{\v j\v i}\psi_{\v i}$, one concludes that the
transformation of mean-field hamiltonian under $T$ can be simply expressed as
$U_{\v i\v j}\rightarrow U_{\v i \v j}^*$, with no fermion transformation. If
$U_{\v i \v j}^*$ and original $U_{\v i \v j}$ can be related by a $SU(2)$
gauge transformation, the system has time-reversal symmetry; otherwise the $T$
was broken.

For our $SU(2)$-linear state, $U_{\v i \v j}^*$ and $U_{\v i \v j}$ are indeed
related by a gauge transformation $W_{T}$.
\begin{align}
W_{T}: W_{\v i}=-(-)^{i_x+i_y}
\end{align} 
After choosing unit cell, in terms of four-component fermion
$\psi$,
\begin{align}
W_T: \left(\begin{array}{c}
\psi_{i0}\\
\psi_{i1}\\
\psi_{i2}\\
\psi_{i3}
\end{array}\right)\rightarrow \left(\begin{array}{cccc}
-1&&&\\
&1&&\\
&&1&\\
&&&-1
\end{array}
\right)\left(\begin{array}{c}
\psi_{i0}\\
\psi_{i1}\\
\psi_{i2}\\
\psi_{i3}
\end{array}\right)
\end{align}
In terms of usual Dirac fermion $\tilde{\psi}$:
\begin{align}
W_T: \tilde{\psi}\rightarrow \left(\begin{array}{cccc}
1&&&\\
&-1&&\\
&&-1&\\
&&&1
\end{array}\right)\tilde{\psi}=\gamma_0\tilde{\psi}
\end{align}
The combined transformation $W_T\circ T$ leaves $U_{\v i\v j}$ invariant. So $W_T\circ T$ is another element if PSG. And we know in terms of Dirac fermion $\tilde{\psi}$, 
\begin{align}
W_T\circ T:\;\; \tilde{\psi}\rightarrow\tilde{\psi}'=(i\tau_2)\gamma_0\tilde{\psi}
\end{align}
This is very similar to the time-reversal transformation in usual Dirac
field theory.

\subsection{Spin rotation: ``charge conjugation"} 

There is another important
symmetry, the global spin rotation. Let us think about rotation around
y-axis by 180 degree:
\begin{align}
R_{spin}: (S_x,S_y,S_z)\rightarrow (-S_x,S_y,-S_z)
\end{align}
In terms of $f$ spinon,
\begin{align}
R_{spin}:f\rightarrow f'=(i\sigma_2)f=\left(
\begin{array}{c}
f_{\downarrow}\\
-f_{\uparrow}
\end{array}
\right)
\end{align}
In terms of $\psi$ fermion,
\begin{align}
R_{spin}:\psi=\left(
\begin{array}{c}
f_{\uparrow}\\
f_{\downarrow}^{\dag}
\end{array}
\right)\rightarrow \psi'=\left(
\begin{array}{c}
f_{\downarrow}\\
-f_{\uparrow}^{\dag}
\end{array}
\right)=(i\tau_2)\psi^*\label{Eq:Rfermion}
\end{align}
It seems Eq.(\ref{Eq:Rfermion}) is identical to Eq.(\ref{Eq:Tfermion}),
which means $T$ and $R_{spin}$ are identical. This is obviously wrong, the
difference here lies in the fact that $T$ is anti-linear but $R_{spin}$ is
linear. 

In terms of 4-component Dirac fermion $\tilde{\psi}$:
\begin{align}
R_{spin}:\tilde{\psi}=\mathbf{R}^{\dag}\psi\rightarrow& \mathbf{R}^{\dag}(i\tau_2)\psi^*=(i\tau_2)\mathbf{R}^{\dag}\mathbf{R}^*\tilde{\psi}^*\notag\\
=& (i\tau_2)(-\gamma_1\gamma_5)\tilde{\psi}^*\label{Eq:Rtilde}
\end{align}

The transformation rule above is in real space. In momentum space, 
\begin{align}
R_{spin}:\tilde{\psi}_k&=\frac{1}{\sqrt{N}}\sum_j e^{-ikj}\tilde{\psi}_j\notag\\
&\rightarrow (-i\tau_2)(\gamma_1\gamma_5)\frac{1}{\sqrt{N}}\sum_j e^{-ikj}\tilde{\psi}_j^{*}\notag\\
&=(-i\tau_2)(\gamma_1\gamma_5)\tilde{\psi}_{-k}^{*}
\end{align}
i.e., the spin rotation transformation $R_{spin}$ also flip the sign of momentum!

Now think about how $\hat{U}_{\v i \v j}$ transforms under $R_{spin}$. The form
of Eq.(\ref{Eq:Uinpsi}) is explicitly $R_{spin}$ invariant, so $\hat{U}_{\v i\v
j}$ is $R_{spin}$ invariant. This is expected since we have spin rotation invariant $\hat{\chi}$ and
$\hat{\eta}$. This is actually the point of
this slave-boson mean-field approach. We choose the mean-field variables to be
spin rotation singlets, such that even it acquires non-zero average value, it
doesn't break spin-rotation symmetry, which describes spin liquid states.
Moreover, since $R_{spin}$ is a linear operator, we know that the average value
of $\hat{U}_{\v i \v j}$ is also $R_{spin}$ invariant. 
\begin{align}
R_{spin}: U_{\v i\v j}\rightarrow R_{spin}U_{\v i\v j}R_{spin}^{-1}=U_{\v i\v j}\label{Eq:RUij}
\end{align}
This is consistent with the usual property of linear operator: it commutes with
$c$-numbers.

What is the transformation on mean-field Hamiltonian? We should put
transformations on fermion Eq.(\ref{Eq:Rfermion}) and $U_{\v i \v j}$
Eq.(\ref{Eq:RUij}) together:
\begin{align}
R_{spin}: \psi_{\v i}^{\dag}U_{\v i\v j}\psi_{\v j}\rightarrow &\psi_{\v i}^T (i\tau_2)^{\dag}U_{\v i\v j}(i\tau_2)\psi_{\v j}^*\notag\\
\mbox{using Eq.(\ref{Eq:u0123})   }=&\psi_{\v i}^T (-U_{\v i\v j}^*)\psi_{\v j}^*\notag\\
=&\psi_{\v j}^{\dag}U_{\v i\v j}^{\dag}\psi_{\v i}=\psi_{\v j}^{\dag}U_{\v j\v i}\psi_{\v i}\notag
\end{align}
This is just another term in the original mean-field Hamiltonian. We conclude
that $H_{mean}$ is invariant under $R_{spin}$. This again is expected since
our spin-liquid states should not break $R_{spin}$. So $R_{spin}$ is also an
element of PSG.

Here we point out that the form of $R_{spin}$ transformation
Eq.(\ref{Eq:Rtilde}) is very similar to the charge-conjugation transformation
$C$ in usual Dirac field theory. In fact we can denote $R_{spin}$ as $C$ in
our later discussion.
We summarize the transformations on Dirac fermion in Table
\ref{Tb:PSGonfermion}.
\begin{table}
\begin{tabular}{|c|c|}
\hline
PSG element&Transformation on fermion\\
\hline
$T_x^{PSG}=W_{T_x}\circ T_x$&$\gamma_3$\\
\hline
$T_y^{PSG}=W_{T_y}\circ T_y$&$\gamma_5$\\
\hline
$P_x^{PSG}=W_{P_x}\circ P_x$&$\gamma_1$\\
\hline
$P_y^{PSG}=W_{P_y}\circ P_y$&$\gamma_2$\\
\hline
$P_{xy}^{PSG}=W_{P_{xy}}\circ P_{xy}$&$\frac{1}{2}(\gamma_1-\gamma_2)(\gamma_3+\gamma_5)$\\
\hline
$T^{PSG}=W_{T}\circ T$&$\tilde{\psi}\rightarrow(i\tau_2)\gamma_0\tilde{\psi}^*$ (anti-linear)\\
\hline
$C^{PSG}=W_{C}\circ C$&$\tilde{\psi}\rightarrow-(i\tau_2)\gamma_1\gamma_5\tilde{\psi}^*$ (linear)\\
\hline
\end{tabular}
\caption{Transformations of Dirac fermion $\tilde{\psi}$, here $C$ is $R_{spin}$ and $W_{C}$ is identity.}
\label{Tb:PSGonfermion}
\end{table}

\subsection{Transformations of fermion bilinears}

Let us focus on $SU(2)$-linear state. We know that the mean-field Hamiltonian
is characterized by a $SU(2)$ doublet of massless Dirac fermions. But can the
mass gap be generated after including quantum fluctuations? If yes then the
mean-field theory cannot describe the real physical system at all since the low
energy behavior is completely different. In this section we will show that this
is not the case.

We know that for a certain mean-field state, it has a certain PSG, which is the
symmetry of the mean-field hamiltonian. After we include fluctuation, this
symmetry will still be respected. Suppose we go through an renormalization
group process to find the low energy theory, any counter-term explicitly
breaking PSG will not be generated when taking care of fluctuations. 

This can be viewed in the following way: let $L(\psi_{\v i},U_{\v i\v
j},a^l_{0\v i})$ in Eq.(\ref{E:single}) be the Lagrangian describing the
dynamics of fermion and gauge fields. In the original theory, we have a huge
"symmetry" group leaving $L$ invariant, for example, the translation along
$x$-axis by one lattice site $T_x$:
\begin{align}
L(\psi_{\v i},\hat{U}_{\v i\v j})&=L(T_x\psi_{\v i}T_x^{-1},T_x \hat{U}_{\v i\v j}T_x^{-1})\notag\\
&=L(\psi_{\v i-\v x},\hat{U}_{\v i-\v x,\v j-\v x})
\end{align}
or an arbitrary local $SU(2)$ gauge ``symmetry" transformation $W$ (here the
meaning of quotion mark is that gauge ``symmetry" is not a physical symmetry, instead it is
just a many-to-one bad labelling.):
\begin{align}
L(\psi_{\v i},\hat{U}_{\v i\v j})&=L(W_{\v i}\psi_{\v i},W_{\v i} \hat{U}_{\v i\v j}W_{\v j}^{\dag})
\end{align}

But after $\hat{U}_{\v i\v j}$ condense, things are different. The above huge
"symmetry" group will be ``spontaneously'' broken (the unbroken state
must have $U_{\v i\v j}=0$). PSG is the remaining unbroken ``symmetry" group 
after this symmetry breaking. PSG is defined as the collection of all
transformations leaving the Lagrangian $L(\psi_{\v i},\hat{U}_{\v i\v
j},a^l_{0\v i})$ invariant and also leaving average value $U_{\v i\v j}$
invariant. Let $P$ be an element of PSG, then 
\begin{align}
L(\psi_{\v i},\hat{U}_{\v i\v j})&=L(P\psi_{\v i}P^{-1},P\hat{U}_{\v i\v j}P^{-1})\label{Eq:PSGonL}\\
U_{\v i\v j}&=PU_{\v i\v j}P^{-1}
\end{align}
We can consider fluctuation around the average value $U_{\v i\v j}$:
\begin{align}
\hat{U}_{\v i\v j}=U_{\v i\v j}+\delta \hat{U}_{\v i\v j}
\end{align}
Plugging into Eq.(\ref{Eq:PSGonL})
\begin{align}
L(\psi_{\v i},U_{\v i\v j}+\delta\hat{U}_{\v i\v j})&=L(P\psi_{\v i}P^{-1},P(U_{\v i\v j}+\delta\hat{U}_{\v i\v j})P^{-1})\notag\\
&=L(P\psi_{\v i}P^{-1},U_{\v i\v j}+P\delta\hat{U}_{\v i\v j}P^{-1})
\end{align}
So the Lagrangian for the fluctuations
$L(\psi_{\v i},\delta\hat{U}_{\v i\v j})$ must be invariant under
PSG transformations.

We want to find out the transformations of fermion bilinears under PSG. Because
if they all transform non-trivially under PSG, they will be all forbidden.
That is why the fermions remain to be massless after including fluctuations.

\begin{table*}
\begin{tabular}{|c|c|c|c|c|c|c|c|}
\hline
&$T_x^{PSG}$&$T_y^{PSG}$&$P_x^{PSG}$&$P_y^{PSG}$&$P_{xy}^{PSG}$&$T^{PSG}$&$C^{PSG}$\\
\hline&&&&&&&\\
$\bar{\tilde{\psi}}\tilde{\psi}$&$-1$&$-1$&$-1$&$-1$&$+1$&$-1$&$-1$ \\
\hline&&&&&&&\\
$\bar{\tilde{\psi}}\gamma_0\tilde{\psi}$&$+1$&$+1$&$+1$&$+1$&$+1$&$-1$&$-1$ \\
\hline&&&&&&&\\
$i\bar{\tilde{\psi}}\gamma_3\gamma_5\tilde{\psi}$&$+1$&$+1$&$-1$&$-1$&$-1$&$-1$&$+1$ \\
\hline&&&&&&&\\
$i\bar{\tilde{\psi}}\gamma_1\gamma_2\tilde{\psi}$&$-1$&$-1$&$+1$&$+1$&$-1$&$-1$&$+1$ \\
\hline\hline
$\left(\begin{array}{cc}
i\bar{\tilde{\psi}}\gamma_{1}\tilde{\psi}\\
i\bar{\tilde{\psi}}\gamma_{2}\tilde{\psi}
\end{array}\right)$
&$+1$&$+1$&
$\left(\begin{array}{cc}
-1&\\
&+1
\end{array}\right)
$&$\left(\begin{array}{cc}
+1&\\
&-1
\end{array}\right)$&$
\left(\begin{array}{cc}
&+1\\
+1&
\end{array}\right)$&$-1$&$-1$\\
\hline
$\left(\begin{array}{cc}
\bar{\tilde{\psi}}\gamma_0\gamma_{1}\tilde{\psi}\\
\bar{\tilde{\psi}}\gamma_0\gamma_{2}\tilde{\psi}
\end{array}\right)$
&$-1$&$-1$&
$\left(\begin{array}{cc}
+1&\\
&-1
\end{array}\right)
$&$\left(\begin{array}{cc}
-1&\\
&+1
\end{array}\right)$&$
\left(\begin{array}{cc}
&+1\\
+1&
\end{array}\right)$&$+1$&$+1$\\
\hline
$\left(\begin{array}{cc}
i\bar{\tilde{\psi}}\gamma_{3}\tilde{\psi}\\
i\bar{\tilde{\psi}}\gamma_{5}\tilde{\psi}
\end{array}\right)$
&$\left(\begin{array}{cc}
-1&\\
&+1
\end{array}\right)
$&$\left(\begin{array}{cc}
+1&\\
&-1
\end{array}\right)
$&
$+1$&$+1$&$
\left(\begin{array}{cc}
&-1\\
-1&
\end{array}\right)$&$+1$&$+1$\\
\hline
$\left(\begin{array}{cc}
\bar{\tilde{\psi}}\gamma_0\gamma_{3}\tilde{\psi}\\
\bar{\tilde{\psi}}\gamma_0\gamma_{5}\tilde{\psi}
\end{array}\right)$
&$\left(\begin{array}{cc}
+1&\\
&-1
\end{array}\right)
$&$\left(\begin{array}{cc}
-1&\\
&+1
\end{array}\right)
$&
$-1$&$-1$&$
\left(\begin{array}{cc}
&-1\\
-1&
\end{array}\right)$&$+1$&$-1$\\
\hline
$\left(\begin{array}{c}
i\bar{\tilde{\psi}}\gamma_1\gamma_3\tilde{\psi}\\
i\bar{\tilde{\psi}}\gamma_2\gamma_5\tilde{\psi}
\end{array}\right)$&$\left(\begin{array}{cc}
+1&\\
&-1
\end{array}\right)
$&$\left(\begin{array}{cc}
-1&\\
&+1
\end{array}\right)
$&
$\left(\begin{array}{cc}
+1&\\
&-1
\end{array}\right)$&$\left(\begin{array}{cc}
-1&\\
&+1
\end{array}\right)$&$
\left(\begin{array}{cc}
&-1\\
-1&
\end{array}\right)$&$-1$&$-1$\\
\hline
$\left(\begin{array}{c}
i\bar{\tilde{\psi}}\gamma_1\gamma_5\tilde{\psi}\\
i\bar{\tilde{\psi}}\gamma_2\gamma_3\tilde{\psi}
\end{array}\right)$&$\left(\begin{array}{cc}
-1&\\
&+1
\end{array}\right)
$&$\left(\begin{array}{cc}
+1&\\
&-1
\end{array}\right)
$&
$\left(\begin{array}{cc}
+1&\\
&-1
\end{array}\right)$&$\left(\begin{array}{cc}
-1&\\
&+1
\end{array}\right)$&$
\left(\begin{array}{cc}
&-1\\
-1&
\end{array}\right)$&$-1$&$-1$\\
\hline
\end{tabular}
\caption{Transformations of 16 fermion bilinears. All transform non-trivially.}
\label{Tb:PSGonbilinear}
\end{table*}

Here we will consider only the fermion bilinears of form ${\tilde\psi}^\dag
\tilde\psi$.  The bilinears of forms $\tilde\psi \tilde\psi$ and
${\tilde\psi}^\dag {\tilde\psi}^\dag$ are not invariant under the spin $S_z$
rotation and are not allowed in the effective Lagrangian.  Since we are using
4-component fermion, there should be $4\times4=16$ different fermion bilinears
of form ${\tilde\psi}^\dag \tilde\psi$.  From Table \ref{Tb:PSGonfermion}, it is quite easy to find the transformations of all fermion bilinears under PSG.  It
turns out that among these 16 bilinears, as shown in Table
\ref{Tb:PSGonbilinear}, there are four 1-dimensional representations and six
2-dimensional representations of PSG.  All the fermion bilinears transform
non-trivially under the PSG.  So perturbatively, the fermions remain massless
after inclusion of fluctuations. 

Now let us consider the fermion bi-linear terms that also
contain a single spatial derivative.
Those terms represent marginal perturbations
when $N_f=\infty$. From the table \ref{Tb:PSGonbilinear}, we see that the only term
that is allowed by the PSG is
$\bar{\tilde \psi} \partial_x \gamma_1 \tilde \psi+\bar{\tilde \psi }\partial_y \gamma_2 \tilde \psi$.
All other terms are forbiden. The reason is as follows: The 1-dimensional representations together with a spatial derivative cannot be Lorentz singlet, so are ruled out. Among 2-dimensional representations together with a spatial derivative, only $\bar{\tilde \psi} \partial_x \gamma_1 \tilde \psi+\bar{\tilde \psi }\partial_y \gamma_2 \tilde \psi$ is invariant under translation $T_x^{PSG}$ (in fact, it is invariant under full PSG); all others are ruled out. But the term
$\bar{\tilde \psi} \partial_1 \gamma_1 \tilde \psi+\bar{\tilde \psi} \partial_2 \gamma_2 \tilde \psi$,
which is already present in Eq.(\ref{Eq:SU2lag}),
only changes the velocity of the fermions. In a RG study, this counter-term means a wavefunction renormalization. The low-energy effective theory Eq.(\ref{Eq:SU2lag}) remains valid.


The next question is, will there be 4-fermion interaction terms? The answer is
yes. For example, we choose a certain 1-dimension representation in Table
\ref{Tb:PSGonbilinear}, say $\bar{\tilde{\psi}}\tilde{\psi}$, then couple this
term to itself to make a 4-fermion interaction. It is obvious that this
4-fermion term is PSG invariant, which is allowed in the Lagrangian. Will this kind
of 4-fermion term change the low energy behavior drastically? The answer is
no. This is because we are in 2+1 space-time dimension, and by power counting
4-fermion terms are of dimension 4, so they are irrelevant couplings. Same argument can be done for fermion bilinear terms with second order derivatives; those terms are also irrelevant.

In summary, we have discussed the possible fermion
self-interactions. Our conclusion is that in perturbative sense, these fermion
self-interactions will not change the low energy behavior from the mean-field
result. We may say that the $SU(2)$-linear mean-field state is stable under
fermion self-interactions.

\subsection{Emergent $Sp(4)$ physical symmetry} 

In this section we will
discuss the emergent symmetry for $SU(2)$-linear phase whose low energy
effective theory is Eq.(\ref{Eq:SU2lag}). We already know that there are two
fermion-chiral symmetry generators $\gamma_3$ and $\gamma_5$, and they are
anticommuting. (Here please note that we are talking about fermion-chiral
symmetry, which is different from the spin-chirality in the later discussion about
chiral spin liquid.) Therefore the symmetry of $SU(2)$-linear phase
contains at least a global $SU(2)$ fermion-chiral Lie group whose generators are
$\gamma_3,\gamma_5$ and $i\gamma_3\gamma_5$. 

We also know that the theory should be global $SU(2)$ spin-rotation
invariant, since we are talking about a spin liquid phase here. Thus there
should be at least another $SU(2)$ spin-rotation symmetry group. However after
the particle-hole transformation we made in Eq.(\ref{Eq:p-h}), this
spin-rotation symmetry was hidden in our formalism.

The full physical symmetry group of $SU(2)$-linear phase should contain both
the $SU(2)$ fermion-chiral and $SU(2)$ spin-rotation as its subgroups. The
naive guess for the full group is $SU(2)\times SU(2)$ but it turns out to be
wrong. We will show in this section that the correct full symmetry group is
$Sp(4)$. We noticed that the same $Sp(4)$ symmetry was found earlier by Tanaka and Hu\cite{PhysRevLett.95.036402} by viewing the $\pi$-flux state as a fermionic mean field state, i.e., ignoring the effect of $SU(2)$ gauge field. Here we clarified the gauge field effect and obtained the same global flavor symmetry. Then we can classify all the fermion bilinears according to
their transformation rules under $Sp(4)$ group.

Later, we will show that after including the $SU(2)$ gauge fluctuations, the
$SU(2)$-linear state remains gapless and the correlations between various
operators remain algebraic. But the exponents of algebraic correlations
may be modified by the $SU(2)$ gauge interaction.  Classifying fermion
bilinears according to their transformation under the $Sp(4)$ group is very
important in understanding the scaling properties of those operators.  The
operators that belong to the same irreducible $Sp(4)$ representation will have
the same scaling dimension.

First we consider the spin rotation group; basically it will mix $\psi$
and $\psi^*$. To make the spin rotation transformation explicit, it is
convenient to reintroduce the $\widehat{\psi}$ formalism in
Eq.(\ref{Eq:hatpsi}):
\begin{align}
\widehat{\psi}=i\sigma_2\psi^*=\left(\begin{array}{c}
f_{\downarrow}\\
-f_{\uparrow}^{\dag}\end{array}\right)
\end{align}
Let us look at lattice fermion $\psi$ at certain site. If we put $\psi$
(2-component, corresponding to spin up and down) and $\widehat{\psi}$
(2-component) together to form a 4-component vector:
\begin{align}
\Psi=\left(
\begin{array}{c}
\psi\\
\widehat{\psi}
\end{array}
\right)
\end{align}
then it is straightforward to write down the spin rotation transformation. For example,
the rotation around $z$-axis:
\begin{align}
\psi&\rightarrow \left(\begin{array}{cc}
e^{i\theta/2}&0\\
0&e^{i\theta/2}
\end{array}\right)\psi&\widehat{\psi}&\rightarrow \left(\begin{array}{cc}
e^{-i\theta/2}&0\\
0&e^{-i\theta/2}
\end{array}\right)\widehat{\psi}
\end{align}
Therefore
\begin{align}
\Psi\rightarrow\mathbf{1}\otimes e^{i\theta\sigma_3/2}\Psi
\end{align}
where the identity matrix labels the internal space of $\psi$ (spin up and
down), while $e^{i\theta\sigma_3/2}$ acts on the space mixing $\psi$ and
$\widehat{\psi}$.

What about rotation along $y$-axis? Suppose we do an infinitesimal
transformation 
\begin{align}
f_{\uparrow}&\rightarrow f_{\uparrow}+\frac{\theta}{2}f_{\downarrow}\notag\\
f_{\downarrow}&\rightarrow f_{\downarrow}-\frac{\theta}{2}f_{\uparrow}
\end{align}
it implies:
\begin{align}
\psi_1&\rightarrow\psi_1+\frac{\theta}{2}\psi_2^*\notag\\
\psi_2&\rightarrow\psi_2-\frac{\theta}{2}\psi_1^*
\end{align}
thus
\begin{align}
\psi&\rightarrow\psi+\frac{\theta}{2}\widehat{\psi}\notag\\
\widehat{\psi}&\rightarrow\widehat{\psi}-\frac{\theta}{2}\psi
\end{align}
in terms of $\Psi$:
\begin{align}
\Psi\rightarrow\mathbf{1}\otimes e^{i\theta\sigma_2/2}\Psi
\end{align}
Similarly the rotation along $x$-axis is:
\begin{align}
\Psi\rightarrow\mathbf{1}\otimes e^{i\theta\sigma_1/2}\Psi
\end{align}
To summarize, we know that the spin rotation is acting on the space mixing $\psi$ and $\widehat{\psi}$.

Let us go to continuous limit, and consider the 4-component Dirac fermion $SU(2)$ doublet $\tilde{\psi}$ in Eq.(\ref{Eq:tildepsi}). Again we write it together with $\widehat{\tilde{\psi}}=(-\gamma_1\gamma_5)i\sigma_2\tilde{\psi}^*$, where the $(-\gamma_1\gamma_5)$ is inserted to make the spin-rotation have a simple form:
\begin{align}
\widetilde{\Psi}=\left(\begin{array}{c}\tilde{\psi}\\\widehat{\tilde{\psi}}\end{array}\right)
\end{align}
Note that actually $\widetilde{\Psi}$ has 16 components and
$16=4\times2\times2$ where 4 is the number of Dirac components, the first 2 is
for $SU(2)$ gauge doublet and the second 2 is for the space mixing
$\tilde{\psi}$ and $\widehat{\tilde{\psi}}$. From the above discussion, the space
mixing $\tilde{\psi}$ and $\widehat{\tilde{\psi}}$ is actually spin rotation
space. A generic transformation $G$ on fermion field can be written as a
transformation on $\widetilde{\Psi}$:
\begin{align}
G=G_{Dirac}\otimes G_{gauge}\otimes G_{spin}
\end{align}
where the transformations with subscripts act on each corresponding space.

The three spin rotation generators are, from above discussion:
\begin{align}
&\mathbf{1}\otimes\mathbf{1}\otimes\sigma_1, &&\mathbf{1}\otimes\mathbf{1}\otimes\sigma_2,&&\mathbf{1}\otimes\mathbf{1}\otimes\sigma_3\label{Eq:spin}
\end{align}

The fermion-chiral generator $\gamma_3$ is acting on $\tilde{\psi}$. One can
easily check that while acting on $\widetilde{\Psi}$, since
$\widehat{\tilde{\psi}}=(-\gamma_1\gamma_5)i\sigma_2\tilde{\psi}^*$, the generator has the form:
$\gamma_3\otimes\mathbf{1}\otimes\sigma_3$. Similarly one can find the other
two generators of fermion-chiral transformation. In summary, they are:
\begin{align}
&\gamma_3\otimes\mathbf{1}\otimes\sigma_3,&&\gamma_5\otimes\mathbf{1}\otimes\sigma_3,&&i\gamma_3\gamma_5\otimes\mathbf{1}\otimes\mathbf{1}\label{Eq:chiral}
\end{align}

Now if we do commutations between Eq.(\ref{Eq:spin}) and Eq.(\ref{Eq:chiral}),
the full set of symmetry generators can be found:
\begin{align}
&\mathbf{1}\otimes\mathbf{1}\otimes\boldsymbol{\sigma},&&\gamma_3\otimes\mathbf{1}\otimes\boldsymbol{\sigma},&&\gamma_5\otimes\mathbf{1}\otimes\boldsymbol{\sigma},&&i\gamma_3\gamma_5\otimes\mathbf{1}\otimes\mathbf{1}\label{sp4}
\end{align}
Totally $3+3+3+1=10$ elements, which satisfy $Sp(4)$ algebraic relation. 

Here one thing we need to mention is that the three gauge transformation
generators:
\begin{align}
\mathbf{1}\otimes\boldsymbol{\tau}\otimes\mathbf{1}
\end{align}
will also keep the Lagrangian Eq.(\ref{Eq:SU2lag}) invariant. But they are
gauge transformations and should not be taken as physical symmetries.

We just showed the emergent $Sp(4)$ global symmetry. Can the emergent continuous symmetry group larger than $Sp(4)$? The answer is no, as one can see in the following. We have totally 8 components of fermions, and they form four $SU(2)$ gauge doublets. For global symmetry we should only consider transformations invariant in the gauge sector, which means we should consider the transformation between the 4 doublets only (i.e., in flavor space but not in gauge space), including the mixing between $\psi$ and $\psi^{\dagger}$. In Majorana fermion representation, it is obvious that the allowed flavor transformations form $SO(8)$ group. The Lorentz transformations $i\gamma_0\gamma_1,i\gamma_0\gamma_2,i\gamma_1\gamma_2$ generate $SO(3)$ group in Euclidean space, and we also know that $Sp(4)=SO(5)$. The flavor symmetry $SO(5)$ and Lorentz symmetry $SO(3)$ actually commute. This $SO(5)$ is the largest continuous subgroup of $SO(8)$ which can commute with $SO(3)$ and has no common element with $SO(3)$ except for identity. Therefore $Sp(4)$ is the largest continuous global symmetry.

If we introduce $N_f$ flavors of fermions, it turns out that the emergent
symmetry group is $Sp(4N_f)$. We should also include the Lorentz symmetry.
Here by Lorentz Group we mean the continuous group $SO(2,1)$, generated by
$\gamma_1\gamma_2,\gamma_0\gamma_1,\gamma_0\gamma_2$. Note that the physical
lattice rotation is not identical to the rotation element in this emergent
Lorentz group. For example, 
according to Table \ref{Tb:PSGonfermion},
the rotation on lattice $R_{90}=P_{xy}\circ
P_{x}$ is given by
\begin{align}
R_{90}&=\frac{1}{2}(\gamma_1-\gamma_2)(\gamma_3+\gamma_5)\gamma_1\notag\\
&=\frac{1}{2}(1+\gamma_1\gamma_2)(\gamma_3+\gamma_5)\notag\\
&=e^{\frac{\pi}{4}\gamma_1\gamma_2}\frac{1}{\sqrt{2}}(\gamma_3+\gamma_5)\notag\\
&=\mbox{Dirac $90^{\circ}$ Rotation }\cdot\frac{1}{\sqrt{2}}(\gamma_3+\gamma_5)
\end{align}
in the continuum limit.
We can see that the physical rotation on lattice is a combination
of the Dirac rotation and an element in the $Sp(4)$:
$\frac{1}{\sqrt{2}}(\gamma_3+\gamma_5)$. This element actually exchanges
$\gamma_3$ and $\gamma_5$.

We should also include certain discrete symmetries such 
as time-reversal $T$, spatial
reflections $P_x,P_y,P_{xy}$,  total parity $-\boldsymbol{1}$ and charge
conjugation. But we know that charge conjugation is related to the spin
rotation, which is included in the $Sp(4)$; $P_{xy}$ is related to Dirac
rotation, $P_x$ and element $\frac{1}{\sqrt{2}}(\gamma_3+\gamma_5)$ in
$Sp(4)$; and $-\boldsymbol{1}$ is included in $Sp(4)$ as well, namely
$e^{i\pi\gamma_3}$. Therefore the full symmetry group of the low energy
effective theory for the $SU(2)$-linear state is
$Sp(4N_f)\times\mbox{Lorentz Group}\times T\times P_x\times P_y$. 
Such a symmetry group is certainly much larger than the symmetry group of the
lattice model.  (The effective theory for the $SU(2)$-linear state does
contain terms that violate the $Sp(4N_f)\times\mbox{Lorentz Group}$.  But all
those terms are irrelevant and have vanishing effects at low energies.)

One can classify the fermion bilinears according to their transformation
rules under $Sp(4)$ and Lorentz group. It is convenient to rewrite the 16
bilinears in terms of $\widehat{\tilde{\psi}}$, then in terms of
$\widehat{\Psi}$:
\begin{align*}
\overline{\tilde{\psi}}\tilde{\psi}&=-\overline{\widehat{\tilde{\psi}}}\widehat{\tilde{\psi}}\notag\\
\overline{\tilde{\psi}}\gamma_0\tilde{\psi}&=-\overline{\widehat{\tilde{\psi}}}\gamma_0\widehat{\tilde{\psi}}\notag\\
\overline{\tilde{\psi}}i\gamma_3\gamma_5\tilde{\psi}&=\overline{\widehat{\tilde{\psi}}}i\gamma_3\gamma_5\widehat{\tilde{\psi}}\notag\\
\overline{\tilde{\psi}}i\gamma_1\gamma_2\tilde{\psi}&=\overline{\widehat{\tilde{\psi}}}i\gamma_1\gamma_2\widehat{\tilde{\psi}}\notag\\
\overline{\tilde{\psi}}i\gamma_1\tilde{\psi}&=-\overline{\widehat{\tilde{\psi}}}i\gamma_1\widehat{\tilde{\psi}}\notag\\
\overline{\tilde{\psi}}i\gamma_2\tilde{\psi}&=-\overline{\widehat{\tilde{\psi}}}i\gamma_2\widehat{\tilde{\psi}}\notag\\
\overline{\tilde{\psi}}\gamma_0\gamma_1\tilde{\psi}&=\overline{\widehat{\tilde{\psi}}}\gamma_0\gamma_1\widehat{\tilde{\psi}}\notag\\
\end{align*}
\begin{align}
\overline{\tilde{\psi}}\gamma_0\gamma_2\tilde{\psi}&=\overline{\widehat{\tilde{\psi}}}\gamma_0\gamma_2\widehat{\tilde{\psi}}\notag\\
\overline{\tilde{\psi}}i\gamma_3\tilde{\psi}&=\overline{\widehat{\tilde{\psi}}}i\gamma_3\widehat{\tilde{\psi}}\notag\\
\overline{\tilde{\psi}}i\gamma_5\tilde{\psi}&=\overline{\widehat{\tilde{\psi}}}i\gamma_5\widehat{\tilde{\psi}}\notag\\
\overline{\tilde{\psi}}\gamma_0\gamma_3\tilde{\psi}&=-\overline{\widehat{\tilde{\psi}}}\gamma_0\gamma_3\widehat{\tilde{\psi}}\notag\\
\overline{\tilde{\psi}}\gamma_0\gamma_5\tilde{\psi}&=-\overline{\widehat{\tilde{\psi}}}\gamma_0\gamma_5\widehat{\tilde{\psi}}\notag\\
\overline{\tilde{\psi}}i\gamma_1\gamma_3\tilde{\psi}&=-\overline{\widehat{\tilde{\psi}}}i\gamma_1\gamma_3\widehat{\tilde{\psi}}\notag\\
\overline{\tilde{\psi}}i\gamma_2\gamma_5\tilde{\psi}&=-\overline{\widehat{\tilde{\psi}}}i\gamma_2\gamma_5\widehat{\tilde{\psi}}\notag\\
\overline{\tilde{\psi}}i\gamma_1\gamma_5\tilde{\psi}&=-\overline{\widehat{\tilde{\psi}}}i\gamma_1\gamma_5\widehat{\tilde{\psi}}\notag\\
\overline{\tilde{\psi}}i\gamma_2\gamma_3\tilde{\psi}&=-\overline{\widehat{\tilde{\psi}}}i\gamma_2\gamma_3\widehat{\tilde{\psi}}
\end{align}
Here if there is no minus sign, it transforms as singlet under spin rotation.
If there is a minus sign after the equal sign, it means the fermion bilinear has a
$\sigma_3$ in spin space, which in turn means the fermion bilinear transforms
as triplet under spin rotation. Triplet should have 3 components, but in our
16 bilinears we only included one of them (the one along $z$-axis). And the
other two are fermion bilinears of form
$\tilde\psi \tilde\psi$ and
${\tilde\psi}^\dag {\tilde\psi}^\dag$.

One can express all the fermion bilinears in terms of
$\widetilde{\Psi}$. In summary, one can organize them as in Table
\ref{Tb:sp4bilinear}. Notice that there are 10 conserved currents of $Sp(4)$, so they all have zero anomalous dimension.

\begin{table*}
\begin{tabular}{|l|c|}
\hline
Dirac Scalar (5 elements)&$\begin{array}{ccc}\overline{\widetilde{\Psi}}\mathbf{1}\otimes\mathbf{1}\otimes\boldsymbol{\sigma}\widetilde{\Psi},&\overline{\widetilde{\Psi}}i\gamma_3\otimes\mathbf{1}\otimes\mathbf{1}\widetilde{\Psi},&\overline{\widetilde{\Psi}}i\gamma_5\otimes\mathbf{1}\otimes\mathbf{1}\widetilde{\Psi}\end{array}$\\
\hline
Dirac Scalar (1 element)&$\overline{\widetilde{\Psi}}i\gamma_3\gamma_5\otimes\mathbf{1}\otimes\boldsymbol{1}\widetilde{\Psi}$
\\
\hline
Dirac Vector (30 elements)&
$\begin{array}{cccc}\overline{\widetilde{\Psi}}\gamma_0\otimes\mathbf{1}\otimes\boldsymbol{\sigma}\widetilde{\Psi},&\overline{\widetilde{\Psi}}\gamma_0\gamma_3\otimes\mathbf{1}\otimes\boldsymbol{\sigma}\widetilde{\Psi},&\overline{\widetilde{\Psi}}\gamma_0\gamma_5\otimes\mathbf{1}\otimes\boldsymbol{\sigma}\widetilde{\Psi},&\overline{\widetilde{\Psi}}i\gamma_1\gamma_2\otimes\mathbf{1}\otimes\boldsymbol{1}\widetilde{\Psi}\\
\overline{\widetilde{\Psi}}i\gamma_1\otimes\mathbf{1}\otimes\boldsymbol{\sigma}\widetilde{\Psi},&\overline{\widetilde{\Psi}}i\gamma_1\gamma_3\otimes\mathbf{1}\otimes\boldsymbol{\sigma}\widetilde{\Psi},&\overline{\widetilde{\Psi}}i\gamma_1\gamma_5\otimes\mathbf{1}\otimes\boldsymbol{\sigma}\widetilde{\Psi},&\overline{\widetilde{\Psi}}\gamma_0\gamma_2\otimes\mathbf{1}\otimes\boldsymbol{1}\widetilde{\Psi}\\
\overline{\widetilde{\Psi}}i\gamma_2\otimes\mathbf{1}\otimes\boldsymbol{\sigma}\widetilde{\Psi},&\overline{\widetilde{\Psi}}i\gamma_2\gamma_3\otimes\mathbf{1}\otimes\boldsymbol{\sigma}\widetilde{\Psi},&\overline{\widetilde{\Psi}}i\gamma_2\gamma_5\otimes\mathbf{1}\otimes\boldsymbol{\sigma}\widetilde{\Psi},&\overline{\widetilde{\Psi}}\gamma_0\gamma_1\otimes\mathbf{1}\otimes\boldsymbol{1}\widetilde{\Psi}\\
\end{array}$\\
\hline
\end{tabular}
\caption{Under $Sp(4)$ and Lorentz group, all fermion bilinears can be
classified into 3 groups. A group of Dirac scalar and $Sp(4)$ 5-dimenion representation, a group of Dirac
scalar and $Sp(4)$ singlet, and a group with Dirac vectors in it. For a
given group of bilinears, they are connected by $Sp(4)$ transformation in each
row, and connected by Lorentz group in each column. Totally there are 36
bilinears. All elements in a given group have the same scaling dimension. In the
group of Dirac vector, we actually have conserved current
corresponding to each column, totally 10 conserved currents. Those are the
conserved $Sp(4)$ currents, and they all have zero anomalous dimension after
inclusion of the $SU(2)$ gauge interaction.}
\label{Tb:sp4bilinear}
\end{table*} 

In Table \ref{Tb:sp4bilinear}, we enumerate all the fermion bilinears. But
what do they correspond to in our original spin model? For example, let us
look at a particular fermion bilinear
$\overline{\widetilde{\Psi}}\gamma_0\gamma_3\otimes\mathbf{1}\otimes\boldsymbol{\sigma}\widetilde{\Psi}$.
We will show that this term corresponds to the spin triplet bond order:
$(-)^{\v i_y}\mathbf{S}_{\v i}\times\mathbf{S}_{\v i+\v x}$. Let us write this
fermion bilinear interaction in terms of the lattice fermion operators in a
unit cell $\psi_i$ $(i=0,1,2,3)$, as shown in Fig.\ref{F:su2linearcell}:
\begin{align}
H_1=&\overline{\widetilde{\Psi}}\gamma_0\gamma_3\otimes\mathbf{1}\otimes\boldsymbol{\sigma}\widetilde{\Psi}\notag\\
&=\left(\psi_0^{\dag},\psi_1^{\dag},\psi_2^{\dag},\psi_3^{\dag}\right)\left(\begin{array}{cccc}
0&-1&0&0\\
-1&0&0&0\\
0&0&0&1\\
0&0&1&0
\end{array}\right)
\left(\begin{array}{c} \psi_0\\ \psi_1 \\ \psi_2\\ \psi_3 \end{array}\right)
\end{align}

Now we can write down the Hamiltonian $H_\text{SU(2)-linear}+\theta H_1$ on
lattice. For example, the interaction between site-0 and site-1, in terms of
$f$-operator, is
\begin{align}
i\chi f_{0\alpha}^{\dag} f_{1\alpha}+\theta f_{0\alpha}^{\dag}\sigma^3_{\alpha\beta}f_{1\beta}+h.c.
\end{align}
which basically tells us that $\left<f_{0\uparrow}^{\dag} f_{1\uparrow}\right>=i\chi+\theta$, and $\left<f_{0\downarrow}^{\dag} f_{1\downarrow}\right>=i\chi-\theta$.

On the other hand, we can write down the spin operator $\mathbf{S}_0\times\mathbf{S}_1$ in terms of $f$ operators. Focusing on the $z$-component:
\begin{align}
&\langle\left(\mathbf{S}_0\times\mathbf{S}_1\right)_z\rangle\notag\\
&=-2i\langle(f_{0\uparrow}^{\dag}f_{1\uparrow}f_{1\downarrow}^{\dag}f_{0\downarrow}-f_{0\downarrow}^{\dag}f_{1\downarrow}f_{1\uparrow}^{\dag}f_{0\uparrow})\rangle\notag\\
&=-2i\left(\left<f_{0\uparrow}^{\dag} f_{1\uparrow}\right>\left<f_{0\downarrow}^{\dag} f_{1\downarrow}\right>^*-\left<f_{0\downarrow}^{\dag} f_{1\downarrow}\right>\left<f_{0\uparrow}^{\dag} f_{1\uparrow}\right>^*\right)\notag\\
&=-8\chi\theta\neq0
\end{align}

Similarly one can show that $\langle\left(\mathbf{S}_2\times\mathbf{S}_3\right)_z\rangle=8\chi\theta$, and our correspondence is established.

In Table \ref{Tb:bilinearspin} we list the correspondence between the field
theory operators and original spin operators. From this table we know that
ferromagnetic order, triplet VBS order and staggered spin chirality order all
have zero anomalous dimension and their correlation function all scale as
$\frac{1}{x^4}$ even after the inclusion of the $SU(2)$ gauge interaction. We
also know that the Neel order and VBS order have the same anomalous dimension which
turns out to be non-zero after inclusion of the $SU(2)$ gauge interaction. We noticed that the same $Sp(4)$ emergent group for $\pi$-flux state which rotates Neel order into VBS order was found\cite{PhysRevLett.95.036402} when ignoring the $SU(2)$ gauge field effect.

\begin{table*}
\begin{tabular}{|l|c|l|}
\hline
Dirac Scalar&
$\overline{\widetilde{\Psi}}\mathbf{1}\otimes\mathbf{1}\otimes\boldsymbol{\sigma}\widetilde{\Psi}$&Neel Order: $(-)^{\v i}\mathbf{S}_{\v i}$\\
\cline{2-3}
&$\overline{\widetilde{\Psi}}i\gamma_3\otimes\mathbf{1}\otimes\mathbf{1}\widetilde{\Psi}$&VBS: $(-)^{{\v i}_y}\mathbf{S}_{\v i}\cdot\mathbf{S}_{{\v i+\v x}}$\\
&$\overline{\widetilde{\Psi}}i\gamma_5\otimes\mathbf{1}\otimes\mathbf{1}\widetilde{\Psi}$&VBS: $(-)^{{\v i}_x}\mathbf{S}_{\v i}\cdot\mathbf{S}_{{\v i+\v y}}$\\
\hline
Dirac Scalar&$\overline{\widetilde{\Psi}}i\gamma_3\gamma_5\otimes\mathbf{1}\otimes\boldsymbol{1}\widetilde{\Psi}$&Uniform Spin Chirality: $\mathbf{S}_{\v i}\cdot(\mathbf{S}_{\v i+\v x}\times\mathbf{S}_{\v i+\v x+\v y})$\\
\hline
Dirac Vector&$\overline{\widetilde{\Psi}}\gamma_0\otimes\mathbf{1}\otimes\boldsymbol{\sigma}\widetilde{\Psi}$&Ferromagnetic Order: $\mathbf{S}_{\v i}$\\
\cline{2-3}
&$\overline{\widetilde{\Psi}}\gamma_0\gamma_3\otimes\mathbf{1}\otimes\boldsymbol{\sigma}\widetilde{\Psi}$&Triplet VBS Order: $(-)^{\v i_y}\mathbf{S}_{\v i}\times\mathbf{S}_{\v i+\v x}$\\
&$\overline{\widetilde{\Psi}}\gamma_0\gamma_5\otimes\mathbf{1}\otimes\boldsymbol{\sigma}\widetilde{\Psi}$&Triplet VBS Order: $(-)^{\v i_x}\mathbf{S}_{\v i}\times\mathbf{S}_{\v i+\v y}$\\
\cline{2-3}
&$\overline{\widetilde{\Psi}}i\gamma_1\gamma_2\otimes\mathbf{1}\otimes\boldsymbol{1}\widetilde{\Psi}$&Staggered Spin Chirality: $(-)^{\v i}\mathbf{S}_{\v i}\cdot(\mathbf{S}_{\v i+\v x}\times\mathbf{S}_{\v i+\v x+\v y})$\\
\hline
\end{tabular}
\caption{The correspondence between fermion field operators and original spin
operators. In the group of Dirac vector, only the density components
of each current is presented, since the other components correspond to the
flow of these densities. From this table, we know that ferromagnetic order,
triplet VBS order and staggered spin chirality order all have zero anomalous
dimension and their correlation functions all scale as $\frac{1}{x^4}$. We
also know that the Neel order and VBS order have same anomalous dimension.}
\label{Tb:bilinearspin}
\end{table*}

Here we should mention the work
done by Hermele et.al\cite{hermele}, where the
$U(1)$-linear spin liquid was discussed and the emergent symmetry is $SU(4)$,
and similar classification of totally 64 fermion bilinears was done. One can
recover their result from our formalism. From our formulation of
$SU(2)$-linear phase, the $U(1)$-linear phase can be regarded as a Higgs phase
where $SU(2)$ gauge field is broken down to $U(1)$. Let us assume the
remaining $U(1)$-gauge symmetry is along $\tau_3$ direction. The only things
one should add to recover their result are: first the gauge invariant transformations not only include those in Eq.(\ref{sp4}); we should also include
\begin{align}
\gamma_3\otimes\tau_3\otimes\mathbf{1},\gamma_5\otimes\tau_3\otimes\mathbf{1},i\gamma_3\gamma_5\otimes\tau_3\otimes\boldsymbol{\sigma}.
\end{align}
So totally 15 elements, and they form a $SU(4)$ algebra. Secondly the bilinear with a
$\tau_3$ in the gauge sector is also gauge invariant, as shown in Table
\ref{Tb:su4bilinear}.

\begin{table*}
\begin{tabular}{|l|c|}
\hline
Dirac Scalar (15 elements)&$\begin{array}{ccc}\overline{\widetilde{\Psi}}\mathbf{1}\otimes\mathbf{1}\otimes\boldsymbol{\sigma}\widetilde{\Psi},&\overline{\widetilde{\Psi}}i\gamma_3\otimes\mathbf{1}\otimes\mathbf{1}\widetilde{\Psi},&\overline{\widetilde{\Psi}}i\gamma_5\otimes\mathbf{1}\otimes\mathbf{1}\widetilde{\Psi}\end{array}$\\
&$\begin{array}{ccc}\overline{\widetilde{\Psi}}\mathbf{1}\otimes\mathbf{\tau_3}\otimes\boldsymbol{1}\widetilde{\Psi},&\overline{\widetilde{\Psi}}i\gamma_3\otimes\mathbf{\tau_3}\otimes\boldsymbol{\sigma}\widetilde{\Psi},&\overline{\widetilde{\Psi}}i\gamma_5\otimes\mathbf{\tau_3}\otimes\boldsymbol{\sigma}\widetilde{\Psi}\end{array}$\\
&$\overline{\widetilde{\Psi}}i\gamma_3\gamma_5\otimes\mathbf{\tau_3}\otimes\boldsymbol{\sigma}\widetilde{\Psi}$\\
\hline
Dirac Scalar (1 element)&$\overline{\widetilde{\Psi}}i\gamma_3\gamma_5\otimes\mathbf{1}\otimes\boldsymbol{1}\widetilde{\Psi}$
\\
\hline
Dirac Vector (45 elements)&
$\begin{array}{cccc}\overline{\widetilde{\Psi}}\gamma_0\otimes\mathbf{1}\otimes\boldsymbol{\sigma}\widetilde{\Psi},&\overline{\widetilde{\Psi}}\gamma_0\gamma_3\otimes\mathbf{1}\otimes\boldsymbol{\sigma}\widetilde{\Psi},&\overline{\widetilde{\Psi}}\gamma_0\gamma_5\otimes\mathbf{1}\otimes\boldsymbol{\sigma}\widetilde{\Psi},&\overline{\widetilde{\Psi}}i\gamma_1\gamma_2\otimes\mathbf{1}\otimes\boldsymbol{1}\widetilde{\Psi}\\
\overline{\widetilde{\Psi}}i\gamma_1\otimes\mathbf{1}\otimes\boldsymbol{\sigma}\widetilde{\Psi},&\overline{\widetilde{\Psi}}i\gamma_1\gamma_3\otimes\mathbf{1}\otimes\boldsymbol{\sigma}\widetilde{\Psi},&\overline{\widetilde{\Psi}}i\gamma_1\gamma_5\otimes\mathbf{1}\otimes\boldsymbol{\sigma}\widetilde{\Psi},&\overline{\widetilde{\Psi}}\gamma_0\gamma_2\otimes\mathbf{1}\otimes\boldsymbol{1}\widetilde{\Psi}\\
\overline{\widetilde{\Psi}}i\gamma_2\otimes\mathbf{1}\otimes\boldsymbol{\sigma}\widetilde{\Psi},&\overline{\widetilde{\Psi}}i\gamma_2\gamma_3\otimes\mathbf{1}\otimes\boldsymbol{\sigma}\widetilde{\Psi},&\overline{\widetilde{\Psi}}i\gamma_2\gamma_5\otimes\mathbf{1}\otimes\boldsymbol{\sigma}\widetilde{\Psi},&\overline{\widetilde{\Psi}}\gamma_0\gamma_1\otimes\mathbf{1}\otimes\boldsymbol{1}\widetilde{\Psi}\\
\end{array}$\\
&
$\begin{array}{cccc}&\overline{\widetilde{\Psi}}\gamma_0\gamma_3\otimes\mathbf{\tau_3}\otimes\boldsymbol{1}\widetilde{\Psi},&\overline{\widetilde{\Psi}}\gamma_0\gamma_5\otimes\mathbf{\tau_3}\otimes\boldsymbol{1}\widetilde{\Psi},&\overline{\widetilde{\Psi}}i\gamma_1\gamma_2\otimes\mathbf{\tau_3}\otimes\boldsymbol{\sigma}\widetilde{\Psi}\\
&\overline{\widetilde{\Psi}}i\gamma_1\gamma_3\otimes\mathbf{\tau_3}\otimes\boldsymbol{1}\widetilde{\Psi},&\overline{\widetilde{\Psi}}i\gamma_1\gamma_5\otimes\mathbf{\tau_3}\otimes\boldsymbol{1}\widetilde{\Psi},&\overline{\widetilde{\Psi}}\gamma_0\gamma_2\otimes\mathbf{\tau_3}\otimes\boldsymbol{\sigma}\widetilde{\Psi}\\
&\overline{\widetilde{\Psi}}i\gamma_2\gamma_3\otimes\mathbf{\tau_3}\otimes\boldsymbol{1}\widetilde{\Psi},&\overline{\widetilde{\Psi}}i\gamma_2\gamma_5\otimes\mathbf{\tau_3}\otimes\boldsymbol{1}\widetilde{\Psi},&\overline{\widetilde{\Psi}}\gamma_0\gamma_1\otimes\mathbf{\tau_3}\otimes\boldsymbol{\sigma}\widetilde{\Psi}\\
\end{array}$\\
\hline
Dirac Vector (3 elements)&$\overline{\widetilde{\Psi}}i\gamma_0\otimes\mathbf{\tau_3}\otimes\boldsymbol{1}\widetilde{\Psi},\overline{\widetilde{\Psi}}i\gamma_1\otimes\mathbf{\tau_3}\otimes\boldsymbol{1}\widetilde{\Psi},\overline{\widetilde{\Psi}}i\gamma_2\otimes\mathbf{\tau_3}\otimes\boldsymbol{1}\widetilde{\Psi}$\\
\hline
\end{tabular}
\caption{Under $SU(4)$ and Lorentz group, the total 64 bilinears can be classified into 4 groups: A group of Dirac scalars and $SU(4)$ adjoint representation with 15 elements, a group of Dirac scalar and $SU(4)$ singlet with one element, a group of Dirac vectors and $SU(4)$ 12-dimension representation with $3\times15=45$ elements, and group of Dirac vectors and $SU(4)$ singlets with 3 elements.}
\label{Tb:su4bilinear}
\end{table*}

\subsection{The effect of $SU(2)$ gauge interaction
on $SU(2)$-linear spin liquid}
\label{gaugefluctuation}

We know that in the continuous limit, the full lagrangian should be
Eq.(\ref{Eq:SU2lag}). The question is, will the $SU(2)$ gauge interaction
change the low energy behavior of mean-field theory drastically? The answer is
complicated. There are two main concerns: spontaneous chiral symmetry breaking
(SCSB) and confinement. SCSB means that fermion mass is generated dynamically.
And confinement means that no excitation with gauge charge can show up in
the spectrum, and gauge interaction is linearly confining. If any of these
happens, the low energy behavior of the system is changed drastically and we
say that the mean-field state is unstable under gauge fluctuation.  In this
case, the mean-field $SU(2)$-linear state does not lead to a stable algebraic
spin liquid.  (Actually, we do not know the low energy properties of the model
beyond the mean-field theory.)

This problem is a famous and difficult problem in QCD, 
since both effects are
non-perturbative. And these two effects are related: if there is a mass gap
generated for the fermion, then below the mass gap there is effectively no
fermion to screen the gauge interaction and we only have pure gauge field. We
know that pure gauge interaction is confining. So logically SCSB will induce
confinement. The other way, whether confinement will induce SCSB, is not clear
yet.
 
Usually it is believed that for a $SU(N)$ gauge theory coupled to $N_f$
flavor of massless fermions, there is a critical
$N_f^c$\cite{appelquist:105017,3622404,5071739}. If $N_f$ is smaller than
$N_f^c$, the system have both SCSB and confinement. However if $N_f$ is larger
than $N_f^c$, there is a conformal invariant IR fixed point. 

In particular, in 2+1 dimension, QCD always 
has a stronger interaction at low energies.
But if $N_f$ is large enough, the fermion screening effect is dominant and 
the renormalization group
flow will terminate at an IR fixed point $g_*^2\sim\frac{1}{N_f}$. The low
energy behavior is governed by that fixed point, which is deconfined and has
no SCSB. When $N_f\rightarrow \infty$, $g_*\rightarrow 0$, we are in the
perturbative region.

\begin{figure}
\includegraphics[width=0.4\textwidth]{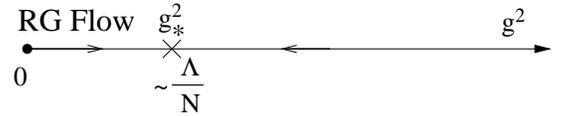}
\caption{The fixed point of many Dirac fermions coupling to $SU(2)$ gauge field}
\label{F:SU2FP}
\end{figure}

To have a quantitative study, we need work within large $N_f$
limit\cite{8105445}. Before we study the $SU(2)$ gauge fluctuation, it is
worthwhile to mention the $U(1)$ case\cite{RWspin}. The main result there was
that in the large $N_f$ expansion, the model remains gapless and the spin-spin
correlation function is a power law with an anomalous dimension $\gamma$. This
anomalous dimension $\gamma$ is calculable in large $N_f$ expansion, and up to the
$1/N_f$ (leading) order, $\gamma$ is found to be $-\frac{32}{3\pi^2
N_f}$. 

Now we look at the $SU(2)$ case in detail. We will study the $SU(2)$-linear
state in particular. The low energy effective theory is Eq.(\ref{Eq:SU2lag}).
Technically there are two ways to do large-$N_f$ limit. The first way is a
complete renormalization group analysis. To have an controlled calculation, one first does an
$\epsilon$-expansion, then studies the renormalization group flow, finds out the IR fixed point, and
the scaling dimension of operators at that fixed point, finally sets $N_f$ to be
large. This way is generally accepted and the result is thought to be reliable.
However the calculation is complicated. 

Here we do the large-$N_f$ calculation in a different way\cite{8105445}. Taking
large-$N_f$ limit:
\begin{align}
N_f\rightarrow \infty, \;\; N_fg^2\rightarrow \mbox{const.}\label{Eq:LargeN}
\end{align}
The fermion contribution to any physical quantity can be expanded in
$\frac{1}{N_f}$ systematically. This is just a way to organize the summation of
Feynman diagrams. For example, the leading order term usually corresponds to
summation of fermion one-loop diagrams. The IR fixed point can be found by
cancellation of leading correction to scaling. We will discuss this in detail
soon. Here we want to discuss whether this approach and the
first approach are equivalent. There is no general proof that these two approaches
are equivalent, but in \cite{8105445} quite a lot of examples are
presented and it was found that these two approaches are equivalent, for
example, in the case of Dirac fermions coupling to $U(1)$
gauge field, the scaling dimension and fixed point found in the two approaches
are consistent. Gracey et al.\cite{4539133,6336423} calculated large-$N_f$
expansion of anomalous dimensions of many quantities in QED$_3$ and QCD$_3$,
including fermion mass, which is gauge invariant. Then they compared the results
with results from usual MS renormalization+dimensional regularization, and
found they are consistent.

\begin{figure} 
\includegraphics[width=0.4\textwidth]{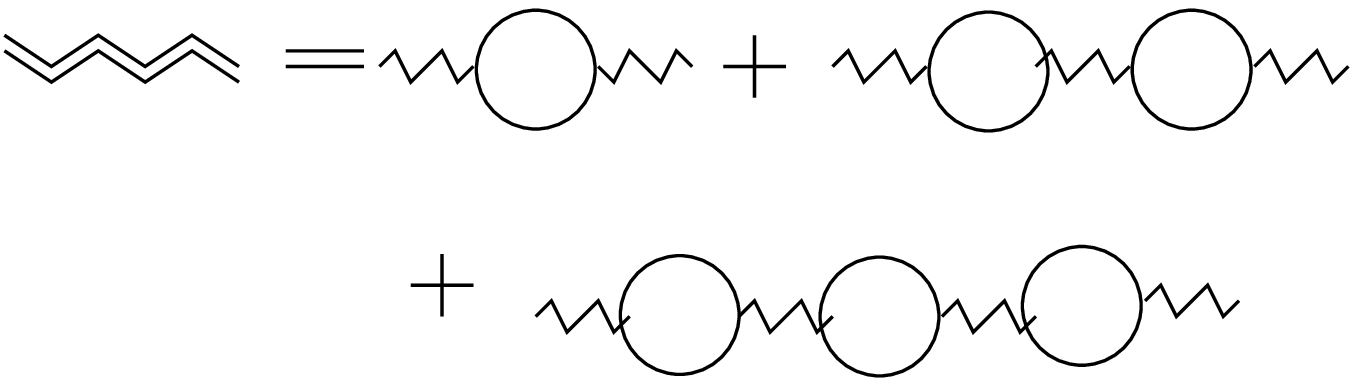}
\caption{Gauge propagator at leading order of $1/N_f$ expansion.}\label{F:gaugeprop}
\end{figure}

First let us take a look at the gauge interaction. From Eq.(\ref{Eq:LargeN}), $g^2$ is
of order $\frac{1}{N_f}$. In Figure \ref{F:gaugeprop}, all leading order
contributions to gauge propagator are  summed together. The double line
represents the leading order dressed gauge propagator.

One can calculate the gauge-gauge two point function in Fig.\ref{F:gaugeprop}. The result is:
\begin{align}
&-\frac{1}{2g^2}A_{\mu}^a(k^2\delta_{\mu\nu}-k_{\mu}k_{\nu})A_{\nu}^a\notag\\
&-N_f \mbox{Tr}\mathbf{1}\mbox{Tr}(\tau^a\tau^b)A^a_{\mu}(k^2\delta_{\mu\nu}-k_{\mu}k_{\nu})\left(\frac{1}{64k}-\frac{1}{48\pi\Lambda}\right)A^b_{\nu}\label{Eq:gaugetwopoint}
\end{align}
Here $\mathbf{1}$ is the identity matrix in Dirac spinor space. In our
$SU(2)$-linear example, $\mbox{Tr}\mathbf{1}=4$. And $\tau^a$ are usual $SU(2)$
Pauli matrices. Note that the term $\frac{1}{64k}$ is independent of
regularization scheme, but $-\frac{1}{48\pi\Lambda}$ is dependent. In different
regularization schemes, the coefficients in front of $\frac{1}{\Lambda}$ are
different. The result here $-\frac{1}{48\pi\Lambda}$ is from the Pauli-Villar
regularization.

One immediately sees that at low energy ($k\rightarrow 0$), the term
$\frac{1}{64k}$ dominates. The low energy two point function of gauge field
is:
\begin{align}
&N_f\mbox{Tr}\mathbf{1}\frac{1}{2}
A^a_{\mu}(k^2\delta_{\mu\nu}-k_{\mu}k_{\nu})\frac{1}{64k}A^a_{\nu}\notag\\
=&A_{\mu}^a\frac{N_f}{32k}(k^2\delta_{\mu\nu}-k_{\mu}k_{\nu})A_{\nu}^a
\end{align}

The IR fixed point is found by cancellation of the leading correction to
scaling. Here it means that the $-\frac{1}{2g^2}$ term and the
$-\frac{1}{48\pi\Lambda}$ term should cancel:`
\begin{align}
\frac{1}{2g_{*}^2}&=\frac{N_f\mbox{Tr}\mathbf{1}}{2}\frac{1}{48\pi\Lambda}\notag\\
\Rightarrow g_{*}^2&=\frac{12\pi\Lambda}{N_f}
\end{align}
The dimensionless coupling $g^*/\La$ describes the strength of gauge interaction
at the energy scale $\La$.  We see that dimensionless coupling $g^*/\La\to 0$
as the cut-off energy scale $\La\to 0$ and $N_f\to \infty$, as we expected in
the renormalization group flow diagram Fig. \ref{F:SU2FP}.  Remember that if the number of flavors
of fermions is small, confinement will happen, renormalization group flow will go to some fixed
point of $g^2_{*}/\La\sim 1$ as $\La\to 0$ and perturbation theory breaks down.
The message is that the fermion screening effect in our large $N_f$ model
drives the gauge interaction to weak limit at low energies.

From Eq.(\ref{Eq:gaugetwopoint}), we know that the gauge field remains massless,
and in deconfined phase. So the $SU(2)$-linear state is also stable under gauge
fluctuation. We can also see that the scaling dimension of gauge field $A$ is
$d_A=1$ at leading order. With $1/N_f$ correction, we expect to have
$d_A=1+O\left(\frac{1}{N}\right)$. Now we can understand why the fixed point
$g^2_{*}$ is IR stable. Suppose we are slightly away from the fixed point:
\begin{align}
L&=\sum_{i=1}^{N_f}\bar{\psi}_i\left( \partial_{\mu}-i a_\mu^l \tau^l \right)\gamma_{\mu}\psi_i+\frac{1}{2g_*^2} \textrm{Tr}\left[f^l_{\mu\nu}f^l_{\mu\nu}\right]\notag\\
&+\delta g \textrm{Tr}\left[f^l_{\mu\nu}f^l_{\mu\nu}\right]
\end{align}
By power counting, the scaling dimension of $\delta g$ is $4+O(\frac{1}{N_f})$, so it is irrelevant. Therefore at low energy $\delta g$ flows to zero and $g^2_{*}$ is IR stable. 

In the above discussion, we have shown that the $SU(2)$-linear mean-field state is stable
under fermion self-interaction and gauge fluctuation in large $N_f$ limit. So it is a stable phase.
Here by ``stable" we mean that the low energy behavior does not change
drastically from mean-field result. When $N_f=\infty$ we are back to the
mean-field result, but if $N_f$ large but finite, the low energy behavior is
changed from the mean-field result, but not drastically. Below we will see what
this change is.

\section{Spin-spin correlation function in $SU(2)$-linear phase}

In this section we investigate the spin-spin correlation function. In a frustrated spin liquid,
there is always a strong trend to antiferromagnetic long range order. To
describe this trend, we can calculate the staggered spin-spin correlation
function in fermion $\tilde{\psi}$ formalism:
\begin{align}
\<(-1)^xS^z(x)S^z(0)\>=\frac{1}{64}
\left(\<\bar{\tilde{\psi}}\tilde{\psi}(x) 
\bar{\tilde{\psi}}\tilde{\psi}(0)\>- 
\<\bar{\tilde{\psi}}\tilde{\psi}\>^2\right)
\end{align}
Here $\tilde{\psi}$ is actually an 8-component fermion operator, with both spinor indices and $SU(2)$ gauge indices.

\begin{figure}
\includegraphics[width=0.1\textwidth]{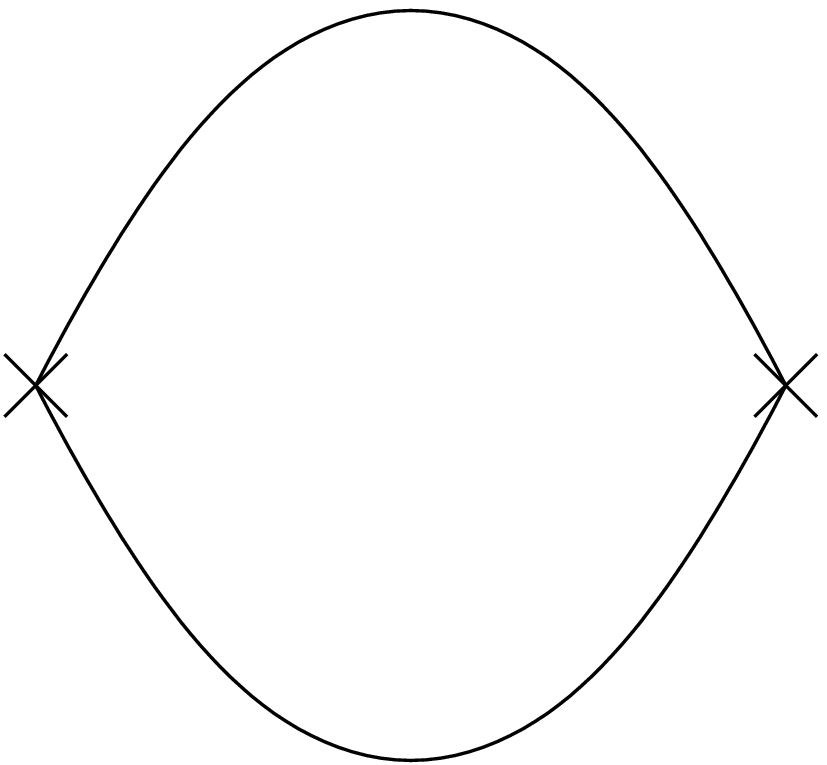}
\caption{Spin-spin correlation function at zeroth order. Cross means spin operator insertion.}\label{F:zero-order}
\end{figure}

One can see that in field theory language, staggered spin-spin correlation function is nothing but the mass operator correlation function. At zeroth order, we have the free fermion Feynman diagram (Fig. \ref{F:zero-order}).
The staggered spin-spin correlation function in momentum space at zeroth order is
\begin{align}
<S^z_s(q)S^z_s(-q)\>_0&=-\frac{1}{64}\int\frac{dp^3}{(2\pi)^3}\mbox{Tr}\left[G_{\alpha\beta}(p)G_{\beta\alpha}(q-p)\right]\notag\\
&=-\frac{\sqrt{q_0^2+\vec{q}^2}}{128} \label{Eq:zero-order}
\end{align}

\begin{figure}
\includegraphics[width=0.4\textwidth]{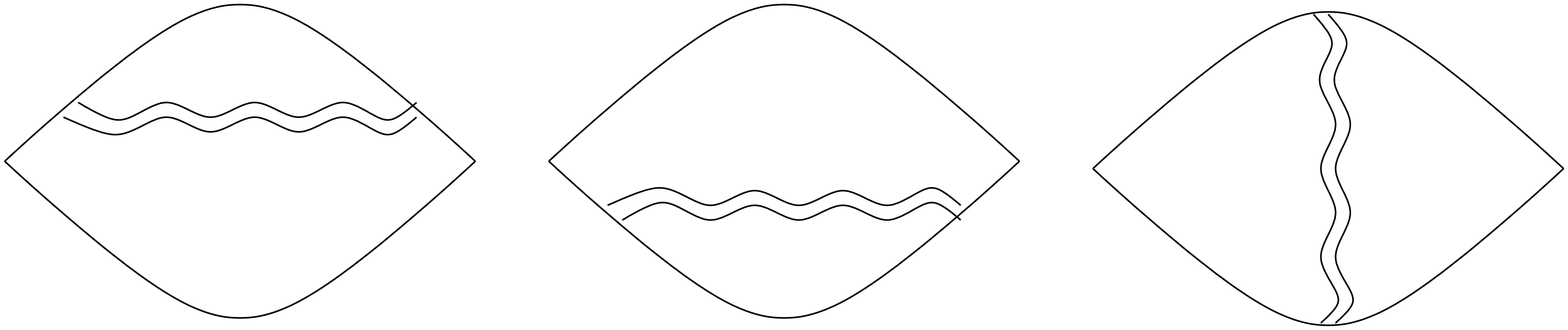}
\caption{Contribution to spin-spin correlation function at order of $\frac{1}{N_f}$}\label{F:firstorder}
\end{figure}

The first order in $\frac{1}{N_f}$ expansion involves diagrams in
Fig.\ref{F:firstorder}. Note that the double line represents the dressed gauge
field propagator:
\begin{align}
D_{\mu\nu}^{ab}=\frac{16}{N_f}\frac{1}{k}\left(\delta_{\mu\nu}-\frac{k_{\mu}k_{\nu}}{k^2}\right)\delta^{ab}
\end{align}

These diagrams will give the spin-spin correlation function an anomalous
dimension. According to Eq.(\ref{Eq:zero-order}), the spin-spin correlation
function in momentum space scales like $\sim |k|$. With $1/N_f$ correction, the
correlation function scales like $\sim |k|^{1+2\gamma}$. Here $\gamma$ is
called the anomalous dimension. The contributions from the three diagrams in
Fig.\ref{F:firstorder} give (See Appendix \ref{ap1}):
\begin{align}
\gamma=-\frac{16}{\pi^2 N_f}\label{Eq:gaugeanormaldim}
\end{align}

We see that in the SU(2)-linear state, spin-spin correlation function remains
gapless and algebraic, but the scaling dimensions of physical operators, for
example spin operator, are unusual due to the gapless gauge interaction.
Actually there is no quasi-particle in this phase.

\section{Phase transition between spin liquids}

\subsection{Phase transition between $SU(2)$-linear phase and $SU(2)$-chiral
phase} \subsubsection{The effective field theory for phase transition} In
this section we study the phase transition between $SU(2)$-linear phase (A in
phase diagram Fig.\ref{phaseJ12}) and $SU(2)$-chiral phase (D in
Fig.\ref{phaseJ12}).

We have argued that the $SU(2)$-linear state is a stable phase. What about
the $SU(2)$-chiral state in Eq.(\ref{SU2csp})? 
One can check that the fermion spectrum
is gapped at mean-field level. As for gauge field, because time-reversal
symmetry is broken in the $SU(2)$-chiral state, a Chern-Simons term is
generated after integrating out the fermions. 
And that will also give gauge field a mass gap. So
the $SU(2)$-chiral state is a fully gapped state. We know quantum fluctuation
cannot kill a gapped system perturbatively, so $SU(2)$-chiral state is also a
stable phase. 

Now we have two stable phases. According to phase diagram Fig.\ref{phaseJ12},
there is a phase transition between the two phases around $J_2=0.35$. It also
seems that the energies of the two phases connect smoothly on the same diagram,
which indicates the phase transition may be continuous. We will study this
phase transition in this section.

\begin{figure}\hspace{0.05\textwidth}
\includegraphics[width=0.15\textwidth]{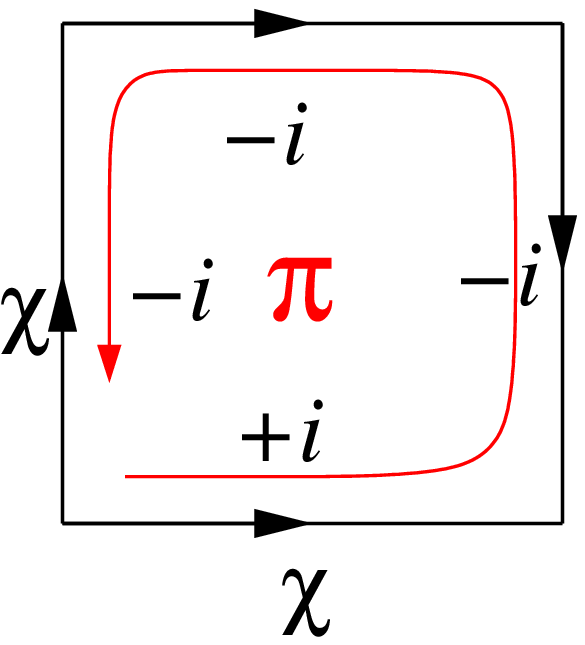}\hfill
\includegraphics[width=0.15\textwidth]{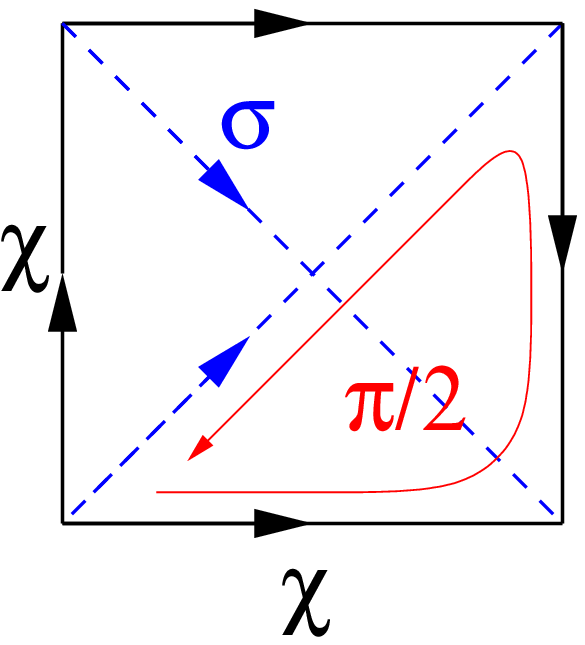}
\caption{$SU(2)$-linear phase\hfill$SU(2)$-chiral phase}\label{linear-chiral}
\end{figure}

Fig.\ref{linear-chiral} shows the plot of  the ansatzs of the two phases on lattice. Our notation here is
\begin{align}
&\mbox{$SU(2)$-linear phase:}&&
\mbox{$SU(2)$-chiral phase:}\notag\\
&\begin{array}{rl}
u_{\v i,\v i+{\v x}} =& i\chi , \\
u_{\v i,\v i+{\v y}} =& i(-)^{i_x} \chi.
\end{array}&&
\begin{array}{rl}
u_{\v i,\v i+{\v x}} =& i\chi , \\
u_{\v i,\v i+ \v y}=& i\chi (-)^{i_x}, \\
u_{\v i,\v i+\v x+\v y}=& -i\sigma (-)^{i_x}, \\ 
u_{\v i,\v i+\v x-\v y}=&  i\sigma (-)^{i_x}.
\end{array}
\end{align}
In Fig.\ref{linear-chiral}, we also show the phases one fermion gains after
hopping around a plaquette. In $SU(2)$-linear case, this phase for the square
plaquette is $\pi$; while in $SU(2)$-chiral case, this phase for the triangle
plaquette is $\frac{\pi}{2}$. Those phases indicate the fluxes through the
plaquettes. After a time-reversal transformation, the flux will change sign. For
$SU(2)$-linear case, we have $-\pi$ flux, and $-\pi$ differs from $\pi$ by
$2\pi$, thus equivalent. This indicates the $SU(2)$-linear phase respects
time-reversal symmetry. However for the $SU(2)$-chiral phase, we have
$-\frac{\pi}{2}$, which is physically different from $\frac{\pi}{2}$. So the
$SU(2)$-chiral phase breaks time-reversal symmetry and the parity symmetry. 
Other than that, one can show
that $SU(2)$-chiral phase has full translation and rotation symmetry:
\begin{align}
&\mbox{Physical symmetry of $SU(2)$-chiral phase}\notag\\
=&\{T_x,T_y,R_{90^\circ},C\}
\end{align}

The two phases involved in the phase transition, $SU(2)$-linear and
$SU(2)$-chiral states, have different PSGs, because even their physical
symmetries are different. The phase transition breaks time-reversal symmetry. 

The low energy physics at mean-field level of the systems can be derived by taking the lattice model into continuous limit:
\begin{align}
&\mbox{$SU(2)$-linear phase:}\notag\\
&L_{mean}=\bar{\psi}\partial_{\mu}\gamma_{\mu}\psi
\end{align}
\begin{align}
&\mbox{$SU(2)$-chiral phase:}\notag\\
&L_{mean}=\bar{\psi}\partial_{\mu}\gamma_{\mu}\psi+\sigma \bar{\psi}[i\gamma_3\gamma_5]\psi
\end{align}
To save notation, we dropped the tilde above the fermion operator $\psi$.

Here we see that the $\sigma$ boson field is driving the phase transition. In $SU(2)$-linear phase, $\<\sigma\>=0$; in $SU(2)$-chiral phase, $\<\sigma\>\neq 0$. Therefore to understand the phase transition, we need to know the dynamics of the $\sigma$ field. Let us include quantum fluctuations of all fields, the low energy effective theory is:
\begin{align}
L=&\sum_{i=1}^{N_f}\bar{\psi}_i\left( \partial_{\mu}-i a_\mu^l \tau^l \right)\gamma_{\mu}\psi_i+\frac{1}{2g^2} \textrm{Tr}\left[f^l_{\mu\nu}f^l_{\mu\nu}\right]\notag\\
&+\sigma \bar{\psi}[i\gamma_3\gamma_5]\psi+\frac{1}{2\rho^2}(\partial_{\mu}\sigma)^2+V(\sigma)
\label{Eq:su2chiraltransition}
\end{align}
Here to have a controlled calculation, we again introduced $N_f$ flavors of
fermion. The first line is the Dirac fermion coupling to $SU(2)$ gauge field.
The second line is the coupling between the fermion and $\sigma$ boson, and the
dynamics of $\sigma$ boson field. The potential $V(\sigma)$ is not known
yet. Nevertheless we know in $SU(2)$-linear phase, the dynamics of $\sigma$ boson gives
$\<\sigma\>=0$; while in $SU(2)$-chiral phase, $\<\sigma\>\neq 0$, and the
time-reversal symmetry is broken. This is similar to the usual formalism of the
phase transition of symmetry breaking, except for the fact that we have
gauge field involved here. 

\subsubsection{The correct effective theory from PSG consideration}

Eq. (\ref{Eq:su2chiraltransition}) is the effective Lagrangian for both phases.
What is the symmetry that the lagrangian should respect? It should respect the full
symmetry of the lattice model.  Before the symmetry breaking the $SU(2)$-linear
phase has a symmetry described by the $SU(2)$-linear PSG.  Thus, the effective
theory for the phase transition should respect the symmetry described by the
$SU(2)$-linear PSG.

We have shown that, under the $SU(2)$-linear PSG, the fermion bilinears
transform in the way described by
Table \ref{Tb:PSGonbilinear}.
For example, the transformation $P_x^{PSG}$:
\begin{align}
P_x^{PSG}:i\bar{\psi}[\gamma_3\gamma_5]\psi\rightarrow -i\bar{\psi}[\gamma_3\gamma_5]\psi
\end{align}
and
\begin{align}
P_x^{PSG}:\sigma\rightarrow -\sigma
\end{align}
In fact, $\si$ is the average of $i\bar{\psi}[\gamma_3\gamma_5]\psi$ and
transforms in the same way as  $i\bar{\psi}[\gamma_3\gamma_5]\psi$ under
the $SU(2)$-linear PSG (see Table \ref{Tb:PSGonbilinear}), thus the term
$\sigma \bar{\psi}[i\gamma_3\gamma_5]\psi$ is invariant under the full
$SU(2)$-linear PSG.  We also see that the other three possible couplings
$\sigma \bar{\psi}\psi$, $\sigma \bar{\psi}\gamma_0\psi$, and $\sigma
\bar{\psi}\gamma_1\gamma_2\psi$ are not invariant under the  $SU(2)$-linear
PSG and hence are not allowed the effective theory.

Similarly, the potential $V(\sigma)$ should also
respect $\si\to -\si$ symmetry and take a form
\begin{align}
V(\sigma)=\frac{m^2}{2}\sigma^2+\lambda \sigma^4
\end{align}
up to quartic order.  There is no cubic term since it breaks the $P_x^{PSG}$.

Here we see that PSG tells us the correct form of low energy effective theory:
\begin{align}
L=&\sum_{i=1}^{N_f}\bar{\psi}_i\left( \partial_{\mu}-i a_\mu^l \tau^l \right)
\gamma_{\mu}\psi_i+\frac{1}{2g^2} 
\textrm{Tr}\left[f^l_{\mu\nu}f^l_{\mu\nu}\right]\notag\\
&+\sigma \bar{\psi}[i\gamma_3\gamma_5]\psi+\frac{1}{2\rho^2}
(\partial_{\mu}\sigma)^2+\frac{m^2}{2}\sigma^2+\lambda \sigma^4
\label{Eq:su2chiraltransition1}
\end{align}

\begin{figure}
\hspace{0.05\textwidth}
\includegraphics[width=0.15\textwidth]{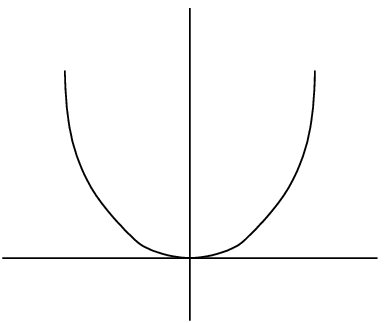}\hspace{0.1\textwidth}
\includegraphics[width=0.12\textwidth]{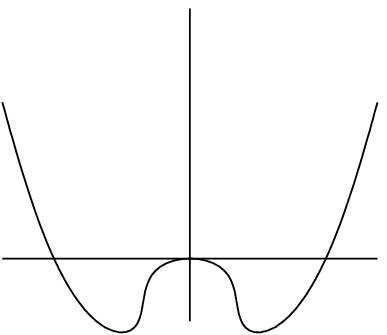}\hspace{0.05\textwidth}
\caption{The behavior of potential $V(\sigma)$ before(left) and after(right) the phase transition from $SU(2)$-linear phase to $SU(2)$-chiral phase.}\label{F:transition}
\end{figure}

At the mean-field level, we already have the picture for this phase
transition, as shown in Fig. \ref{F:transition}. We see that when $m^2>0$,
$\<\sigma\>=0$, we are in the $SU(2)$-linear phase; when $m^2<0$,
$\<\sigma\>\neq0$, we are in the $SU(2)$-chiral phase. $m^2=0$ is the phase
transition point.

At this level the phase transition is second-order, $\<\sigma\>$ changes
continuously from zero to nonzero. The next question is, will this transition
be second-order after including quantum fluctuations?

\subsubsection{The $T$ breaking phase transition does not belong to the Ising
class}\label{sigmascaling} 

To answer the above question, we need to count
the number of relevant coupling constants at the phase transition fixed point in
the renormalization group  sense. If there is only one relevant coupling
$m^2$, that means the phase transition is indeed second-order, and $m^2=0$ is
the critical point. 

\begin{figure}
\includegraphics[width=0.4\textwidth]{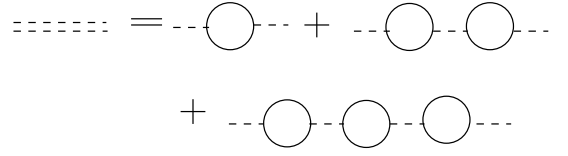}
\caption{The leading order $\sigma$ boson propagator in $\frac{1}{N_f}$ expansion.}\label{F:sigmadoubleline}
\end{figure}

We can estimate the scaling dimension $d_{\sigma}$ of $\sigma$ field.
In tree level, power counting gives us $d_{\sigma}=\frac{1}{2}$. But in
large-$N_f$ limit, the fermion dressing changes $d_{\sigma}$ strongly. The
leading order $\sigma$ propagator in $\frac{1}{N_f}$ expansion is shown in
Fig.\ref{F:sigmadoubleline}.
the $\sigma$ boson two point function at leading order is
\begin{align}
\frac{N_f}{16}\sigma \sqrt{\partial^2} \sigma
\end{align}
which indicates that the scaling dimension $d_{\sigma}=1+O(\frac{1}{N_f})$.
Therefore the scaling dimension of $(\partial\sigma)^2$ and $\sigma^4$ are both
$4+O(\frac{1}{N_f})$, which is larger than space-time dimension $3$ and thus
irrelevant. As for the gauge coupling $g$, the argument in the end of section
\ref{gaugefluctuation} is still valid. So $\delta g$ is also irrelevant. Now we
can safely say that the only relevant coupling is $\frac{m^2}{2}\sigma^2$,
whose scaling dimension is $2+O(\frac{1}{N_f})$.  

\begin{figure}
\includegraphics[width=0.4\textwidth]{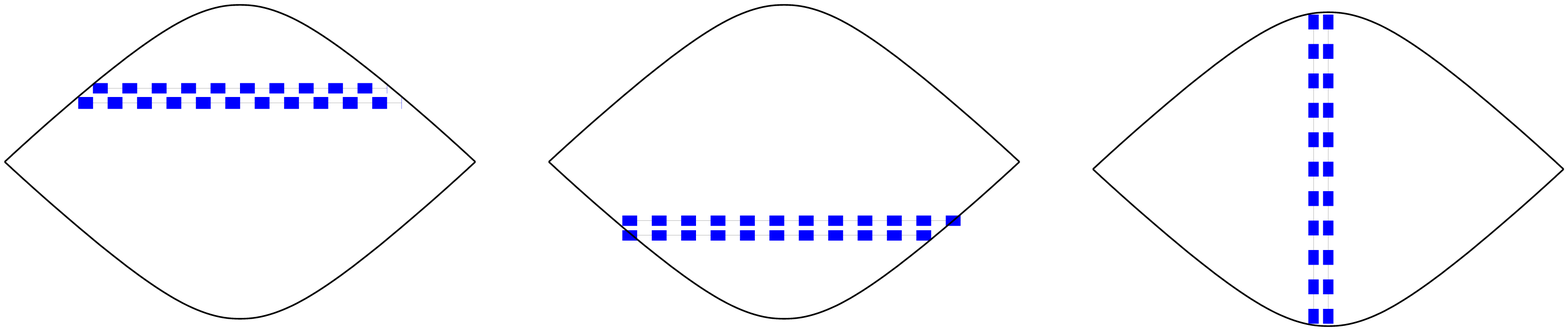}
\caption{$\sigma$ boson contribution to spin-spin correlation function at critical point at the first order of $\frac{1}{N_f}$ expansion.}\label{F:bosontospin}
\end{figure}

We can calculate the scaling behavior at the critical point where $\sigma$
boson is also massless. For example, we again compute the staggered spin-spin
correlation function. At the critical point, apart from the contribution from
massless gauge field in Fig.\ref{F:firstorder}, we have the contribution
from massless $\sigma$ boson in Fig.\ref{F:bosontospin} as well.

So at the critical point, the staggered spin-spin correlation function not only
receives an anomalous dimension from gauge field $\gamma=-\frac{16}{\pi^2
N_f}$ (Eq.(\ref{Eq:gaugeanormaldim})), but also receives an anomalous dimension
from the gapless $\sigma$ boson $\gamma'$. Detailed calculation shows that at
order of $\frac{1}{N_f}$, $\gamma'=\frac{4}{3\pi^2 N_f}$ (See Appendix
\ref{ap1}).

The total anomalous dimension $\gamma_{total}$ is the sum of $\gamma$ and
$\gamma'$:
\begin{align}
\gamma_{total}=\gamma+\gamma'=-\frac{44}{3\pi^2 N_f}
\end{align}

\begin{figure}
\includegraphics[width=0.45\textwidth]{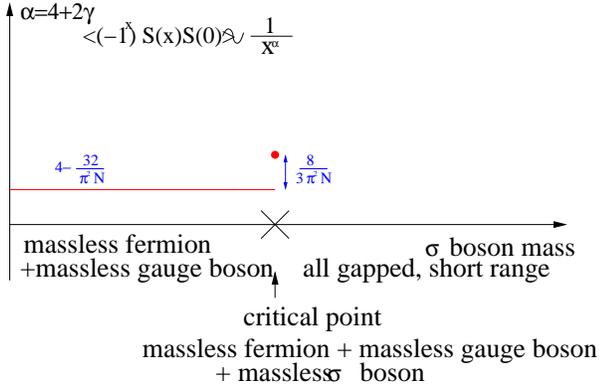}
\caption{Change of scaling dimension of staggered spin-spin correlation function during phase transition.}\label{F:chiraltransition}
\end{figure}

Fig.\ref{F:chiraltransition} shows the change of scaling dimension of staggered spin-spin correlation function during the phase transition. Note that in terms of symmetry breaking, this phase transition is quite normal: it simply breaks the time-reversal, which is a $Z_2$-like symmetry. In our usual understanding of phase transition, the Landau-Ginzburg paradigm, the symmetry determines the universality of the phase transition. Therefore our usual understanding for this phase transition should be a $Z_2$-like or Ising-like phase transition. However, our study shows it is not the case. Although the phase transition is characterized by the breaking of a $Z_2$-like symmetry, it is obviously not Ising-like. For example, an Ising like transition is gapless only at the critical point, whereas in our transition the $SU(2)$-linear phase is also gapless. The scaling behavior is very different from the Ising-universality. Actually with different value of $N_f$, the scaling exponent can have infinite number of values, which indicates infinite number of universalities.

\begin{figure}
\includegraphics[width=0.45\textwidth]{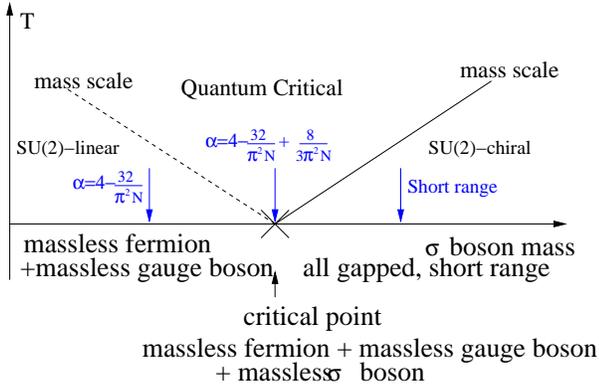}
\caption{At finite temperature $T$, system shows crossover behavior between $SU(2)$-linear phase and quantum critical region (dashed line). But the $SU(2)$-chiral phase and quantum critical region are still separated by a phase transition since there is a physical symmetry breaking (solid line).}\label{F:crossover}
\end{figure}

The above discussion is at zero temperature, but in experiments, people can only measure the system at finite temperature. What people can see in experiments actually should be crossover behavior between disordered phase and quantum critical region, as shown in Fig.\ref{F:crossover}.

\subsection{Phase transition between $SU(2)$-linear phase and $Z_2$-linear phase.}
\subsubsection{$Z_2$-linear phase.}
\begin{figure}\hspace{0.05\textwidth}
\includegraphics[width=0.15\textwidth]{su2linear-1-lattice.eps}\hfill
\includegraphics[width=0.15\textwidth]{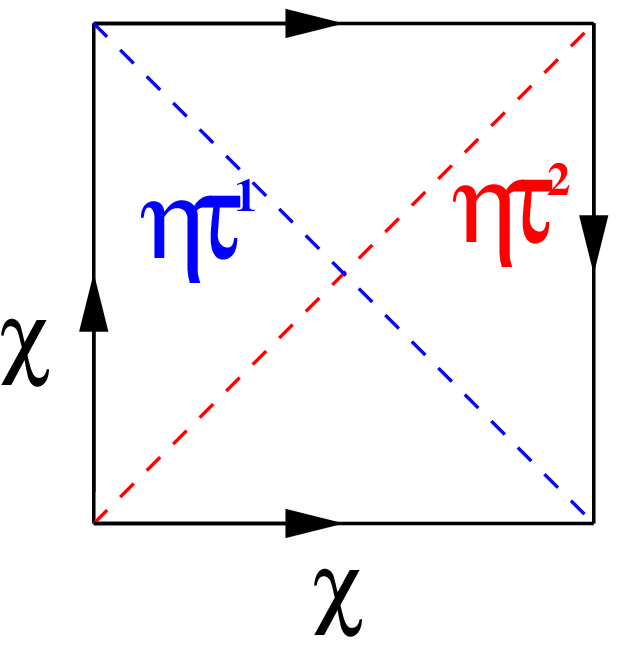}
\caption{$SU(2)$-linear phase\hfill$Z_2$-linear phase}\label{F:z2linear}
\end{figure}

In this section we study the phase transition between $SU(2)$-linear phase (A in phase diagram Fig.\ref{phaseJ12}) and $Z_2$-linear phase (G in Fig.\ref{phaseJ12}). The following is the ansatz of the $Z_2$-linear state (Fig.\ref{F:z2linear}).
\begin{align}
u_{\v i,\v i+{\v x}} =& i\chi , \notag\\
u_{\v i,\v i+{\v y}} =& i(-)^{i_x} \chi\notag\\
u_{\v i,\v i+{\v x}+{\v y}} =& \eta\tau^1, \notag\\
u_{\v i,\v i-{\v x}+{\v y}} =& \eta\tau^2.\label{Eq:z2linear}
\end{align}

\begin{figure}
\includegraphics[width=0.2\textwidth]{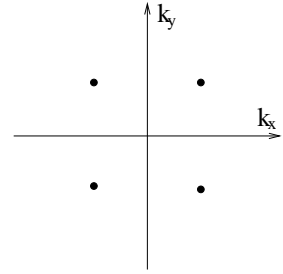}
\caption{The 4 fermi points of $Z_2$-linear state.}\label{F:4fermipoint}
\end{figure}

The fermion energy spectrum of $Z_2$-linear state at mean-field level is characterized by 4 fermi points as shown in Fig.\ref{F:4fermipoint}.

The low energy effective theory is the massless fermion coupling to $Z_2$ gauge field. At mean-field level where we do not include gauge fluctuation yet, after taking continuous limit, we have
\begin{align}
L_{mean}=&\bar{\psi}\partial_{\mu}\gamma_{\mu}\psi+\bar{\psi}\left[\eta\tau^1 i\gamma_1\gamma_5+\eta\tau^2 i\gamma_2\gamma_3\right]\psi\label{z2linear}
\end{align}

Once again we need to argue that the $Z_2$-linear state is a stable phase. Here by
stable we still mean that the low energy behavior is not changed after
including quantum fluctuations. Similar to what we did in Section
\ref{su2stable}, we can discuss the stability of $Z_2$-linear state. 

Through a PSG analysis we can show that the fermion bilinear term is not
allowed and the fermions remain massless after including quantum
fluctuation\cite{Wqoslpub}. Therefore the PSG protects the masslessness 
of fermion in
the $Z_2$-linear state. In addition, $Z_2$ gauge fluctuation is always gapped.
For pure $Z_2$ gauge theory on lattice in 2+1 dimension, it can be in
deconfined phase or confined phase\cite{W7159}. Here we have massless fermion
coupling to $Z_2$ gauge field, fermion dressing effect should drive the system
even more likely to be deconfined. We assume that the gauge field is
deconfined, then the gapped $Z_2$ gauge fluctuation should be irrelevant.
Therefore $Z_2$-linear state is also a stable phase.

In Eq.(\ref{z2linear}), $\eta\tau^1$ and $\eta\tau^2$ describe the fluxes
through the red and blue triangle plaquettes in Fig.\ref{F:z2linear}. Since the
two fluxes are not colinear, the $SU(2)$ gauge group breaks down to $Z_2$ gauge
group. In Eq.(\ref{z2linear}), one can say that the $\eta$ field is driving the
phase transition. When $\<\eta\>=0$, we go back to $SU(2)$-linear phase, and
when $\<\eta\>\neq 0$, we are in $Z_2$-linear phase. 

We notice that $\eta$ field is not gauge invariant. After a local $SU(2)$ gauge
transformation, the direction of $\tau^1$ and $\tau^2$ in Eq.(\ref{z2linear})
will be rotated.  Thus we should use a vector field $\vec n$ to describe the
fluctuations of $\eta$.  Furthermore, since the $\eta$ in front of $\gamma_1\gamma_5$
and the $\eta$ in front of $\gamma_2\gamma_3$ should be able to fluctuate
independently, we should have two vector fields, say $\vec{n}_1$ and
$\vec{n}_2$, to describe them. So the gauge invariant way to write
Eq.(\ref{z2linear}) would be:
\begin{align}
L_{mean}=\bar{\psi}\partial_{\mu}
\gamma_{\mu}\psi+\bar{\psi}\left[\left(\vec{n}_1\cdot\vec{\tau}\right)i\gamma_1
\gamma_5+\left(\vec{n}_2\cdot\vec{\tau}\right)i\gamma_2\gamma_3\right]
\psi\label{z2gaugeinvariant}
\end{align}
where $\vec{n}_1$ and $\vec{n}_2$ transform as the adjoint representation of
$SU(2)$ gauge group. Now we include dynamics of $\vec{n}$ fields. To have
controlled calculation, we introduce $N_f$ flavors of fermions, too. The
following is the low energy effective theory of the system:
\begin{align}
L=&\sum_{i=1}^{N_f}\bar{\psi}_i\partial_{\mu}\gamma_{\mu}
\psi_i+\frac{1}{2\kappa^2}\left((D_{\mu}\vec{n}_1)^2+(D_{\mu}\vec{n}_2)^2\right)
\notag\\
+&\sum_{i=1}^{N_f}\bar{\psi}_i
\left[\left(\vec{n}_1\cdot\vec{\tau}\right)i\gamma_1\gamma_5+
\left(\vec{n}_2\cdot\vec{\tau}\right)i\gamma_2\gamma_3\right]
\psi_i+V(\vec{n}_1,\vec{n}_2) 
\label{z2effective}
\end{align}
where $D_{\mu}$ is the covariant derivative of $SU(2)$ gauge theory, and the
form of potential $V(\vec{n}_1,\vec{n}_2)$ is unknown yet. The phase transition
is described by a Higgs mechanism, $\vec{n}_1$ and $\vec{n}_2$ are Higgs
bosons. When $\<\vec{n}_1\>=\<\vec{n}_2\>=0$, we are in the $SU(2)$-linear phase;
when $\<\vec{n}_1\>\neq0$, $\<\vec{n}_2\>\neq0$ and
$\<\vec{n}_1\>\perp\<\vec{n}_2\>$, we are in the $Z_2$-linear phase.

\subsubsection{The low energy effective theory from PSG consideration}

What is the symmetry that the low energy effective Lagrangian
Eq.(\ref{z2effective}) should respect? Again it should be the full symmetry
described by the PSG of the $SU(2)$-linear state. We simply need to review
Table \ref{Tb:PSGonbilinear} again to see how $\vec{n}$ fields transform under
PSG.  For example:
\begin{align}
T_x^{PSG}:\begin{array}{rcl}
\bar{\psi}[\gamma_1\gamma_5]\psi&\rightarrow&-\bar{\psi}[\gamma_1\gamma_5]\psi\\
\bar{\psi}[\gamma_2\gamma_3]\psi&\rightarrow&\bar{\psi}[\gamma_2\gamma_3]
\end{array}\\
T_y^{PSG}:\begin{array}{rcl}
\bar{\psi}[\gamma_1\gamma_5]\psi&\rightarrow&\bar{\psi}[\gamma_1\gamma_5]\psi\\
\bar{\psi}[\gamma_2\gamma_3]\psi&\rightarrow&-\bar{\psi}[\gamma_2\gamma_3]
\end{array}\\
P_{xy}^{PSG}:\begin{array}{rcl}
\bar{\psi}[\gamma_1\gamma_5]\psi&\rightarrow&-\bar{\psi}[\gamma_2\gamma_3]\\
\bar{\psi}[\gamma_2\gamma_3]\psi&\rightarrow&-\bar{\psi}[\gamma_1\gamma_5]
\end{array}
\end{align}
To have the term $\bar{\psi}\left[\left(\vec{n}_1\cdot\vec{\tau}\right)i\gamma_1\gamma_5+\left(\vec{n}_2\cdot\vec{\tau}\right)i\gamma_2\gamma_3\right]\psi$ in Eq.(\ref{z2effective}) invariant under PSG, we should have:
\begin{align}
T_x^{PSG}:\begin{array}{rcl}\vec{n}_1 &\rightarrow &-\vec{n}_1\\
\vec{n}_2 &\rightarrow &\vec{n}_2
\end{array}\\
T_y^{PSG}:\begin{array}{rcl}\vec{n}_1 &\rightarrow &\vec{n}_1\\
\vec{n}_2 &\rightarrow &-\vec{n}_2
\end{array}\\
P_{xy}^{PSG}:\begin{array}{rcl}\vec{n}_1 &\rightarrow &-\vec{n}_2\\
\vec{n}_2 &\rightarrow &-\vec{n}_1
\end{array}
\end{align}
In summary, the following three transformations of $\vec{n}$ should be the symmetry
of the potential $V(\vec n_1,\vec n_2)$:
\begin{align}
\vec{n}_1 &\rightarrow -\vec{n}_1&\vec{n}_2 &\rightarrow -\vec{n}_2&\vec{n}_1 &\leftrightarrow \vec{n}_2
\end{align}
which strongly constrains the form of potential $V(\vec{n}_1,\vec{n}_2)$. The
only allowed gauge invariant form of $V(\vec{n}_1,\vec{n}_2)$ up to quartic
order is:
\begin{align}
V(\vec{n}_1,\vec{n}_2)=&\frac{1}{2}m^2(\vec{n}_1^2+\vec{n}_2^2)+a(\vec{n}_1^4+\vec{n}_2^4)+b(\vec{n}_1^2\vec{n}_2^2)\notag\\
&+c(\vec{n}_1\cdot\vec{n}_2)^2
\end{align}
Terms like $\vec{n}_1\cdot\vec{n}_2$ and $(\vec{n}_1\cdot\vec{n}_2)(\vec{n}_1)^2$
are forbidden since they break PSG.

\subsubsection{A phase transition with no breaking symmetry.}

One can show that the two phases involved in the phase transition,
the $SU(2)$-linear phase and the $Z_2$-linear phase, have different PSGs. In
\cite{Wqoslpub}, PSG of the $SU(2)$-linear phase was labelled as $SU2Bn0$, whereas
PSG of $Z_2$-linear phase was labelled as $Z2Azz13$. But after projection, the
physical symmetries of the two phases are identical. They both have the full
symmetry of translation, rotation and time-reversal:
\begin{align}
&\mbox{Physical symmetry of $SU(2)$-linear and $Z_2$-linear states}\notag\\
&=\{T_x,T_y,P_x,P_y,P_{xy},T,C\}
\end{align}

We are investigating a phase transition with no breaking of physical symmetry. Here
the introduction of quantum order, or PSG is inevitable. Otherwise we do not
know what is changed during the phase transition.

\begin{figure}
\includegraphics[width=0.4\textwidth]{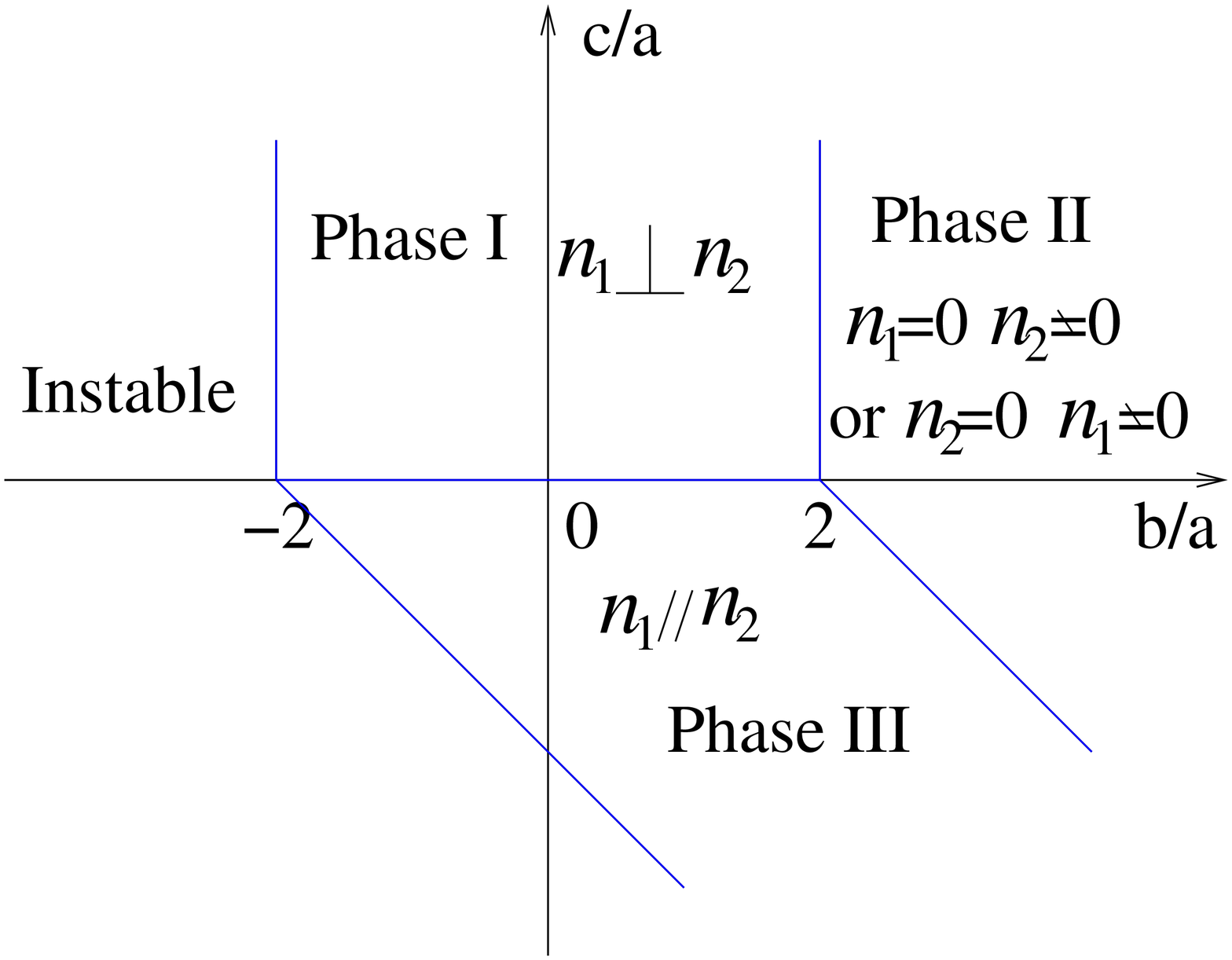}
\caption{Various Higgs condensed phases with different values of $a,b$ and $c$.
In phase I, $|\<\vec{n}_1\>|=|\<\vec{n}_2\>|$ and
$\<\vec{n}_1\>\perp\<\vec{n}_2\>$, it is the $Z_2$-linear phase. In phase II,
$\<\vec{n}_1\>=0 \mbox{ and } \<\vec{n}_2\>\neq0$ or vice versa. It is a
$U(1)$-linear phase which breaks translation and rotation symmetry. In phase
III, $|\<\vec{n}_1\>|=|\<\vec{n}_2\>|$ and
$\<\vec{n}_1\>\parallel\<\vec{n}_2\>$. It is another $U(1)$-linear phase which
breaks rotation and translation symmetry.} \label{F:higgscondensed}
\end{figure}

At mean-field level, we already have the picture for the phase transition. With
different values of coupling constant in potential $V(\vec{n}_1,\vec{n}_2)$, the
Higgs bosons $\vec{n}_1,\vec{n}_2$ may or may not condense. If they do not
condense, we are in the $SU(2)$-linear phase. If they condense in such a
fashion that $\<\vec{n}_1\>\perp\<\vec{n}_2\>$, we are in the $Z_2$-linear phase.
Detailed study of the potential shows that if $m^2>0$, Higgs bosons do not
condense. If $m^2<0$, Higgs bosons do condense, and the way of condensation is
determined by the value of parameters $a,b$ and $c$ as shown in Fig.
\ref{F:higgscondensed}. There are three different Higgs condensed phases,
labelled by I, II and III.

Our $Z_2$-linear phase is phase I. On the mean-field level, the phase
transition from the $SU(2)$-linear phase to the $Z_2$-linear phase can be described by
changing $m^2$ from positive to negative, and $m^2=0$ is the phase transition
point. 

The next question is whether this mean-field picture survives after
inclusion of quantum fluctuations. If there is only one relevant coupling
constant $m^2$, our mean-field picture remains valid, otherwise it would
fail.

We now estimate the scaling dimension of $\vec{n}$ field. Again the
fermion dressing is strong in the large-$N_f$ limit. Similar to our argument
for $\sigma$ boson in section \ref{sigmascaling}, we know that the scaling
dimension of $\vec{n}$ field is $d_{\vec{n}}=1+O(\frac{1}{N_f})$. Therefore by
power counting, the terms $(\partial_{\mu}\vec{n})^2$ and $\vec{n}^4$(those $a,b,c$
terms) are both of scaling dimension $4+O(\frac{1}{N_f})$, which are irrelevant.
However since the Lagrangian Eq.(\ref{z2effective}) is not Lorentz invariant,
calculating the $\frac{1}{N_f}$ correction to $d_{\vec{n}}$ would be
complicated. 

\begin{figure}
\includegraphics[width=0.3\textwidth]{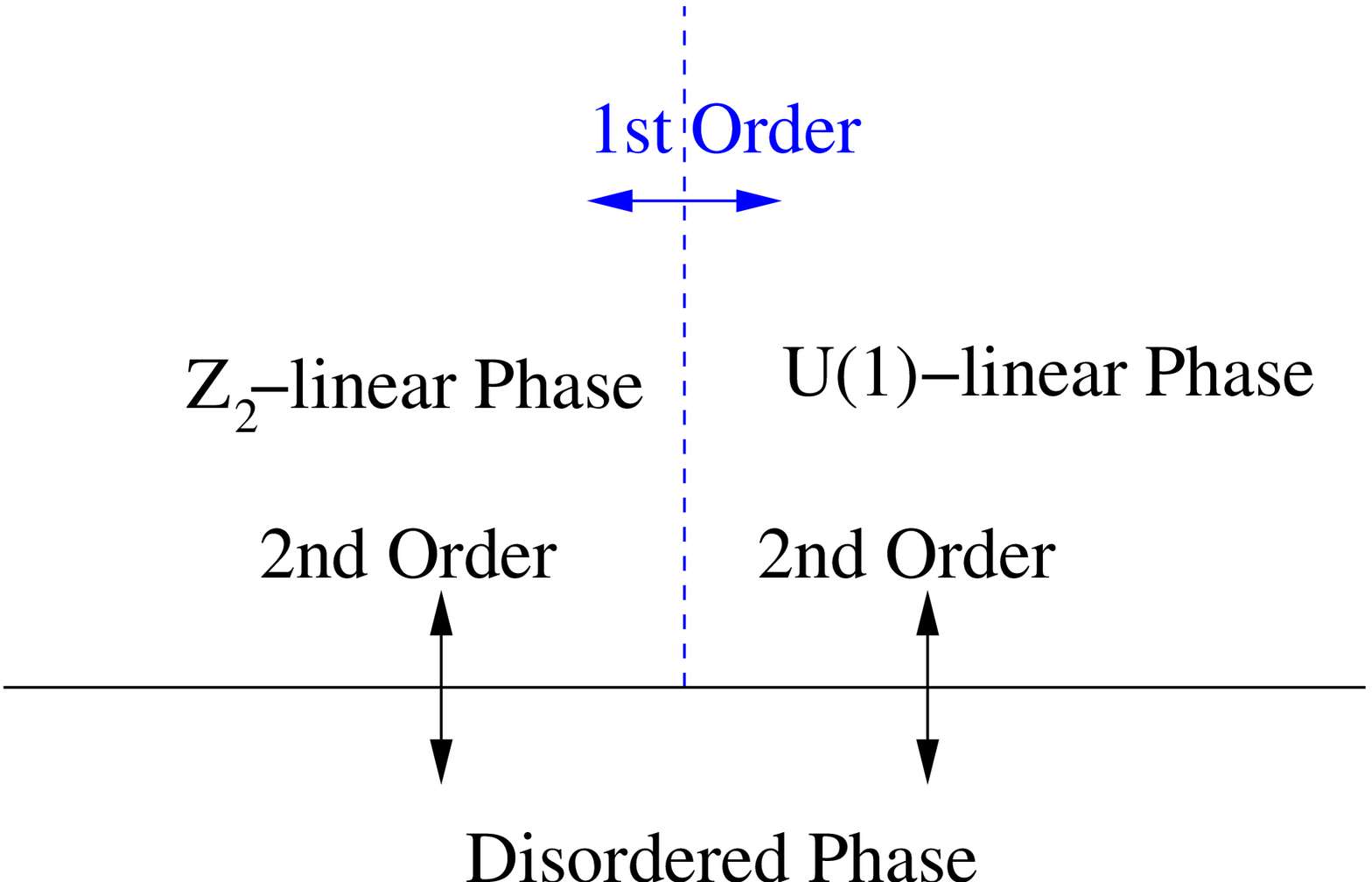}
\caption{The critical theories are the same along the whole solid line.}\label{F:samecritical}
\end{figure}

We have just concluded that there is only one relevant coupling $m^2$, so the phase
transition is second-order and $m^2=0$ is the critical point. Although $a,b,c$
couplings are irrelevant at the critical point, they are important to determine
which Higgs condensed phase the system would end up. Therefore they are
dangerous irrelevant couplings. This can be seen from Fig.\ref{F:samecritical}.
Although the critical theories for the phase transition from the $SU(2)$-linear
phase to Higgs condensed phases are the same, the system may change into
different Higgs condensed phases depending the values of couplings $a,b,c$. Different Higgs phases
separate from each other by first order transition boundaries.

\begin{figure}
\includegraphics[width=0.4\textwidth]{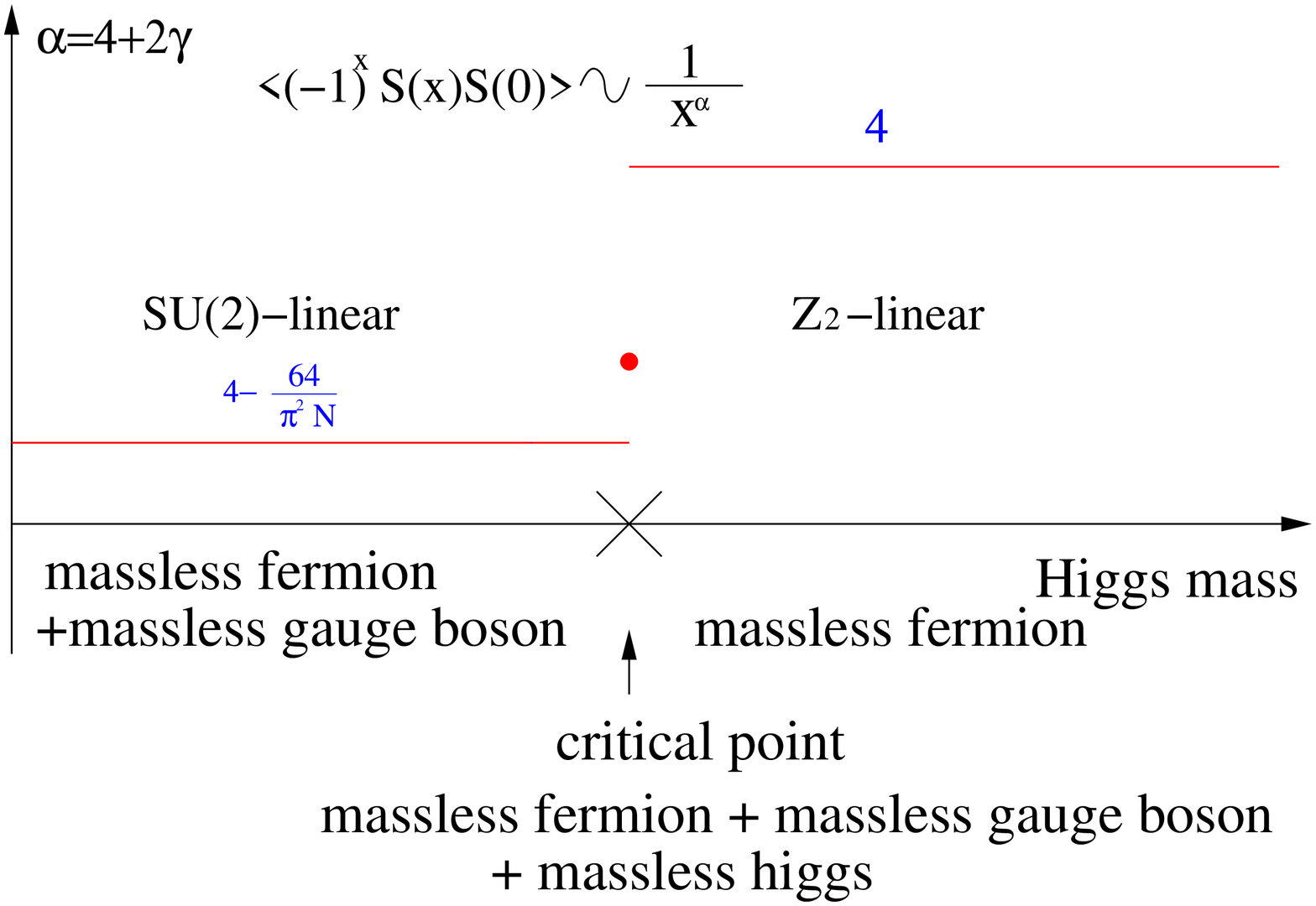}
\caption{The change of scaling exponent of staggered spin-spin correlation
function during the phase transition from $SU(2)$-linear phase to $Z_2$-linear
phase.}\label{F:z2spin}
\end{figure}

The transition from the $SU(2)$-linear state to the $Z_2$-linear state is a
phase transition without breaking of any symmetry. What are the changes in
physically measurable quantities during the phase transition? Let us think
about the staggered spin-spin correlation function again. On both sides of
the phase transition, the fermions are massless so the spin-spin correlation
functions are of power law. But the values of power are different. As shown in
Fig.\ref{F:z2spin}, in the $Z_2$-linear phase, since $Z_2$ gauge field is
gapped, spin-spin correlation does not receive anomalous dimension. At the
critical point, due to the existence of the massless Higgs fields, the
correlation function will have another scaling exponent.

\subsubsection{Phase transition from spin liquids to ordered phases} 

Spin liquids phases can also experience phase transitions into ordered phases.
Our discussion of phase transition between $SU(2)$-linear phase and chiral spin
liquid is an example where the time-reversal is broken, but no space translation
or spin rotation symmetries are broken. Here we focus on the phase transition from
spin liquids to phases breaking space translation or spin rotation
symmetries. For example, it can go into ferromagnetic phase, anti-ferromagnetic
phase or VBS phase. 

Suppose our starting point is the $SU(2)$-linear phase (or $\pi$-flux
phase). Table \ref{Tb:bilinearspin} is very useful. For example, a phase
transition from $SU(2)$-linear phase to Neel phase can be easily understood as
the opening of a mass gap $\overline{\tilde{\psi}}\psi$. Then because at energy
scale below the mass gap there is no fermion, we are left with a pure $SU(2)$
gauge field which is confined in 2+1 dimension. This phase transition then have
two ingredients in it: fermion-chiral symmetry breaking and confinement.
Because of the confinement effect the spinons are always bounded together and do not
appear in physical excitations.

Similarly one can study the phase transition from $SU(2)$-linear phase to VBS
order and ferromagnetic order. The features for these phase transitions are all
similar: opening of a mass gap and the confinement.

\section{Conclusion} 

In this article, we studied the stability of various spin liquids, including
gapless spin liquid (also called algebraic spin liquid).  Spin liquids are
defined as the disordered phases of a spin system. They have the full
space-time translation and spin rotation symmetries.  Different spin liquids
may have the same physical symmetry. To understand the physics of these phases,
first one needs to understand why they are different despite they have the same
symmetry. That means we have to classify quantum states in greater detail than
those achieved through usual symmetry group. This is the main motivation to
introduce the idea of quantum order. The PSG is just one attempt to
characterize quantum order mathematically. 

We find that PSG is very important in understanding the stability of algebraic
spin liquid.  Without PSG, it is hard to understand why fermions can remain
massless after including fluctuations around the mean-field state.  After
considering certain PSG transformations originated from lattice symmetry, we
find that such PSG transformations turns into chiral symmetry in continuous
limit, which guarantees the masslessness of fermions.  Using this idea, we show
the existence of an algebraic spin liquid -- $SU(2)$-linear state -- whose low
energy effective theory is a QCD$_3$ with $SU(2)$ gauge group.  The spin-spin
correlation function is also calculated.  We find that the $SU(2)$-linear state
has a large emergent symmetry -- $Sp(4)\times\mbox{Lorentz Group}\times T\times
P_x\times P_y$ (or $Sp(4N)\times\mbox{Lorentz Group}\times T\times P_x\times
P_y$ for the large $N$ model) -- at low energies.  The lattice model does
contain terms that violate the $Sp(4)\times\mbox{Lorentz Group}$ symmetry.  But
all those terms are shown to be irrelevant with the help of the PSG analysis.

We also discussed the continuous quantum phase transition between spin liquids.
Again, the PSG plays a key role here.  The first transition we studied is a
quantum phase transition that breaks time reversal and parity symmetries, which
is the transition between the $SU(2)$-linear state and the $SU(2)$-chiral
state.  Such a $Z_2$ symmetry breaking transition has a well defined order
parameter. However, as one can see from the calculated critical exponents, the
critical point at the transition does not belong to the Ising universality
class.  It is interesting to see that even some symmetry breaking continuous
phase transitions are beyond Landau's symmetry breaking paradigm in the sense
that critical properties are different for those obtained from Ginzburg-Landau
theory.

The second transition that we studied is a continuous quantum phase
transition between the $SU(2)$-linear state and the $Z_2$-linear state.  The
two states have the same symmetry.  Hence we show that continuous phase
transitions even exist between two phases with the \emph{same}
symmetry\cite{WWtran,CFW9349,SMF9945,RG0067,Wctpt}.  

The third transition that we studied is a continuous quantum phase transition
between the $SU(2)$-linear state and a $U(1)$-linear state.  Such a transition
breaks the lattice translation and lattice rotation symmetry.  Amazingly, we
found that the third transition and the second transition are described by the
same critical theory with the same set of critical exponents.  So it is
possible for transitions between very different states to have the same
critical point.

All above phenonmena are beyond Landau-Ginzburg paradigm for phase and phase
transition.  Those discoveries suggest that we need rethink Landau-Ginzburg
symmetry breaking approach to phase and phase transition.  We know that stable
phases and critical points can all be viewed as fixed point in the
renormalization group sense.  If a fixed point has no relevant operator that is
allowed by the symmetry (or PSG), the fixed point will represent a stable
phase.  If a fixed point has only one relevant operator that is allowed by the
symmetry (or PSG), the fixed point will represent a critical point between two
phases.  In this paper, we found that one cannot use symmetry to characterize all
the possible fixed points. New kind of order beyond the symmetry description
exists.  We showed that how to use the PSG analysis to capture the new physics
beyond Landau-Ginzburg symmetry breaking paradigm.

The phase transitions studied here are characterized by a change of quantum
order in addition to a possible change of symmetry.  This is why those phases
and phase transitions are beyond Landau-Ginzburg paradigm of breaking symmetry.
We emphasize that quantum order, or PSG, is necessary to understand the correct
low energy effective theory and the critical phenomena.  Also, an
experimental discovery of a new critical point (with unusual critical
exponents) implies a discovery of new quantum orders.  Thus it is very
important to measure critical exponents even for seemingly ordinary symmetry
breaking transitions.

This research is supported by NSF grant No.  DMR-0433632 and ARO grant No. W911NF-05-1-0474.


\appendix 

\section{Detailed calculation of anomalous dimension $\gamma$}\label{ap1} 

The scaling dimension of staggered spin-spin correlation function $\<(-)^{\v
x}\mathbf{S}(\v x)\mathbf(S)(\v 0)\>$ is calculated by the large-$N_f$
expansion of quantum field theory. In our formalism, one can show that the
staggered spin-spin correlation function is just the fermion mass operator
$\<\bar{\psi}\psi(\v x)\bar{\psi}\psi(\v 0)\>$ correlation function in the
effective theory Eq.(\ref{Eq:su2chiraltransition}). By power counting, the
scaling behavior should be $\<\bar{\psi}\psi(\v x)\bar{\psi}\psi(\v
0)\>=\frac{1}{x^4}$, but quantum fluctuations change it into
$\<\bar{\psi}\psi(\v x)\bar{\psi}\psi(\v
0)\>=\frac{1}{x^{4+2\gamma_{\bar{\psi}\psi}}}$, where $\gamma_{\bar{\psi}\psi}$
is called the anomalous dimension of fermion mass operator. It turns out that
the easiest way of calculating $\gamma_{\bar{\psi}\psi}$ is not to calculate
$\<\bar{\psi}\psi(\v x)\bar{\psi}\psi(\v 0)\>$ directly, but to calculate the
correlation function of fermion field $\psi$: $\<\psi(\v x)\bar{\psi}(\v 0)\>$,
and the three-point correlation function $\<\bar{\psi}\psi(\v x)\bar{\psi}(\v
y)\psi(\v 0)\>$. 

\begin{figure}
\includegraphics[width=0.4\textwidth]{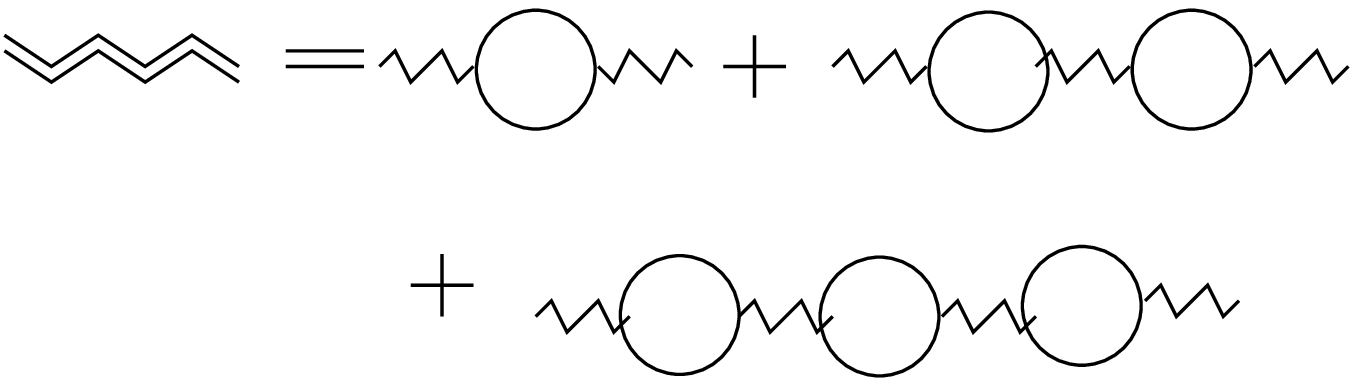}
\caption{The dressed gauge propagator in the leading order of large $N_f$ limit, which is nothing but the one-loop correction in polarization.}\label{bosondoubleline}
\end{figure}

Let us firstly calculate the staggered spin-spin correlation
function in $SU(2)$-linear phase, where the low energy effective theory is
Eq.(\ref{Eq:SU2lag}). We need to understand the gauge interaction. In the large-$N_f$ limit, the gauge field is strongly screened by fermions, and under renormalization group the coupling $g$ will flow to an IR stable conformal invariant fixed point $g^2_*\sim\frac{\Lambda}{N_f}$. Here $\Lambda$ is the UV cut-off of our theory. To the leading order of $\frac{1}{N_f}$, the dressed gauge propagator is as Fig.\ref{bosondoubleline}. Let us work within Euclidean space and Landau gauge, where the bare gauge propagator is:
\begin{align}
G^{ab}_{\mu\nu}(x,y)&=\left<A_{\mu}^a(x)A_{\nu}^b(y)\right>\\\notag
&=\int\frac{dk^3}{(2\pi)^3}e^{ik\cdot(x-y)}\frac{g^2\delta^{ab}}{k^2}(\delta_{\mu\nu}-\frac{k_{\mu}k_{\nu}}{k^2})
\end{align}
The bare fermion propagator is:
\begin{align}
\left<\psi_i(x)\bar{\psi}_j(y)\right>=\int\frac{dp^3}{(2\pi)^3}e^{ip\cdot(x-y)}\frac{-i{p\slh\delta_{ij}}}{p^2}
\end{align}
where $i,j$ label the gauge components. The dressed gauge propagator can be calculated as:
\begin{align}
&G^{ab}_{\mu\nu,dressed}(k)=\frac{g^2\delta^{ab}}{k^2(1+\Pi)}\left(\delta_{\mu\nu}-\frac{k_{\mu}k_{\nu}}{k^2}\right)\\
&\mbox{where if we do Pauli-Villar regularization,}\notag\\
&(k^2\delta_{\mu\nu}-k_{\mu}k_{\nu})\Pi=N_f g^2T_F\int\frac{dq^3}{(2\pi)^3}\frac{\mbox{Tr}\left[\gamma_{\mu}{q\slh}\gamma_{\nu}({q\slh}-{k\slh})\right]}{q^2(q-k)^2}\notag\\
=&(k^2\delta_{\mu\nu}-k_{\mu}k_{\nu})N_fg^2T_F(\frac{1}{8k}-\frac{1}{6\pi\Lambda})
\end{align}
in which
\begin{align}
\mbox{Tr}[\tau^a\tau^b]&=T_F\delta^{ab}&\Rightarrow &&T_F&=\frac{1}{2}
\end{align}
At the fixed point, where $g^2_*=\frac{6\pi\Lambda}{N_fT_F}$, the dressed gauge propagator is:
\begin{align}
G^{ab}_{\mu\nu,dressed}(k)=\frac{8\delta^{ab}}{N_fT_Fk}\left(\delta_{\mu\nu}-\frac{k_{\mu}k_{\nu}}{k^2}\right)
\end{align}

\begin{figure}
\includegraphics[width=0.2\textwidth]{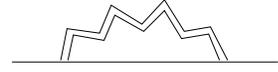}
\caption{Gauge dressed fermion propagator at first order of $\frac{1}{N_f}$.}\label{gauge_dressed_fermion}
\end{figure}
we now study the fermion correlation function with first order correction in $\frac{1}{N_f}$ expansion, as shown in Fig.\ref{gauge_dressed_fermion}. The dressed fermion propagator is:
\begin{align}
&S_{ij}(k)=\frac{-ik\slh\delta_{ij}}{k^2}(1+\Sigma)\\
&\mbox{where }\notag\\
p\slh\Sigma&=i\int\frac{dq^3}{(2\pi)^3}\frac{\gamma_{\mu}(-i)(k\slh+q\slh)\gamma_{\nu}}{(k+q)^2}\frac{C_F8}{N_fT_Fq}\left(\delta_{\mu\nu}-\frac{q_{\mu}q_{\nu}}{q^2}\right)\notag\\
&=-p\slh\frac{8C_F}{3\pi^2N_fT_F}log(\frac{k}{\Lambda})
\end{align}
and
\begin{align}
\tau^a\tau^a&=C_F\mathbf{I}&\Rightarrow&&C_F=\frac{3}{4}
\end{align}
Thus we know that the anomalous dimension of $\psi$ is:
\begin{align}
\gamma_{\psi}=-\frac{1}{2}\frac{8C_F}{3\pi^2N_fT_F}
\end{align}
\begin{figure}
\smallskip\smallskip
\hspace{\stretch{0.5}}\includegraphics[width=0.1\textwidth]{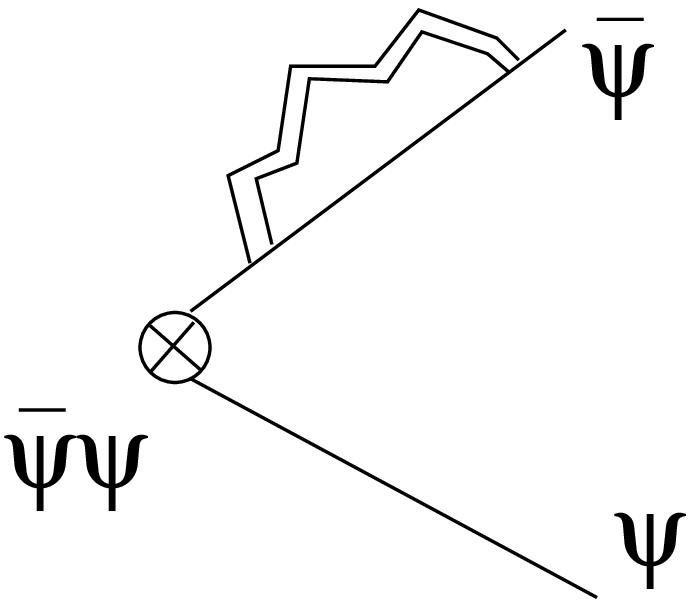}\hspace{\stretch{1}}\includegraphics[width=0.1\textwidth]{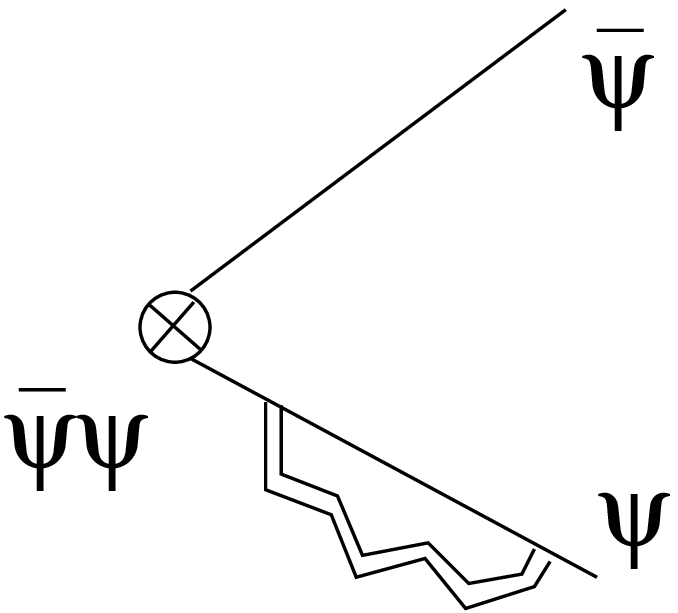}\hspace{\stretch{1}}\includegraphics[width=0.1\textwidth]{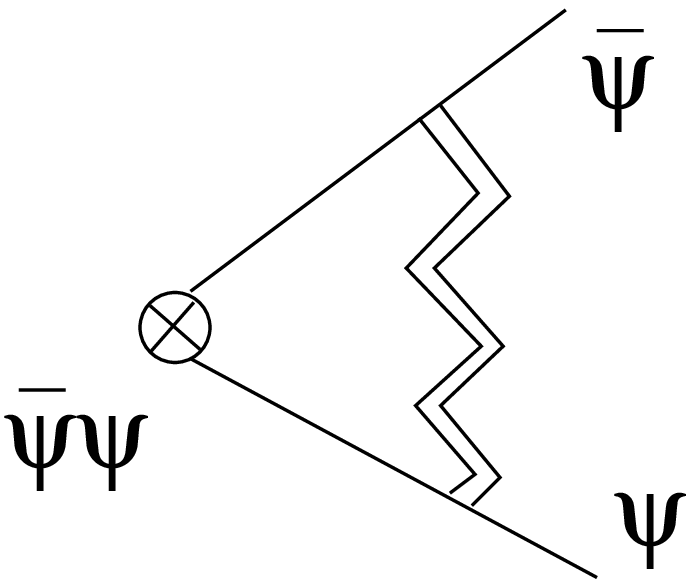}\hspace{\stretch{0.5}}\\
\hspace{\stretch{0.5}}A\hspace{\stretch{1}}B\hspace{\stretch{1}}C\hspace{\stretch{0.5}}
\caption{
Gauge dressed three point correlation function at order of $\frac{1}{N_f}$.}\label{gauge_dressed_three_point}
\end{figure}

Then we look at the dressed three-point correlation function $\<\bar{\psi}\psi(\v x)\bar{\psi}(\v y)\psi(\v 0)\>$ at the order of $\frac{1}{N_f}$, as shown in Fig.\ref{gauge_dressed_three_point}. Suppose we fix the momentum of $\bar{\psi}\psi$ to be $2k$, while $\bar{\psi}$ and $\psi$ each carry momentum $k$, then the tree level three point correlation function will be $G_3(2k,k,k)=\frac{-ik\slh}{k^2}\frac{-i(-k\slh)}{k^2}=\frac{1}{k^2}$. From the contributions of diagrams in Fig.\ref{gauge_dressed_three_point}, we will have the dressed three point correlation function:
\begin{align}
G_{3}(2k,k,k)=\frac{1}{k^2}\left(1+(A+B+C)log(\frac{k}{\Lambda})\right)
\end{align}
where $A,B,C$ are the contributions from each corresponding diagram. Actually we know that $A+B+C=\gamma_{\bar{\psi}\psi}+2\gamma_{\psi}$. So by calculating $A+B+C$, we will know the anomalous dimension of fermion mass operator $\gamma_{\bar{\psi}\psi}$.

It is easy to see that $A,B$ are just from the dressed fermion propagator, which has been calculated above: $A=B=2\gamma_{\psi}$. New calculation needs to be done for vertex correction in C. 
\begin{align}
&Clog(\frac{k}{\Lambda})\notag\\
=&\frac{8C_F}{N_fT_F}\int\frac{dq^3}{(2\pi)^3}\frac{\gamma_{\mu}(q\slh+k\slh)(q\slh-k\slh)\gamma_{\nu}}{(q+k)^2(q-k)^2q}\left(\delta_{\mu\nu}-\frac{k_{\mu}k_{\nu}}{k^2}\right)\notag\\
=&-\frac{8C_F}{\pi^2N_fT_F}log(\frac{k}{\Lambda})
\end{align}

Now we can compute $\gamma_{\bar{\psi}\psi}$:
\begin{align}
\gamma_{\bar{\psi}\psi}=&A+B+C-2\gamma_{\psi}=C+2\gamma_{\psi}\notag\\
=&-\frac{32C_F}{3\pi^2N_fT_F}=-\frac{16}{\pi^2N_f}
\end{align}

\begin{figure}
\hspace{\stretch{1}}\includegraphics[width=0.15\textwidth]{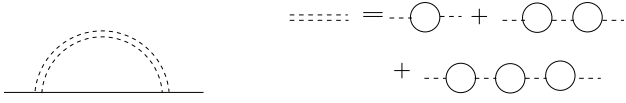}\hspace{\stretch{3}}\includegraphics[width=0.25\textwidth]{bosondoubleline.eps}
\caption{The contribution of $\sigma$-boson to fermion propagator at order of $\frac{1}{N_f}$, where the double dashed line is the dressed $\sigma$-boson propagator at leading order.}\label{boson_dressed_fermion}
\end{figure}

\begin{figure}
\smallskip\smallskip
\hspace{\stretch{0.5}}\includegraphics[width=0.1\textwidth]{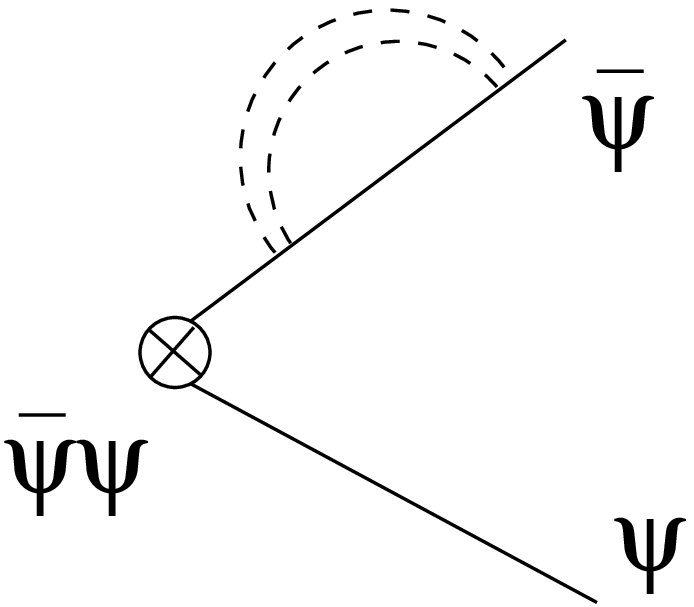}\hspace{\stretch{1}}\includegraphics[width=0.1\textwidth]{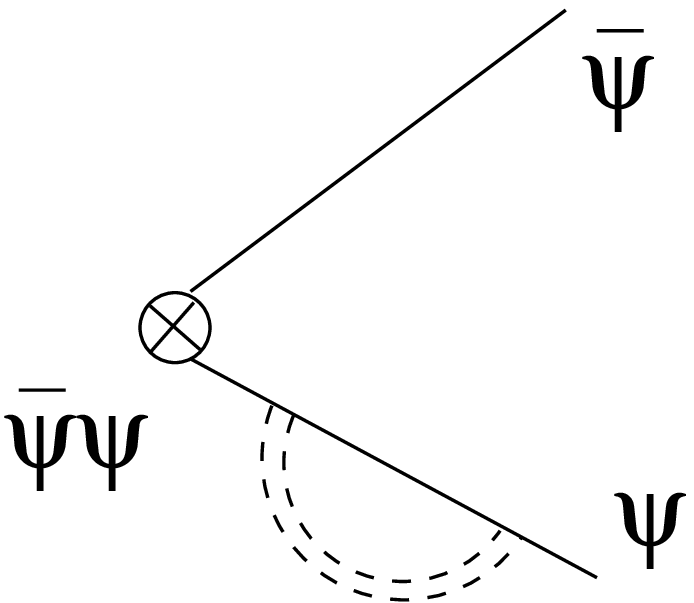}\hspace{\stretch{1}}\includegraphics[width=0.1\textwidth]{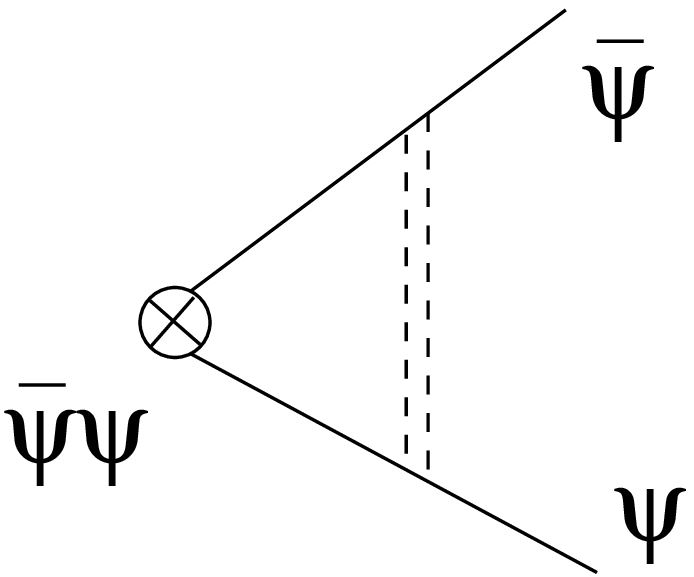}\hspace{\stretch{0.5}}\\
\hspace{\stretch{0.5}}D\hspace{\stretch{1}}E\hspace{\stretch{1}}F\hspace{\stretch{0.5}}
\caption{
Contributions of $\sigma$-boson to three point correlation function at order of $\frac{1}{N_f}$.}\label{boson_dressed_three_point}
\end{figure}
We can also calculate the spin-spin correlation function at the critical point in a similar fashion. The only difference is that the $\sigma$ boson becomes massless at critical point and contributes to the anomalous dimension of correlation functions. The contribution of $\sigma$ boson to fermion propagator and three-point correlation function are shown in Fig.\ref{boson_dressed_fermion} and Fig.\ref{boson_dressed_three_point}. After similar calculation, we find that at the critical point, 
\begin{align}
\gamma_{\bar{\psi}\psi}=-\frac{16}{\pi^2N_f}+\frac{4}{3\pi^2N_f}
\end{align}
where the second term comes from the contribution of massless $\sigma$-boson. The
change of scaling behavior during this phase transition is shown in
Fig.\ref{F:chiraltransition}.

\bibliographystyle{apsrev}
\bibliography{/home/ranying/downloads/reference/simplifiedying,/home/ranying/downloads/reference/wen/wencross,/home/ranying/downloads/reference/wen/all,/home/ranying/downloads/reference/wen/misc,/home/ranying/downloads/reference/wen/publst}

\end{document}